\RequirePackage{fix-cm}

\documentclass[smallextended]{svjour3}       % onecolumn (second format)
\smartqed  % flush right qed marks, e.g. at end of proof
\usepackage{caption}
\usepackage[top=25truemm,bottom=20truemm,left=20truemm,right=20truemm]{geometry}
\usepackage[pdftex]{graphicx}
\usepackage{amsmath,amssymb,ascmac,jumoline}
\usepackage{mediabb}
\usepackage{theorem} \theoremstyle{break} \theorembodyfont{\rmfamily}
\usepackage{framed}
\usepackage{color}
\usepackage{bm}
\usepackage{tikz}
\usepackage{here}
 \usepackage{hhline} 
 \usepackage{url}
 \usepackage{wrapfig}
\usepackage{threeparttable}
\usetikzlibrary{positioning}
\newtheorem{thm}{Theorem}[section]

\newtheorem{lem}{Lemma}[section]

\newtheorem{tyu}{Remark}[section]
\newtheorem{prf}{Proof}

\newtheorem{assum}{Assumption}

\newcommand{\eps}{\varepsilon}
\newcommand{\thus}{\raisebox{.2ex}{.}\raisebox{1.2ex}{.}\raisebox{.2ex}{.}}

\newcommand{\E}{{\rm E}}
\newcommand{\V}{{\rm Var}}
\newcommand{\cov}{{\rm Cov}}
\newcommand{\corr}{{\rm Corr}}
\newcommand{\ii}{{\rm i}\hspace{-.1em}{\rm i}}

\newcommand{\iid}{{\rm i.i.d.}}

\newcommand{\indepe}{\rotatebox{90}{$\models$}}\makeatletter
\newcommand{\figcaption}[1]{\def\@captype{figure}\caption{#1}}
\newcommand{\tblcaption}[1]{\def\@captype{table}\caption{#1}}
    
    \@addtoreset{equation}{section}
\makeatother
\allowdisplaybreaks[4]
\begin{document}

\title{State Space Model of Realized Volatility under the
Existence of Dependent Market Microstructure Noise%\thanks{Grants or other notes
}

\titlerunning{The model of RV under the Existence of Dependent MN}        % if too long for running head

\author{Toru Yano %etc.
}


\institute{T. Yano \at
              Japan \\
               \email{mf17068@shibaura-it.ac.jp}   
}

\date{Received: date / Accepted: date}

\maketitle

\begin{abstract}
Volatility means the degree of variation of a stock price which is  important in finance. Realized Volatility (RV) is an estimator of the volatility calculated using high-frequency observed prices. RV has lately attracted considerable attention of econometrics and mathematical finance. However, it is known that high-frequency data includes observation errors called market microstructure noise (MN).
Nagakura and Watanabe[2015] proposed a state space model that resolves RV into true volatility and influence of MN. In this paper, we assume a dependent MN that autocorrelates and correlates with return as reported by Hansen and Lunde[2006] and extends the results of Nagakura and Watanabe[2015]  and compare models by simulation.
\keywords{realized volatility \and market microstructure noise \and integrated volatility}
\end{abstract}
\section{Introduction}
Volatility means the degree of variation of a stock price. Integrated volatility (IV) is an integral of continuously changing instantaneous volatility over a specified period, for example, a day. IV is  important in finance, for example, option pricing, risk management, and optimal portfolio construction. Volatility is an unobservable variable and needs to be estimated from observable data such as stock price. There are classical volatility models, Generalized Autoregressive Conditional Heteroskedastic (GARCH) model and Stochastic Volatility model. These are models of autocovariance structure of volatility and can estimate the volatility based on the model from daily return. For these, see Bollerslev et al[2010] for example.

In recent years, Realized volatility (RV) has attracted attention as a method of estimating IV. RV is calculated using high frequency data such as seconds and minutes return, and is a model-free estimator that does not assume volatility model like GARCH or SV model. IV can be observed by RV, and modeling that can describe the statistical properties of RV and developments of improvement statistics of RV closer to IV are studied. Research on RV has dramatically increased in recent years due to expansion of availability of high frequency data. In regard to the survey of RV modeling, for example, see MacAleer and Medeiros[2008] and its references.

Since RV is based on the quadratic variation, in order to make estimation that is closer to IV, we have to increase the frequency data used for RV calculation. However, it is actually confirmed that when the RV is calculated using the ultra high frequency data, the RV value skyrocket. The reason why this phenomenon occurs is explained as follows. The theoretical correctness of RV is assumed that there is no measurement error in the data, but the actual data includes observation error called market microstructure noise (MN),  whereby RV includes bias. (Zhou[1996]) An improved estimator of RV that converges to IV under the existence of MN assumed to be independent noise. However, empirical studies of Hansen and Lunde[2006] have reported that MN is not independent: auto-correlate and correlate with return. Under this ``dependent'' MN, bias remains in the estimator of Zhou[1996]. This was solved by the Realized Kernel (RK) of Barndorff-Nielsen et al[2011], and thereafter, study to improve this convergence rate and empirical research using RK are continuing. For the survey of the improved estimators of RV, see  Mukherjee et al[2019] for example.

Barndorff-Nielsen and Shephard[2002] proposed a model approach of IV estimation with RV as observation. In other words, by formulating the RV as a state space time series model representing the sum of IV and observation error, it is possible to estimate and predict IV  based on the model from RV as data by Kalman filter. Meddahi[2003] generalizes this model and assumes a general class continuous time SV model (Square root stochastic autoregressive volatility model) for  instantaneous volatility and derives model of  IV and clarifies the relationship between SV model and IV parameters.

Nagakura and Watanabe[2009, 2015] expanded this model to the case where the observation price includes MN which is an independent noise and derives the covariance structure of the bias of RV due to MN. As a result, the bias follows the MA$(1)$ process. They further derive the relationships between this bias and the MN parameters, uniquely estimate the parameters of the SV model and MN from the RV data and state space model and propose a method to estimate and predict IV and bias due to MN.

In this paper, we will relax the assumption of Nagakura and Watanabe[2009, 2015], and try to extend their model assuming Hansen and Lunde [2006] type dependent MN.

\section{Integrated Volatility and Realized Volatility}
Let $p (t)$ be the logarithmic value of the asset price at time $t$ and let this follow the next Ito process.
\begin{align}
\label{brownreturn2}
dp(t)=\mu(t)dt+\sigma(t)dW(t),
\end{align}
where $W(t)$ is standard Brownian motion and  $\mu(t), \sigma(t)$ are  c\`adl\`ag and $\mathcal{F}(t)$-adapted process. 

Here, the Integrated Volatility (IV) at time $t$ is defined as
\begin{align*}
IV_t=\int^t_{t-1}\sigma^2(s)ds, t=1,2,\ldots,
\end{align*}
where the unit of $t$ is determined depending on research objective. In this paper, $t$ is interpreted as a day.  $IV_t$  is the theoretical true volatility, it represents the price variability at time $t$, it is an important risk indicator in the finance. However, this is unobservable. Our aim is to estimate/predict this IV from observable data.

As data, $m$ intraday returns $r_t$  discretely (equal intervals) observed at time $t$. where
\begin{align}
\label{obsret}
r_t^{(m)}&=p(t)-p(t-1/m).
\end{align}
By intraday returns, the Realized Volatility (RV) at time $t$ is calculated as
\begin{align}
\label{rv}
RV^{(m)}_t=\sum_{i=1}^mr^{(m)2}_{t-1+\frac{i}{m}}.
\end{align}
Under the above, from the classical stochastic calculus, we can see that the following holds. (see, e.g. Barndorff-Nielsen and Shephard[2002].)
\begin{align*}
RV^{(m)}_t\to IV_t\ {\rm in\ P}\qquad(m\to\infty).
\end{align*}
\section{Market Microstructure Noise}
On the above, RV is the consistent estimator of IV, but it is confirmed that if RV is actually calculated in a situation where $ m $ is quite large, the value of RV will skyrocket.

The following figure shows the time interval set as 30 seconds$(m=2880)$ , 1 minute$(m=1440)$, 2 minute$(m=720)$, 5 minute$(m=288)$, 10 minute$(m=144)$, 20 minute$(m=72)$, 25 minute$(m=57.6)$ and 30 minute$(m=48)$ respectively, using the dollar/yen rate data from May 1, 2009 to April 29, 2016, calculate the RV and plot the average.
\begin{figure}[H]
\begin{center}
 \includegraphics[width=9cm]{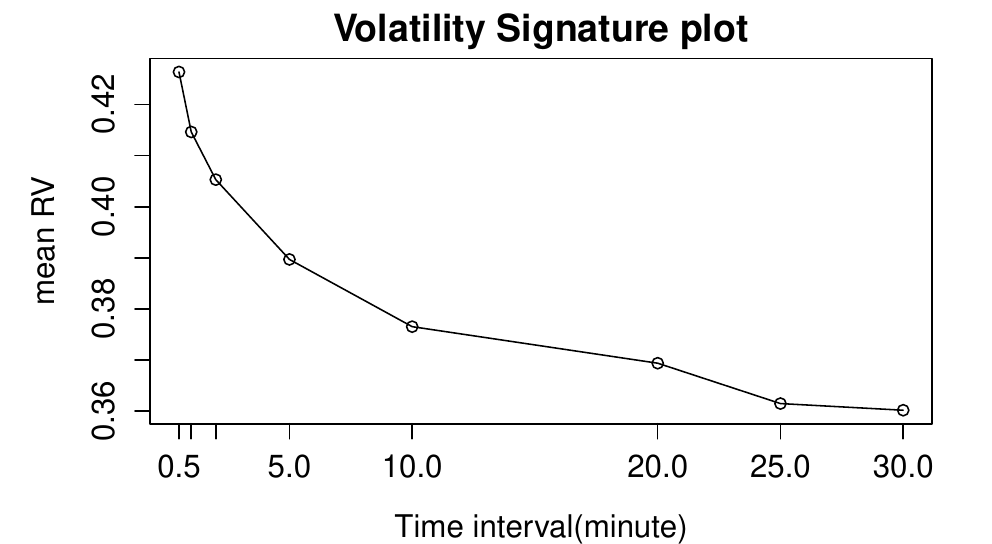}
\caption{Average change in RV due to length of time interval}
\label{mnacf}
\end{center}
\end{figure} 
Since the average of RV increases in proportion to $m$, the upward bias against RV can be confirmed, suggesting a gap between the theory of the previous section and the reality. Zhou[1996] proposed market microstructure noise (MN) as a model to explain this phenomenon. In this model, the actual observation price $p^*_t$ deviates from true log price $p(t)$ of the equation (\ref{brownreturn2}) by MN as follows.
\begin{align*}
p^*_t=p(t)+\eps_t,
\end{align*}
where $\eps_t$ is MN and is explained as observation error due to market microstructure. The observation intraday return $r^{*(m)}_t$ calculated from $p^*_t $ is as follows:
\begin{align*}
r^{*(m)}_t=p^*_t-p^*_{t-1/m}=r^{(m)}_t+e^{(m)}_t,
\end{align*}
where $\eps_t-\eps_{t-\frac{1}{m}}=e^{(m)}_t$. $RV^{*(m)}_t$ calculated by observation intraday return $r^{*(m)}_t$ is expressed as follows:
\begin{align}
RV^{*(m)}_t=\sum^m_{i=1}(r^{(m)}_{t-1+i/m}+e^{(m)}_{t-1+i/m})^2=RV_t^{(m)}+u_t^{(m)},
\end{align}
where $\displaystyle u_t^{(m)}=2\sum^m_{i=1}r^{(m)}_{t-1+i/m}e^{(m)}_{t-1+i/m}+\sum^m_{i=1}e^{(m)2}_{t-1+i/m}.$

This $RV^{*(m)}_t$ is called noise-contaminated RV (NCRV). As mention above, when $m\to\infty$ is $RV^{(m)}_t \to IV_t \ {\rm in \ P}$, but $u_t^{(m)} \nearrow \infty$ can also be proved. So if $m$ is large, the influence of bias $u_t^{(m)}$ by MN will be large. On the other hand, if $m$ is small, then the variance of the estimation error becomes large due to the decrease in the number of samples.
\section{Assumption of asset price}

\begin{assum}[true price process]
\label{assum_true_log_price}
Let the logarithmic price $ p (t) $ follow the next stochastic differential equation.
\begin{align}
dp(t)&=\sigma(t)dW(t),\\
\sigma^2(t)&=\sigma^2+\sum^p_{i=1}\omega_iP_i(f(t)),
\end{align}
where $W(t)$ is standard Brownian motion, $f(t)$ is a state-variable process independent of $W(t)$, and the functions $P_i(f(t)), (i=1,2,\ldots, p)$ satisfy 
\begin{align}
\label{sarv1}
&\E[P_i(f(t))]=0,\\
\label{sarv2}
&\V[P_i(f(t))]=1,\\
\label{sarv3}
&\cov[P_i(f(t)), P_j(f(t))]=0\qquad(i\neq j),\\
\label{sarv4}
&\E[P_i(f(t+h))\mid f(s), p(s), s\leq t]=\exp{(-\lambda_ih)P_i(f(t))}\ (\forall h>0, \lambda_i>0).
\end{align}
This model called the $p$-factor Square root stochastic autoregressive variance model (SR-SARV$(p)$).
\end{assum}

The asset price model of Assumption \ref{assum_true_log_price} is a general model and includes representative multi factor stochastic volatility models e.g. Heston model and GARCH diffusion model. (see, Meddahi[2003])

Meddahi[2003] showed that under the Assumption \ref{assum_true_log_price}, $IV_t$ follows the ARMA$(p, p)$ process:
\begin{align}
\label{IVprocess}
IV_t=c_{IV}+\sum^p_{i=1}\phi_iIV_{t-i}+\eta_t+\sum^p_{i=1}\theta_i\eta_{t-i}, \eta_t\sim {\rm WN}(\sigma_\eta^2),
\end{align}
where ${\rm WN}(\sigma_\eta^2)$ is a white noise of variance $\sigma_\eta^2$\footnote{$\eta_t$ is white noise is that: $\E[\eta_t]=0, \V[\eta_t]=const<\infty , \cov[\eta_t,\ \eta_{t-s}]=0,\ (s\neq 0).$}.

RV has an estimated error $d_t^{(m)}$ for IV. Barndorff-Nielsen and Shephard[2002] suggested that $d_t^{(m)}$ is white noise and proposed a model to estimate and predict IV from RV as follows.
From the equation (\ref{IVprocess}),
\begin{align}
\label{bsm1}
RV_t^{(m)}&=IV_t+d_t^{(m)},\\
\label{bsm2}
IV_t&=c_{IV}+\sum^p_{i=1}\phi_iIV_{t-i}+\eta_t+\sum^p_{i=1}\theta_i\eta_{t-i},
\end{align}
where $d_t^{(m)}, \eta_t$ are the white noise of variance $\sigma_d^{(m)2}, \sigma_\eta^2$.
Equation(\ref{bsm1}), (\ref{bsm2}) are called BSM($p$) model.
The parameters $(c_{IV}, \{\phi_i\}_1^p, \{\theta_i\}_1^p, \sigma^{(m)2}_d, \sigma^2_\eta)$ can be estimated by the Kalman filter and the quasi-maximum likelihood method, and  we can estimate and predict $IV_ {1: T}$ from $RV^{(m)}_{1: T}$ by Kalman smoother and Kalman filter.

\section{Assumption of MN}
\subsection{Previous study (Nagakura and Watanabe[2015])}
Nagakura and Watanabe[2015] assumed the following for MN $\eps_t$.
\begin{enumerate}
\item For all $t$, $\eps_t$ is IID noise\footnote{$\eps_t$ is IID noise is that: $\forall t, \E[\eps_t]=0, \V[\eps_t]=const<\infty , \eps_s\indepe \eps_t, (s\neq t)$. Therefore, if it is independent noise, it is white noise.}  with
$
\E[\eps_t]=0, \V[\eps_t]=\sigma_\eps^2, \V[\eps_t^2]=\omega_\eps^2<\infty.
$
\item $\eps_t$ is independent of $W(s), f(s)$ for any $s, t$.

\end{enumerate}
Under this assumption, Nagakura and Watanabe[2009, 2015] proved the following.
\begin{itemize}
\item Properties of $u_t^{(m)}$:
\begin{align*}
\E\left[u_t^{(m)}\right]&=2m\sigma_\eps^2,\ \V\left[u_t^{(m)}\right]=8\sigma^2\sigma_\eps^2+2(2m-1)\omega_\eps^2+4m\sigma_\eps^4,\\
 \cov\left[u_t^{(m)}, u_{t\pm n}^{(m)}\right]&=
\begin{cases}
\omega_\eps^2,&(n=1),\\
0,&(n\geq 2).
\end{cases}
\end{align*}
$u_t^{(m)}$ follows the next MA$(1)$ process.
\begin{align}
\label{MNprocess}
u_t^{(m)}&=c_u^{(m)}+\xi_t^{(m)}+\theta^{(m)}_u\xi_{t-1}^{(m)}, \xi_t^{(m)}\sim {\rm WN}(\sigma^{2(m)}_\xi),
\intertext{where}
c_u^{(m)}&=2m\sigma^2_\eps, \sigma^{2(m)}_\xi=\cfrac{\omega^2_\eps}{\theta^{(m)}_u},\ \theta^{(m)}_u=A-\sqrt{A^2-1},\nonumber\\
 A&=\cfrac{4\sigma^2\sigma_\eps^2}{\omega_\eps^2}+2m-1+2m\cfrac{\sigma_\eps^4}{\omega_\eps^2}.\nonumber
\end{align}
\item Each state variable is uncorrelated:$\cov\left[u_s^{(m)}, d_t^{(m)}\right]=\cov\left[u_s^{(m)}, IV_t\right]=0, \forall s,t$.
\item From the data of $RV_t^{*(m)}$, it is possible to estimate the parameters \\
$(\sigma^2, \{\omega^2_i\}_1^p, \{\lambda_i\}_1^p, \sigma^2_\eps, \omega^2_\eps)$ of SR-SARV model and MN by quasi-maximum likelihood estimation (QMLE) method.
\end{itemize}
From the above, the observed $RV_t^{*(m)}$ is decomposed into a true process $IV_t$, its estimation error $d_t^{(m)}$ and the bias due to MN $u_t^{(m)}$. From the equations(\ref{IVprocess}), (\ref{MNprocess}),  it is represented by the following linear state space model.
\begin{align}
\label{nw1}
RV_t^{*(m)}&=IV_t+u_t^{(m)}+d_t^{(m)},\\
\label{nw2}
IV_t&=c_{IV}+\sum^p_{i=1}\phi_iIV_{t-i}+\eta_t+\sum^p_{i=1}\theta_i\eta_{t-i},\\
\label{nw3}
u_t^{(m)}&=c_u^{(m)}+\xi_t^{(m)}+\theta^{(m)}_u\xi_{t-1}^{(m)},\\
\label{nw4}
\left(
\begin{array}{c}
d_t^{(m)}\\
\eta_t\\
\xi_t^{(m)}\\
\end{array}
\right)&\sim \left(
 \left[
\begin{array}{c}0\\0\\0
\end{array}\right], 
 \left[
\begin{array}{ccc}
\sigma^{2(m)}_d&0&0\\
0&\sigma^2_\eta&0\\
0&0&\sigma^{2(m)}_\xi\\
\end{array}
\right]\right).
\end{align}
This is called the NW$(p)$ model. The parameters\\ $(c_{IV}, \{\phi_i\}_1^p, \{\theta_i\}_1^p, c_u^{(m)}, \theta^{(m)}_u, \sigma^{2(m)}_d, \sigma^{2(m)}_\eta, \sigma^{2(m)}_\xi)$ can be estimated by the Kalman filter and quasi-maximum likelihood method.  
We can estimate and predict of $IV_{1:T}, u_{1:T}^{(m)}$ from $RV^{*(m)}_{1:T}$ by Kalman smoother and Kalman filter.

Nagakura and Watanabe[2009] did an empirical study using high frequency data of dollar/yen rate and found that, in the case of $p=1,2$,  they compared  the  IV estimation accuracy and the 1-ahead prediction accuracy of the NW$(p)$ model and BSM$(p)$ model, and NW(2) was the best in their results.%%%%%%%%%%%
\subsection{Dependent MN}
According to Hansen and Lunde[2006], MN has the following properties:
\begin{description}
\item[HL.1] MN correlates with return $r_t$:$\cov[r_t, \eps_t]\neq 0.$
\item[HL.2] MN has auto-correlation.
\item[HL.3]  The properties of MN have changed substantially over time.
\end{description}
So, MN is NOT independent noise. We assume MN satisfying HL.1 and HL.2  and derive the properties of $u_t^{(m)}$.

\begin{assum}[properties of MN]
\label{assum_ma_corp}
$\eps_t$ follows the following process:$\eps_t=\eps_t^{(1)}+\eps_t^{(2)}+\delta_t,$\\
where $\displaystyle \eps_t^{(1)}=f(m)\int^t_{t-\frac{1}{m}}dW(s)$ and $f(m):\mathbb{R}^1\to\mathbb{R}^1$ is a deterministic function depending on $m$.\\
$\eps_t^{(2)}$ follows  0-mean normal MA$(q)$ process:$\displaystyle \eps_t^{(2)}=\sum^q_{i=0}\Psi_i \zeta_{t-i}, \Psi_0=1,\ \zeta_t\sim\iid N(0, \sigma_\zeta),\ \forall s,t, \zeta_s\indepe W(t), \zeta_s\indepe f(t).$\\
$\delta_t$ is IID noise:$\delta_t\sim \iid(0, \sigma_\delta^2),\ \forall s,t, \delta_s\indepe W(t), \delta_s\indepe f(t), \delta_s\indepe \zeta_t.$ Suppose further:$\forall t, \omega_\delta^2:=\E\left[\delta_t^4\right]<\infty.$\\
Assume  $\sigma(t)$ is weak stationary process.
\end{assum}

\begin{tyu}
\begin{itemize}
\item $\eps_t^{(1)}$ correlates with $r_t$. (HL.1) 
\item $\eps_t^{(2)}$ has lag $q$ auto-correlation. (HL.2)
\item  We put $\delta_t$ as a variation of MN which can not be explained by $\eps_t^{(1)}$ and $\eps_t^{(2)}$. Also, if $f(m)=0$ and $\sigma_\zeta=0$, then $\eps_t=\delta_t$, so it is the same as the assumption of Nagakura and Watanabe[2009, 2015] with MN as IID noise.
\end{itemize}
\end{tyu}

\begin{tyu}
From Assumption \ref{assum_ma_corp},
\begin{align*}
\E[\eps_t]&=0,\V[\eps^{(1)}_t]=\cfrac{(f(m))^2}{m},\\
 \Omega_n:&=\cov[\eps^{(2)}_t, \eps^{(2)}_{t\pm n}]=
 \begin{cases}
 \sigma_\zeta^2\sum_{i=0}^{q-n}\Psi_i\Psi_{i+n}&(n=0,1, \ldots, q),\\
 0&(n\geq q+1),\\
 \end{cases}
\intertext{Because $\eps^{(1)}_s\indepe \eps^{(2)}_t\indepe \delta_u, \forall s,t,u$,}
\V[\eps_t]&=\cfrac{(f(m))^2}{m}+\Omega_0+\sigma_\delta^2, \cov[\eps_t, \eps_{t\pm n}]=
\begin{cases}
\Omega_n&(n=1,\ldots, q),\\
0&(n\geq q+1).
\end{cases}
\end{align*}
\end{tyu}
\begin{tyu}
Under the Assumption \ref{assum_ma_corp}, the following holds.
\begin{align}
\label{covreps}
\cov[r_t, \eps_s]=
\begin{cases}
\cfrac{f(m)\E[\sigma(t)]}{m}&(t=s),\\
0&(t\neq s).
\end{cases}
\end{align}
\end{tyu}
%%%%%%%%%

\begin{thm}
\label{ret_property}
Under the Assumption \ref{assum_true_log_price} and \ref{assum_ma_corp},
\begin{align*}
\cov\left[r^*_t, r^*_{t\pm\frac{n}{m}}\right]=
\begin{cases}
\cfrac{\sigma^2}{m}+2\left(\cfrac{(f(m))^2}{m}+\Omega_0-\Omega_1+\sigma_\delta^2\right)+\cfrac{f(m)\E[\sigma(t)]}{m}&(n=0),\\
\cfrac{f(m)\E[\sigma(t)]}{m}-\cfrac{(f(m))^2}{m}+2\Omega_1-\Omega_0-\Omega_2-\sigma_\delta^2&(n=1),\\
  2\Omega_n-\Omega_{n-1}-\Omega_{n+1}&(n\geq 2).
  \end{cases}
\end{align*}
As a result of the above, it can be seen that $r_t^*\sim{\rm MA}(q+1)$ since $\Omega_n=\Psi_n=0, (n\geq q+1)$.
\end{thm}

\begin{prf}[Theorem \ref{ret_property}]
After that, omit the superscript $(m)$.

Since the expected value of Ito integral is 0, $\E[p(t)]=\E[r_t]=0.$

Hereafter, for calculation, $e_t$ is expressed as follows.  From Assumption \ref{assum_ma_corp},  $e_t=e^{(1)}_t+e^{(2)}_t+\delta_t'$ where $\displaystyle e^{(1)}_t=f(m)\left[\int^t_{t-\frac{1}{m}}dW_(s)-\int^{t-\frac{1}{m}}_{t-\frac{2}{m}}dW_(s)\right],
e^{(2)}_t=\sum^{q+1}_{i=0}(\Psi_i-\Psi_{i-1})\zeta_{t-i}, \Psi_{-1}=\Psi_{q+1}=0,
\delta_t'=\delta_t-\delta_{t-\frac{1}{m}}.$
\begin{align}
\label{vare}
\V[e^{(1)}_t]&=\cfrac{2(f(m))^2}{m},
 \V[e^{(2)}_t]=2(\Omega_0-\Omega_1),
  \V[\delta_t']=2\sigma_\delta^2,\\
  \label{cove1}
  \E[e^{(1)}_te^{(1)}_{t\pm\frac{n}{m}}]&=
  \begin{cases}
  -\cfrac{(f(m))^2}{m}&(n=1),\\
  0&(n\geq 2),
  \end{cases}\\
  \label{cove2}
  \E[e^{(2)}_te^{(2)}_{t\pm\frac{n}{m}}]&=\sum^{q+1-n}_{i=0}(\Psi_i-\Psi_{i-1})(\Psi_{i+n}-\Psi_{i+n-1})\sigma_\zeta^2=2\Omega_n-\Omega_{n-1}-\Omega_{n+1},\\
  \label{cove3}
  \E[\delta_t'\delta_{t\pm\frac{n}{m}}']&=
   \begin{cases}
  -\sigma_\delta^2&(n=1),\\
  0&(n\geq 2).
  \end{cases}
\end{align}

From Equation (\ref{covreps}),
$
\cov[r_s, e_t]=
\begin{cases}
\cfrac{f(m)\E[\sigma(t)]}{m}&(t=s),\\
0&(t\neq s).
\end{cases}
$
And $\E[p^*(t)]=\E[r_t^*]=\E[e_t]=0.$ From Equation (\ref{vare}),
\begin{align*}
\V[e_t]&=\E\left[e_t^2\right]=\E[e^{(1)2}_t]+\E[e^{(2)2}_t]+\E[e^{(3)2}_t]=2\left(\cfrac{(f(m))^2}{m}+\Omega_0-\Omega_1+\sigma_\delta^2\right),\\
\V[r_t^*]&=\V[r_t+e_t]=\V[r_t]+\V[e_t]+2\cov[r_t, e_t]\\
&=\E\left[\left(\int^t_{t-1/m}\sigma(s)dW(s)\right)^2\right]+\V[e_t]+2\cov[r_t, e_t]\\
&=\cfrac{\sigma^2}{m}+2\left(\cfrac{(f(m))^2}{m}+\Omega_0-\Omega_1+\sigma_\delta^2\right)+\cfrac{f(m)\E[\sigma(t)]}{m}.
\end{align*}

Since $r_s\indepe e^{(2)}_t \delta_t, \forall s,t$, from Equation (\ref{covreps}),
\begin{align*}
\E\left[r_te_{t-\frac{n}{m}}\right]&=\E\left[r_te^{(1)}_{t-\frac{n}{m}}\right]=0,\ (n\geq 1), \\
\E\left[r_te_{t+\frac{n}{m}}\right]&=\E\left[r_te^{(1)}_{t+\frac{n}{m}}\right]=
\begin{cases}
\cfrac{f(m)\E[\sigma(t)]}{m}&(n=1),\\
0&(n\geq 2).
\end{cases}
\end{align*}

From Equation (\ref{cove1}),(\ref{cove2}) and (\ref{cove3}),
\begin{align*}
\cov[e_t, e_{t\pm\frac{n}{m}}]=
\begin{cases}
  -\cfrac{(f(m))^2}{m}+2\Omega_1-\Omega_0-\Omega_2-\sigma_\delta^2&(n=1),\\
  2\Omega_n-\Omega_{n-1}-\Omega_{n+1}&(n\geq 2).
  \end{cases}
\end{align*}

From above, the autocovariance of observation return $r^*_t$ is obtained as follows.
\begin{align}
\label{acvr*}
\cov\left[r^*_t, r^*_{t-n/m}\right]&=\E\left[r^*_tr^*_{t-n/m}\right]-\E\left[r^*_t\right]\E\left[r^*_{t-n/m}\right]\nonumber\\
&=\E\left[r_{t}e_{t+n/m}\right]+\E\left[e_{t}e_{t-n/m}\right] \nonumber\\
&=
\begin{cases}
  \cfrac{f(m)\E[\sigma(t)]}{m}-\cfrac{(f(m))^2}{m}+2\Omega_1-\Omega_0-\Omega_2-\sigma_\delta^2&(n=1),\\
  2\Omega_n-\Omega_{n-1}-\Omega_{n+1}&(n\geq 2).
  \end{cases}
\end{align}
\qed
\end{prf}

\begin{thm}
\label{thm1}
Under the Assumption \ref{assum_true_log_price} and \ref{assum_ma_corp},
\begin{align*}
\E \left[u_t^{(m)}\right]&=2m(\Omega_0-\Omega_1+\sigma_\delta^2)+2(f(m))^2+2f(m)\E\left[\sigma(t)\right],\\
\V \left[u_t^{(m)}\right]&=4mC_0^{(1)}+8\sum^{m-1}_{k=1}(m-k)C_k^{(1)}+4mC_0^{(2)}+4\sum^{m-1}_{k=1}(m-k)C_k^{(2)}\\
&\qquad+mC_0^{(3)}+2\sum^{m-1}_{k=1}(m-k)C_k^{(3)},\\
\cov\left[u_t^{(m)}, u_{t\pm n}^{(m)}\right]&=4mC_{mn}^{(1)}+4\sum^{m-1}_{k=1}(m-k)(C_{mn+k}^{(1)}+C_{mn-k}^{(1)})\\
&\qquad+2mC_{mn}^{(2)}+2\sum^{m-1}_{k=1}(m-k)(C_{mn+k}^{(2)}+C_{mn-k}^{(2)})\\
&\qquad+mC_{mn}^{(3)}+\sum^{m-1}_{k=1}(m-k)(C_{mn+k}^{(3)}+C_{mn-k}^{(3)}),
\end{align*}
where
\begin{align*}
%C1
C^{(1)}_n:&=\cov\left[r_te_t, r_{t\pm\frac{n}{m}}e_{t\pm\frac{n}{m}}\right]\\
&=
\begin{cases}
\cfrac{2\sigma^2}{m}(\Omega_0-\Omega_1+\sigma_\delta^2)\\
\qquad+(f(m))^2\left(\cfrac{2\sigma^2}{m^2}+2\E\left[\left(\int^t_{t-\frac{1}{m}}\sigma(s)ds\right)^2\right]-\cfrac{\E[\sigma(t)]}{m}\right)&(n=0),\\
(f(m))^2\cov\left[\int^t_{t-\frac{1}{m}}\sigma(s)ds, \int^{t\pm\frac{n}{m}}_{t\pm\frac{n\pm1}{m}}\sigma(s)ds\right]&(n\geq  1),
\end{cases}
\end{align*}
\begin{align*}
%C2
C^{(2)}_n:&=\cov\left[r_te_t, e^2_{t-\frac{n}{m}}\right]\\
&=
\begin{cases}
\cfrac{2f(m)}{m}\E\left[\sigma(t)\right]\left(-\cfrac{(f(m))^2}{m}+2(\Omega_0-\Omega_1+\sigma_\delta^2)\right)&(n=0),\\
\cfrac{2f(m)}{m}\E\left[\sigma(t)\right]\left(-\cfrac{(f(m))^2}{m}-(\Omega_0-\Omega_1+\sigma_\delta^2)\right)&(n=1),\\
0&(n\geq 2, -1\geq n),
\end{cases}
\end{align*}
\begin{align*}
%C3
C^{(3)}_n:&=\cov\left[e_t^2, e_{t\pm \frac{n}{m}}^2\right]\\
&=
\begin{cases}
8\left((\Omega_0-\Omega_1+\sigma_\delta^2)\left(\Omega_0-\Omega_1+\cfrac{(f(m))^2}{m}\right)+\cfrac{(f(m))^4}{m^2}
\right)\\
\qquad+2(\omega_\delta^2+\sigma_\delta^4)&(n=0),\\
\gamma_1+\omega_\delta^2-\sigma_\delta^4+\cfrac{4(f(m))^2}{m}&(n=1),\\
\gamma_n&(2\leq n\leq q-1),\\
2(4\Omega_q^2+\Omega_{q-1}^2-4\Omega_{q-1}\Omega_q)&(n=q),\\
2\Omega_q^2&(n=q+1),\\
0&(n\geq q+2),
\end{cases}
\intertext{where}
\gamma_n&=2(4\Omega_{n}^2+\Omega_{n-1}^2+\Omega_{n+1}^2-4\Omega_{n}\Omega_{n+1}-4\Omega_{n-1}\Omega_{n}+2\Omega_{n-1}\Omega_{n+1}).
\end{align*}

\end{thm}

\begin{lem}
\label{mainlem}
\begin{enumerate}
\item 
\begin{align*}
\epsilon_t=\sum^\infty_{i=0}\varrho_i w_{t-i}, w_t\sim {\rm WN}(\sigma_w^2),  \sum^\infty_{i=0}|\varrho_i|<\infty.
\end{align*}
\item For all $t$,
\begin{align*}
&\E\left[w_t^4\right]=:\omega_w^2<\infty, \cov\left[w_s^2, w_t^2\right]=\cov\left[w_s, w_t^3\right]=0, (s\neq t),\\
&\E\left[w_sw_tw_u^2\right]=0, (s\neq t\neq u), \E\left[w_sw_tw_uw_v\right]=0, (s\neq t\neq u\neq v).
\end{align*}
\end{enumerate}
Under the above 1,2, the following holds.
Let $a\geq b\geq c\geq d\in \mathbb{Z}$.
\begin{align*}
&\E\left[\epsilon_{t+a}\epsilon_{t+b}\epsilon_{t+c}\epsilon_{t+d}\right]\\
=&\left(\omega_w^2-3\sigma_w^4\right)\sum^\infty_{i=0}\varrho_{i+a-d}\varrho_{i+b-d}\varrho_{i+c-d}\varrho_{i}+\Omega_{c-d}\Omega_{a-b}+\Omega_{b-d}\Omega_{a-c}+\Omega_{a-d}\Omega_{b-c}
\end{align*}
where  $\Omega_k:=\cov[\epsilon_t, \epsilon_{t\pm n}]=\sigma_w^2\sum^\infty_{i=0}\varrho_{i+k}\varrho_i.$
\end{lem}

\begin{prf}[Lemma\ref{mainlem}]
\begin{align*}
&\E\left[\epsilon_{t+a}\epsilon_{t+b}\epsilon_{t+c}\epsilon_{t+d}\right]\\
=&\E\left[\left(\sum^\infty_{i=-a}\varrho_{i+a}w_{t-i}\right)\left(\sum^\infty_{j=-b}\varrho_{j+b}w_{t-j}\right)\left(\sum^\infty_{k=-c}\varrho_{k+c}w_{t-k}\right)\left(\sum^\infty_{l=-d}\varrho_{l+d}w_{t-l}\right)\right]
\end{align*}
From the above 1 and 2, consider only the terms $({\rm i})w_t^4, (\ii)w_t^2w_s^2$.
\begin{description}
\item[({\rm i})] Coefficients of $w_t^4$\\
When $i=j=k=l$, the coefficients are $\varrho_{i+a}\varrho_{i+b}\varrho_{i+c}\varrho_{i+d} \qquad(i\geq-d)$
\item[(\ii)]Coefficients of $w_t^2w_s^2$\\
\begin{itemize}
\item When $i=j\neq k=l$, the coefficients are $\varrho_{i+a}\varrho_{i+b}\varrho_{k+c}\varrho_{k+d} \qquad(i\geq-b, k\geq-d)$
\item When $i=k\neq l=j$, the coefficients are $\varrho_{i+a}\varrho_{i+c}\varrho_{j+b}\varrho_{j+d} \qquad(i\geq-c, j\geq-d)$
\item When $i=l\neq k=j$, the coefficients are $\varrho_{i+a}\varrho_{i+d}\varrho_{j+b}\varrho_{j+c} \qquad(i\geq-d, j\geq-c)$
\end{itemize}
\end{description}
Therefore,
\begin{align*}
&\E\left[\left(\sum^\infty_{i=-a}\varrho_{i+a}w_{t-i}\right)\left(\sum^\infty_{j=-b}\varrho_{j+b}w_{t-j}\right)\left(\sum^\infty_{k=-c}\varrho_{k+c}w_{t-k}\right)\left(\sum^\infty_{l=-d}\varrho_{l+d}w_{t-l}\right)\right]\\
&=\E\left[\sum^\infty_{i=-d}\varrho_{i+a}\varrho_{i+b}\varrho_{i+c}\varrho_{i+d}w_{t-i}^4\right]\\
&\qquad+\E\left[\sum^\infty_{k=-d}\varrho_{k+c}\varrho_{k+d}\left(\sum^\infty_{i=-b}\varrho_{i+a}\varrho_{i+b}-\varrho_{k+a}\varrho_{k+b}\right)w_{t-i}^2w_{t-k}^2 \right.\\
&\qquad\qquad+\sum^\infty_{i=-d}\varrho_{i+b}\varrho_{i+d}\left(\sum^\infty_{j=-c}\varrho_{j+a}\varrho_{j+v}-\varrho_{i+a}\varrho_{i+c}\right)w_{t-j}^2w_{t-i}^2\\
&\left.\qquad\qquad+\sum^\infty_{i=-d}\varrho_{i+a}\varrho_{i+d}\left(\sum^\infty_{j=-c}\varrho_{j+b}\varrho_{j+c}-\varrho_{i+b}\varrho_{i+c}\right)w_{t-i}^2w_{t-j}^2\right]\\
&=\left(\omega_w^2-3\sigma_w^4\right)\sum^\infty_{i=-d}\varrho_{i+a}\varrho_{i+b}\varrho_{i+c}\varrho_{i+d}\\
&\qquad+\sigma_w^4\left[\sum^\infty_{k=-d}\varrho_{k+c}\varrho_{k+d}\sum^\infty_{i=-b}\varrho_{i+a}\varrho_{i+b}+\sum^\infty_{i=-d}\varrho_{i+b}\varrho_{i+d}\sum^\infty_{j=-c}\varrho_{j+a}\varrho_{j+c}\right.\\
&\qquad\qquad\qquad \left.+\sum^\infty_{i=-d}\varrho_{i+a}\varrho_{i+d}\sum^\infty_{j=-c}\varrho_{j+b}\varrho_{j+c}\right]\\
&=\left(\omega_w^2-3\sigma_w^4\right)\sum^\infty_{i=0}\varrho_{i+a-d}\varrho_{i+b-d}\varrho_{i+c-d}\varrho_{i}+\Omega_{c-d}\Omega_{a-b}+\Omega_{b-d}\Omega_{a-c}+\Omega_{a-d}\Omega_{b-c}.
\end{align*}
\qed
\end{prf}
\begin{lem}
\label{lem2}
Let $\Delta_i(t)$ be adapted to filtration $\mathcal{F}(t)$, and assume $\displaystyle \E\left[\int^t_0\Delta_i^2(s)ds\right]<\infty,\  (i=1,2,3,4)$.
$s\leq t-h_2$ and $h_1,h_2>0$.
\begin{align*}
&\E\left[\int^s_{s-h_1}\Delta_1(u)dW(u)\int^s_{s-h_1}\Delta_2(u)dW(u)\int^t_{t-h_2}\Delta_3(u)dW(u)\int^t_{t-h_2}\Delta_4(u)dW(u)\right]\\
=&\E\left[\int^s_{s-h_1}\Delta_1(u)\Delta_2(u)du\int^t_{t-h_2}\Delta_3(u)\Delta_4(u)du\right].\\
\intertext{Assume $h>0$.}
&\E\left[\int^t_{t-h}\Delta_1(u)dW(u)\int^t_{t-h}\Delta_2(u)dW(u)\int^t_{t-h}\Delta_3(u)dW(u)\int^t_{t-h}\Delta_4(u)dW(u)\right]\\
=&\E\left[\int^t_{t-h}\Delta_1(u)\Delta_2(u)du\int^t_{t-h}\Delta_3(u)\Delta_4(u)du+\int^t_{t-h}\Delta_1(u)\Delta_3(u)du\int^t_{t-h}\Delta_2(u)\Delta_4(u)du\right.\\
&\qquad\left.+\int^t_{t-h}\Delta_1(u)\Delta_4(u)du\int^t_{t-h}\Delta_2(u)\Delta_3(u)du\right].
\end{align*}
\end{lem}
\begin{prf}
The integration interval $[s-h_1, s], [t-h_2, t]$ is divided by $s-h_1=s_0\leq s_1\leq \cdots \leq s_n=s, t-h_2=t_0\leq t_1\leq \cdots \leq t_n=t$. And let $D_{t_i}=W(t_{i+1})-W(t_{i})$,
\begin{align*}
&\E\left[\int^s_{s-h_1}\Delta_1(u)dW(u)\int^s_{s-h_1}\Delta_2(u)dW(u)\int^t_{t-h_2}\Delta_3(u)dW(u)\int^t_{t-h_2}\Delta_4(u)dW(u)\right]\\
=&\lim_{\max_{i}|s_{i+1}-s_i|\to 0\atop \max_{j}|t_{j+1}-t_j|\to 0}\E\left[\sum^{n-1}_{i=0}\Delta_1(s_i) D_{s_i}\sum^{n-1}_{j=0}\Delta_2(s_j) D_{s_j}\sum^{n-1}_{k=0}\Delta_3(t_k) D_{t_k}\sum^{n-1}_{l=0}\Delta_4(t_l) D_{t_l}\right]\\
\intertext{Since $D_{s_j}, D_{t_j}$ is independent from $\mathcal{F}(s_j), \mathcal{F}(t_i)$,}
=&\lim_{\max_{i}|s_{i+1}-s_i|\to 0\atop \max_{j}|t_{j+1}-t_j|\to 0}\E\left[\sum^{n-1}_{i=0}\sum^{n-1}_{j=0}\Delta_1(s_i)\Delta_2(s_j)\Delta_3(t_i)\Delta_4(t_j)\right]\E\left[D_{s_i}^2\right]\E\left[D_{t_j}^2\right]\\
=&\E\left[\int^s_{s-h_1}\Delta_1(u)\Delta_2(u)du\int^t_{t-h_2}\Delta_3(u)\Delta_4(u)du\right].\\
&\E\left[\int^t_{t-h}\Delta_1(u)dW(u)\int^t_{t-h}\Delta_2(u)dW(u)\int^t_{t-h}\Delta_3(u)dW(u)\int^t_{t-h}\Delta_4(u)dW(u)\right]\\
=&\lim_{\max_{j}|t_{j+1}-t_j|\to 0}\E\left[\sum^{n-1}_{i=0}\sum^{n-1}_{j=0}\Delta_1(t_i)\left\{\Delta_2(t_i)\Delta_3(t_j)\Delta_4(t_j)+\Delta_3(t_i)\Delta_2(t_j)\Delta_4(t_j)\right.\right.\\
&\left.\left.\hspace{6cm}+\Delta_4(t_i)\Delta_2(t_j)\Delta_3(t_j)\right\}D_{t_i}^2D_{t_j}^2\right].\\
=&\E\left[\int^t_{t-h}\Delta_1(u)\Delta_2(u)du\int^t_{t-h}\Delta_3(u)\Delta_4(u)du+\int^t_{t-h}\Delta_1(u)\Delta_3(u)du\int^t_{t-h}\Delta_2(u)\Delta_4(u)du\right.\\
&\qquad\left.+\int^t_{t-h}\Delta_1(u)\Delta_4(u)du\int^t_{t-h}\Delta_2(u)\Delta_3(u)du\right].
\end{align*}
\qed
\end{prf}

\begin{prf}[Theorem \ref{thm1}]
First, calculate $\E[u_t]$ using these, and calculate $\cov[r_se_s, r_te_t]$ for this purpose.
\begin{align*}
\cov[r_te_t, r_te_t]&=\E[r_t^2e^{(1)2}_t]+\E[r_t^2]\E[e^{(2)2}_t+\delta^2_t]-\left(\E[r_te^{(1)}_t]\right)^2\nonumber \\
\intertext{From Lemma \ref{lem2},}
&=(f(m))^2\E\left[\cfrac{2}{m}\int^t_{t-\frac{1}{m}}\sigma^2(u)du+2\left(\int^t_{t-\frac{1}{m}}\sigma(u)du\right)^2\right]\\
&\qquad+\E[r_t^2]\E[e^{(2)2}_t+\delta^2_t]-\left(\E[r_te^{(1)}_t]\right)^2\nonumber \\
&=\cfrac{2\sigma^2}{m}(\Omega_0-\Omega_1+\sigma_\delta^2)+(f(m))^2\left(\cfrac{2\sigma^2}{m^2}+2\E\left[\left(\int^t_{t-\frac{1}{m}}\sigma(s)ds\right)^2\right]-\cfrac{\E[\sigma(t)]}{m}\right).
\end{align*}
Next, let $n\geq 1$, calculate $\cov\left[r_te_t, r_{t\pm\frac{n}{m}}e_{t\pm\frac{n}{m}}\right]$.
\begin{align*}
\cov\left[r_te_t, r_{t-\frac{n}{m}}e_{t-\frac{n}{m}}\right]&=\E\left[r_tr_{t-\frac{n}{m}}e^{(1)}_te^{(1)}_{t-\frac{n}{m}}\right]-\E[r_te^{(1)}_t]\E\left[r_{t-\frac{n}{m}}e^{(1)}_{t-\frac{n}{m}}\right]\nonumber\\
\intertext{From Lemma \ref{lem2},}
&=(f(m))^2\cov\left[\int^t_{t-\frac{1}{m}}\sigma(s)ds, \int^{t-\frac{n}{m}}_{t-\frac{n+1}{m}}\sigma(s)ds\right].
\end{align*}
Similarly,
\begin{align}
\label{alpha}
C_n^{(1)}:=\cov\left[r_te_t, r_{t\pm\frac{n}{m}}e_{t\pm\frac{n}{m}}\right]=(f(m))^2\cov\left[\int^t_{t-\frac{1}{m}}\sigma(s)ds, \int^{t\pm\frac{n}{m}}_{t\pm\frac{n\pm1}{m}}\sigma(s)ds\right].
\end{align}

From the above, the average of $u_t$ is as follows.
\begin{align}
\E[u_t]&=2\sum^m_{i=1}\E[r_{t-1+i/m}e_{t-1+i/m}]+\sum^m_{i=1}\E\left[e^2_{t-1+i/m}\right]\nonumber\\
&=2m(\Omega_0-\Omega_1+\sigma_\delta^2)+2(f(m))^2+2f(m)\E\left[\sigma(t)\right]
\end{align}

Next, calculate $\V\left[u_t\right], \cov\left[u_t, u_{t\pm n}\right]$. First, calculate $\cov[r_te_t, e_s^2], \cov[e_s^2, e_t^2]$.
%C2
\begin{align*}
\cov\left[r_te_t, e_t^2\right]&=\E\left[r_te_t^3\right]-\E\left[r_te_t\right]\E\left[e_t^2\right]\\
&=\E\left[r_te_t^{(1)3}\right]-\E\left[r_te_t^{(1)}\right]\E\left[e_t^{(1)2}\right]+3\E\left[r_te_t^{(1)}\right]\left(\E\left[e_t^{(2)2}\right]+\E\left[\delta_t^2\right]\right)\\
\intertext{From Lemma \ref{lem2},}
&=\cfrac{2f(m)}{m}\E\left[\sigma(t)\right]\left(-\cfrac{(f(m))^2}{m}+2(\Omega_0-\Omega_1+\sigma_\delta^2)\right),
\end{align*}
\begin{align*}
\cov\left[r_{t-\frac{1}{m}}e_{t-\frac{1}{m}}, e_t^2\right]&=\cfrac{2f(m)}{m}\E\left[\sigma(t)\right]\left(-\cfrac{(f(m))^2}{m}-(\Omega_0-\Omega_1+\sigma_\delta^2)\right).
\end{align*}
Let $n\geq 2$,
\begin{align*}
\cov\left[r_{t-\frac{n}{m}}e_{t-\frac{n}{m}}, e_t^2\right]&=\E\left[r_{t-\frac{n}{m}}e_{t-\frac{n}{m}}^{(1)}e_t^{(1)2}\right]-\E\left[r_{t-\frac{n}{m}}e_{t-\frac{n}{m}}^{(1)}\right]\E\left[e_t^{(1)2}\right]=0.
\end{align*}
Therefore,
\begin{align}
\label{beta}
C_n^{(2)}:&=\cov\left[r_{t-\frac{n}{m}}e_{t-\frac{n}{m}}, e^2_t\right]\nonumber\\
&=
\begin{cases}
\cfrac{2f(m)}{m}\E\left[\sigma(t)\right]\left(-\cfrac{(f(m))^2}{m}+2(\Omega_0-\Omega_1+\sigma_\delta^2)\right)&(n=0),\\
\cfrac{2f(m)}{m}\E\left[\sigma(t)\right]\left(-\cfrac{(f(m))^2}{m}-(\Omega_0-\Omega_1+\sigma_\delta^2)\right)&(n=1),\\
0&(n\geq 2, -1\geq n),
\end{cases}
\end{align}

Next, we will calculate $\cov[e_t^2, e_{t\pm\frac{n}{m}}^2]$. $\cov[e_t^{(2)2}, e_{t\pm\frac{n}{m}}^{(2)2}]$ can be calculated by considering  the case of  $ \varrho_i=\Psi_i, (1\leq i\leq q), \varrho_i=0, (i\geq q+1), w_t=\zeta_t\sim N(0,\sigma_\zeta^2)$ in Lemma \ref{mainlem}. Note that since $\zeta_t$ follows the 0-mean normal distribution, $\omega_w^2=3\sigma_w^4$, so, the first term of Lemma \ref{mainlem} disappears.

From Lemma \ref{mainlem},
\begin{align*}
\E\left[\eps^{(2)4}_t\right]&=3\Omega_0^2, \E\left[\eps^{(2)2}_t\eps^{(2)2}_{t-1}\right]=\Omega_0^2+2\Omega_1^2,\\
  \E\left[\eps^{(2)3}_t\eps^{(2)}_{t-1}\right]&=3\Omega_0\Omega_1, \E\left[\eps^{(2)}_t\eps^{(2)3}_{t-1}\right]=3\Omega_0\Omega_1.
\end{align*}
Thus,
\begin{align*}
&\cov[e_t^{(2)2}, e_t^{(2)2}]\\
=&\E[\eps^{(2)4}_t]-4\E[\eps^{(2)3}_t\eps^{(2)}_{t-1}]+6\E[\eps^{(2)2}_t\eps^{(2)2}_{t-1}]-4\E[\eps^{(2)}_t\eps^{(2)3}_{t-1}]+\E[\eps^{(2)4}_{t-1}]-4(\Omega_0-\Omega_1)^2\\
=&8(\Omega_0-\Omega_1)^2.
\end{align*}

From Lemma \ref{mainlem},
\begin{align*}
&\E\left[\eps^{(2)2}_t\eps^{(2)2}_{t\pm n}\right]=\Omega_0^2+2\Omega_n^2, \E\left[\eps^{(2)2}_t\eps^{(2)}_{t+n}\eps^{(2)}_{t+n-1}+\eps^{(2)}_t\eps^{(2)}_{t-1}\eps^{(2)2}_{t+n-1}\right]=2\left(\Omega_0\Omega_1+2\Omega_{n-1}\Omega_n\right), \\
& \E\left[\eps^{(2)}_t\eps^{(2)}_{t-1}\eps^{(2)2}_{t+n}+\eps^{(2)}_{t+n}\eps^{(2)}_{t+n-1}\eps^{(2)2}_{t-1}\right]=2\left(\Omega_0\Omega_1+2\Omega_{n+1}\Omega_n\right),\\
&  \E\left[\eps^{(2)}_t\eps^{(2)}_{t-1}\eps^{(2)}_{t+n}\eps^{(2)}_{t+n-1}\right]=\Omega_1^2+\Omega_n^2+\Omega_{n+1}\Omega_{n-1}.
\end{align*}
\begin{align}
\label{turai}
\thus \gamma_n:=&\cov\left[e_t^{(2)2}, e_{t\pm \frac{n}{m}}^{(2)2}\right]\nonumber\\
=&2(4\Omega_{n}^2+\Omega_{n-1}^2+\Omega_{n+1}^2-4\Omega_{n}\Omega_{n+1}-4\Omega_{n-1}\Omega_{n}+2\Omega_{n-1}\Omega_{n+1}).
\end{align}
\begin{align*}
&\cov\left[e_t^2, e_t^2\right]\\
=&8\left(\sigma_\delta^2(\Omega_0-\Omega_1)+(\Omega_0-\Omega_1)^2+\cfrac{(f(m))^4}{m^2}+\cfrac{(f(m))^2}{m}(\Omega_0-\Omega_1+\sigma_\delta^2)+
\right)+2(\omega_\delta^2+\sigma_\delta^4).
\end{align*}
\begin{align}
\label{gamma}
C^{(3)}_n:&=\cov\left[e_t^2, e_{t\pm \frac{n}{m}}^2\right]=\E\left[e_t^2e_{t\pm \frac{n}{m}}^2\right]-\V\left[e_t^2\right]\nonumber\\
&=
\begin{cases}
\cov\left[e_t^{(1)2}, e_{t\pm \frac{1}{m}}^{(1)2}\right]+\cov\left[e_t^{(2)2}, e_{t\pm \frac{1}{m}}^{(2)2}\right]+\cov\left[e_t^{(3)2}, e_{t\pm \frac{1}{m}}^{(3)2}\right]&(n=1),\\
\cov\left[e_t^{(2)2}, e_{t\pm \frac{n}{m}}^{(2)2}\right]&(n\geq 2).
\end{cases}\nonumber\\
\thus C^{(3)}_n&=
\begin{cases}
\gamma_1+\omega_\delta^2-\sigma_\delta^4+\cfrac{4(f(m))^2}{m}&(n=1),\\
\gamma_n&(n\geq 2).
\end{cases}
\end{align}

From above, $\V\left[u_t\right], \cov\left[u_t, u_{t\pm n}\right]$ are calculated as follows.
\begin{align}
\V\left[u_t\right]&=\V\left[2\sum^m_{i=1}r_{t-1+\frac{i}{m}}e_{t-1+\frac{i}{m}}+\sum^m_{i=1}e^2_{t-1+\frac{i}{m}}\right]\nonumber \\
&=4\sum^m_{i=1}\sum^m_{j=1}\cov\left[r_{t-1+\frac{i}{m}}e_{t-1+\frac{i}{m}}, r_{t-1+\frac{j}{m}}e_{t-1+\frac{j}{m}}\right]\nonumber\\
&\qquad+\sum^m_{i=1}\sum^m_{j=1}\cov\left[e_{t-1+\frac{i}{m}}^2, e^2_{t-1+\frac{j}{m}}\right]+4\sum^m_{i=1}\sum^m_{j=1}\cov\left[r_{t-1+\frac{i}{m}}e_{t-1+\frac{i}{m}}, e^2_{t-1+\frac{j}{m}}\right]\nonumber\\
&=4mC_0^{(1)}+8\sum^{m-1}_{k=1}(m-k)C^{(1)}_k\nonumber\\
&\qquad+4mC_0^{(2)}+4\sum^{m-1}_{k=1}(m-k)C_k^{(2)}+mC^{(3)}_0+2\sum^{m-1}_{k=1}(m-k)C^{(3)}_k.\\
\cov\left[u_t, u_{t\pm n}\right]&=\cov\left[2\sum^m_{i=1}r_{t-n+\frac{i}{m}}e_{t-n+\frac{i}{m}}+\sum^m_{i=1}e^2_{t-n+\frac{i}{m}}, 2\sum^m_{i=1}r_{t+\frac{i}{m}}e_{t+\frac{i}{m}}+\sum^m_{i=1}e^2_{t+\frac{i}{m}}\right]\nonumber \\
&=4\cov\left[\sum^m_{i=1}r_{t-n+\frac{i}{m}}e_{t-n+\frac{i}{m}}, \sum^m_{i=1}r_{t+\frac{i}{m}}e_{t+\frac{i}{m}}\right]
+2\cov\left[\sum^m_{i=1}r_{t+\frac{i}{m}}e_{t+\frac{i}{m}}, \sum^m_{i=1}e^2_{t-n+\frac{i}{m}}\right]
\nonumber\\
&\qquad+2\cov\left[\sum^m_{i=1}r_{t-n+\frac{i}{m}}e_{t-n+\frac{i}{m}}, \sum^m_{i=1}e^2_{t+\frac{i}{m}}\right]
+\cov\left[\sum^m_{i=1}e^2_{t-n+\frac{i}{m}}, \sum^m_{i=1}e^2_{t+\frac{i}{m}}\right]\nonumber\\
&=+4mC^{(1)}_{mn}+4\sum^{m-1}_{k=1}(m-k)(C^{(1)}_{mn+k}+C^{(1)}_{mn-k})\nonumber\\
&\qquad +2mC^{(2)}_{mn}+2\sum^{m-1}_{k=1}(m-k)(C^{(2)}_{mn+k}+C^{(2)}_{mn-k})\nonumber\\
&\qquad+mC^{(3)}_{mn}+\sum^{m-1}_{k=1}(m-k)(C^{(3)}_{mn+k}+C^{(3)}_{mn-k}).
\end{align}
\qed
\end{prf}
\begin{thm}
\label{thm2}
Under the Assumption \ref{assum_true_log_price} and \ref{assum_ma_corp},
$\displaystyle
\cov[IV_s, d_t]=0,\ (\forall s, t),\\
\cov[IV_t, u_{t\pm n}^{(m)}]=f(m)\sum^m_{i=1}\sum^m_{j=1}\cov\left[\int^{t-1+\frac{i}{m}}_{t-1+\frac{i-1}{m}}\sigma^2(u)du, \int^{t-1\pm n+\frac{j}{m}}_{t-1\pm n+\frac{i\pm1}{m}}\sigma(u)du\right],$
\begin{align}
&\cov[d_t, u_{t\pm n}^{(m)}]\nonumber\\
=&-f(m)\sum^m_{i=1}\sum^m_{j=1}\cov\left[\int^{t-1+\frac{i}{m}}_{t-1+\frac{i-1}{m}}\sigma^2(u)du, \int^{t-1\pm n+\frac{j}{m}}_{t-1\pm n+\frac{i\pm1}{m}}\sigma(u)du\right]\nonumber\\
&\qquad+m\left(C^{(4)}_{mn}+2C^{(5)}_{mn}\right)+\sum^{m-1}_{k=1}(m-k)\left(C^{(4)}_{mn+k}+2C^{(5)}_{mn+k}+C^{(4)}_{mn-k}+2C^{(5)}_{mn-k}\right),
\end{align}
where
\begin{align}
C^{(4)}_n:=\cov\left[r_t^2, e^2_{t+\frac{n}{m}}\right]&=
\begin{cases}
2(f(m))^2\E\left[\left(\int^t_{t-\frac{1}{m}}\sigma(s)ds\right)^2\right]&(n=0, 1),\\
0&(n\geq2),
\end{cases}\nonumber\\
C^{(5)}_{n}:=\cov\left[r_{t\pm\frac{n}{m}}^2, r_te_t\right]&=
\begin{cases}
f(m)\left(3\E\left[\int^t_{t-\frac{1}{m}}\sigma^2(u)du\int^t_{t-\frac{1}{m}}\sigma(u)du\right]-\cfrac{\sigma^2\E[\sigma(t)]}{m^2}\right)&(n=0),\\
f(m)\cov\left[\int^{t\pm\frac{n}{m}}_{t\pm\frac{n\pm1}{m}}\sigma^2(u)du, \int^t_{t-\frac{1}{m}}\sigma(u)du\right]&(n\geq 1).
\end{cases}\nonumber
\end{align}
\end{thm}

\begin{prf}[Theorem \ref{thm2}]
First, calculate the following with $ \cov[IV_t, u_s]$ as the target.
$\forall s,t, \Delta>0, \Delta'>0$,
\begin{align*}
&\cov\left[\int^{t}_{t-\Delta}\sigma^2(u)du, e_s\int^{s}_{s-\Delta'}\sigma(u)dW(u)\right]\\
&=\E\left[\int^{t}_{t-\Delta}\sigma^2(u)due_s\int^{s}_{s-\Delta'}\sigma(u)dW(u)\right]-\E\left[\int^{t}_{t-\Delta}\sigma^2(u)du\right]\E\left[e_s\int^{s}_{s-\Delta'}\sigma(u)dW(u)\right]\\
&=f(m)\left(\E\left[\int^{t}_{t-\Delta}\sigma^2(u)du\int^s_{s-\Delta'}dW(u)\int^{s}_{s-\Delta'}\sigma(u)dW(u)\right]\right.\\
&\qquad\qquad\qquad\left.-\E\left[\int^{t}_{t-\Delta}\sigma^2(u)du\right]\E\left[\int^s_{s-\Delta'}dW(u)\int^{s}_{s-\Delta'}\sigma(u)dW(u)\right]\right)\\
&=f(m)\left(\E\left[\int^{t}_{t-\Delta}\sigma^2(u)du\int^s_{s-\Delta'}\sigma(u)du\right]-\E\left[\int^{t}_{t-\Delta}\sigma^2(u)du\right]\E\left[\int^{s}_{s-\Delta'}\sigma(u)du\right]\right)\\
&=f(m)\cov\left[\int^{t}_{t-\Delta}\sigma^2(u)du, \int^s_{s-\Delta'}\sigma(u)du\right].
\end{align*}
\begin{align*}
\cov\left[\int^{t}_{t-\Delta}\sigma^2(u)du, e_s^2\right]&=0. \qquad\because \sigma^2(t)\indepe e_s
\end{align*}
Therefore,
\begin{align}
\label{ivu}
&\cov[IV_t, u_{t\pm n}]\nonumber\\
=&\cov\left[\sum^m_{i=1}\int^{t-1+\frac{i}{m}}_{t-1+\frac{i-1}{m}}\sigma^2(u)du, \sum^m_{i=1}\left(e_{t\pm n-1+\frac{i}{m}}\int^{t\pm n-1+\frac{i}{m}}_{t\pm n-1+\frac{i-1}{m}}\sigma(u)dW(u)\right)+\sum^m_{i=1}e_{s-1+\frac{i}{m}}^2\right]\nonumber\\
=&f(m)\sum^m_{i=1}\sum^m_{j=1}\cov\left[\int^{t-1+\frac{i}{m}}_{t-1+\frac{i-1}{m}}\sigma^2(u)du, \int^{t-1\pm n+\frac{j}{m}}_{t-1\pm n+\frac{i\pm1}{m}}\sigma(u)du\right]
\end{align}
Next, calculate $\cov\left[r_t^2, e^2_{t+\frac{n}{m}}\right], \cov\left[r_t^2, r_{t\pm\frac{n}{m}}e_{t\pm\frac{n}{m}}\right]$ with $\cov[RV_s, u_t]$ as the target.
\begin{align*}
%LEM
C^{(4)}_n:=\cov\left[r_t^2,e^2_{t\pm\frac{n}{m}}\right]&=\E\left[r_t^2e^2_{t\pm\frac{n}{m}}\right]-\E\left[r_t^2\right]\E\left[e^2_{t\pm\frac{n}{m}}\right]=\E\left[r_t^2e^{(1)2}_{t\pm\frac{n}{m}}\right]-\E\left[r_t^2\right]\E\left[e^{(1)2}_{t\pm\frac{n}{m}}\right]\\
\intertext{From Lemma \ref{lem2},}
&=
\begin{cases}
2(f(m))^2\E\left[\left(\int^t_{t-\frac{1}{m}}\sigma(s)ds\right)^2\right]&(n=0, 1),\\
0&(n\geq2),
\end{cases}
\end{align*}
\begin{align*}
%LEM
\cov\left[r_t^2, r_{t+\frac{n}{m}}e_{t+\frac{n}{m}}\right]&=\E\left[r_t^2r_{t+\frac{n}{m}}e^{(1)}_{t+\frac{n}{m}}\right]-\E\left[r_t^2\right]\E\left[r_{t+\frac{n}{m}}e^{(1)}_{t+\frac{n}{m}}\right]\\
&=f(m)\cov\left[\int^{t+\frac{n}{m}}_{t+\frac{n-1}{m}}\sigma^2(u)du, \int^t_{t-\frac{1}{m}}\sigma(u)du\right].
\end{align*}
Similarly, $
\cov\left[r_{t+\frac{n}{m}}^2, r_te_t\right]=f(m)\cov\left[\int^{t-\frac{n}{m}}_{t-\frac{n-1}{m}}\sigma^2(u)du, \int^t_{t-\frac{1}{m}}\sigma(u)du\right], $\\
$\cov\left[r_t^2, r_te_t\right]=f(m)\left(3\E\left[\int^t_{t-\frac{1}{m}}\sigma^2(u)du\int^t_{t-\frac{1}{m}}\sigma(u)du\right]-\cfrac{\sigma^2\E[\sigma(t)]}{m^2}\right).
$
To summarize the above,
\begin{align*}
C^{(5)}_n:&=\cov\left[r_{t+\frac{n}{m}}^2, r_te_t\right]\\
&=
\begin{cases}
f(m)\left(3\E\left[\int^t_{t-\frac{1}{m}}\sigma^2(u)du\int^t_{t-\frac{1}{m}}\sigma(u)du\right]-\cfrac{\sigma^2\E[\sigma(t)]}{m^2}\right)&(n=0),\\
f(m)\cov\left[\int^{t\pm\frac{n}{m}}_{t\pm\frac{n\pm1}{m}}\sigma^2(u)du, \int^t_{t-\frac{1}{m}}\sigma(u)du\right]&(n\geq 1).
\end{cases}
\end{align*}

From the above,
\begin{align}
\label{rvu}
\cov[RV_t, u_{t+n}]&=\cov\left[\sum^m_{i=1}r^2_{t-1+\frac{i}{m}}, 2\sum^m_{i=1}r_{t+n-1+\frac{i}{m}}e_{t+n-1+\frac{i}{m}}+\sum^m_{i=1}e^2_{t+n-1+\frac{i}{m}}\right]\nonumber\\
&=2\sum^m_{i=1}\sum^m_{j=1}\cov\left[r^2_{t-1+\frac{i}{m}}, r_{t+n-1+\frac{j}{m}}e_{t+n-1+\frac{j}{m}}\right]\nonumber\\
&\qquad+\sum^m_{i=1}\sum^m_{j=1}\cov\left[r^2_{t-1+\frac{i}{m}}, e^2_{t+n-1+\frac{j}{m}}\right]\nonumber\\
&=\sum^m_{i=1}\sum^m_{j=1}\left(C^{(4)}_{n+j-i}+2C^{(5)}_{n+j-i}\right)\nonumber\\
&=m\left(C^{(4)}_{mn}+2C^{(5)}_{mn}\right)+\sum^m_{k=1}(m-k)\left(C^{(4)}_{mn+k}+2C^{(5)}_{mn+k}+C^{(4)}_{mn-k}+2C^{(5)}_{mn-k}\right).
\end{align}
Quote Nagakura and Watanabe[2009], $\forall s, t$, $\cov[IV_s, d_t]=0.$ Since $d_t=IV_t-RV_t$, from Equation (\ref{ivu}) and (\ref{rvu}),
\begin{align}
\label{du}
\cov[d_t, u_{t+n}]&=\cov[RV_t, u_{t+n}]-\cov[IV_t, u_{t+n}]\nonumber\\
&=\cov[RV_t, u_{t+n}]\nonumber\\
&\qquad-f(m)\sum^m_{i=1}\sum^m_{j=1}\cov\left[\int^{t-1+\frac{i}{m}}_{t-1+\frac{i-1}{m}}\sigma^2(u)du, \int^{t-1\pm n+\frac{j}{m}}_{t-1\pm n+\frac{i\pm1}{m}}\sigma(u)du\right].
\end{align}
\qed
\end{prf}

\begin{tyu}
In the Theorem \ref{thm1} and \ref{thm2}, when $f(m)=0, \sigma_\zeta=0$, it coincides with the result of Nagakura and Watanabe[2015] so it can be said that this is their generalization.
\end{tyu}
\subsection{Model identification}
\subsubsection{Estimation of $\E\left[u_t^{(m)}\right], \sigma^2$}
From Theorem \ref{ret_property},  $r_t^*\sim{\rm MA}(q+1)$. Let its auto-covariance be $G_n$.
\begin{align}
\label{G0}
G_0&:=\V[r_t^*]=\cfrac{\sigma^2}{m}+2(\Omega_0-\Omega_1+\sigma_\delta^2)+\cfrac{2f(m)}{m}(\E[\sigma(t)]+f(m)),\\
\label{G1}
G_1&:=\cov\left[r^*_t, r^*_{t-1/m}\right]=2\Omega_1-\Omega_{2}-\Omega_{0}-\sigma_\delta^2-\cfrac{f(m)}{m}(\E[\sigma(t)]+f(m)),\\
\label{Gn}
G_n&:=\cov\left[r^*_t, r^*_{t-n/m}\right]=2\Omega_n-\Omega_{n+1}-\Omega_{n-1}, (2\leq n\leq q+1).
\end{align}
From these, 
\begin{align}
\label{r2sigma}
\sigma^2=m\left(G_0+2\sum^{q+1}_{i=1}G_i\right)
\end{align}
Also, the following holds.
\begin{align}
\label{r2Eu}
\E[u_t]&=\E[RV_t^*]-\sigma^2.
\end{align}
From these, we can obtain $\sigma^2, \E[u_t]$ from $r_t^*$ and $RV_t^*$.
\subsubsection{Estimation of $\cov[\eps^{(2)}_t, \eps^{(2)}_{t\pm n}]$}
From equation (\ref{Gn}), $\Omega_i, (i=1,\ldots, q)$ can be calculated in order, and the following can be known inductively.
\begin{align}
\label{Omega_i}
\Omega_i&=-\sum^{q+1-i}_{j=1}jG_{i+j}, (i=1,\ldots, q).
\intertext{From equation (\ref{G1}),}
\label{Omega_0}
\Omega_0+\sigma_\delta^2&=2\Omega_1-\Omega_2-G_1-\cfrac{f(m)}{m}(\E[\sigma(t)]+f(m))\\
&=-\sum^{q+1}_{j=1}jG_{j}-\cfrac{f(m)}{m}(\E[\sigma(t)]+f(m)).\nonumber
\end{align}
\subsubsection{Model identification and parameter estimation of $u_t^{(m)}$ in case of $f(m)=0$}
When $f(m)=0, \sigma_\delta=0$, that is, when MN follows normal MA$(q)$ process, from Theorem \ref{thm2} $\cov\left[u_s^{(m)}, d_t^{(m)}\right]=\cov\left[u_s^{(m)}, IV_t\right]=0, \forall s,t$ and $\cov\left[u_t^{(m)}, u_{t\pm n}^{(m)}\right]=mC_{mn}^{(3)}+\sum^{m-1}_{k=1}(m-k)(C_{mn+k}^{(3)}+C_{mn-k}^{(3)})$
. Next, we'll calculate $n$ that $\cov\left[u_t^{(m)}, u_{t\pm n}^{(m)}\right]=0$. In the above equation, when $k=m-1$, $(m-m+1)C_{mn-m+1}^{(3)}=C_{m(n-1)+1}^{(3)}$ is the last term which becomes $0$ with respect to $n$ and since $C^{(3)}_n=0, n\geq q+2$ holds, $m(n-1)+1\geq q+2,$  thus, $n\geq \cfrac{q+1}{m}+1.$ So, $\cov\left[u_t^{(m)}, u_{t\pm n}^{(m)}\right]$ becomes $0$ for the first time $n=\left\lceil \cfrac{q+1}{m}\right\rceil+1$. ($\lceil\cdot\rceil$ is round up.) that is, 
\begin{align*}
\cov\left[u_t^{(m)}, u_{t\pm n}^{(m)}\right]=
\begin{cases}
0,&\left(n\geq \left\lceil \cfrac{q+1}{m}\right\rceil+1\right)\\
\displaystyle mC_{mn}^{(3)}+\sum^{m-1}_{k=1}(m-k)(C_{mn+k}^{(3)}+C_{mn-k}^{(3)}).
&({\rm otherwise})
\end{cases}
\end{align*}
Therefore,  $u_t^{(m)}\sim$MA$\left(\left\lceil \cfrac{q+1}{m}\right\rceil\right)$.

Nagakura and Watanabe[2015] proved $u_t^{(m)}\sim$MA$(1)$, assuming MN is IID noise. This is the case when $q=0$ in Assumption \ref{assum_ma_corp},which is  $\left\lceil \cfrac{q+1}{m}\right\rceil=1.$ This is true when $q\leq m-1$. 
On the contrary,  $u_t^{(m)}\sim$MA$(Q), Q\geq 2$ is when  $q> m-1$, which means that the autocorrelation of MN continues for more than one day. 
In the empirical study of Ubukata and Oya[2009], autocorrelation of MN continues for about 60 seconds long, so in reality it can be considered to be $q\leq m-1$.
 
 Now if $q\leq m-1$ then $u_t^{(m)}\sim{\rm MA}(1)$ so, there are parameters $c_u^{(m)}, \theta^{(m)}_u, \sigma^{2(m)}_\xi$ that can be expressed as
\begin{align*}
u_t^{(m)}&=c_u^{(m)}+\xi_t^{(m)}+\theta^{(m)}_u\xi_{t-1}^{(m)}, \xi_t^{(m)}\sim {\rm WN}(\sigma^{2(m)}_\xi).
\end{align*}
Under this expression, the following holds.
\begin{align}
\label{u_rep}
\E\left[u_t^{(m)}\right]=c_u^{(m)},  \V\left[u_t^{(m)}\right]=(1+\theta^{(m)2}_u)\sigma^{2(m)}_\xi, \nonumber\\
\cov\left[u_t^{(m)}, u_{t\pm n}^{(m)}\right]=
\begin{cases}
\theta^{(m)}_u\sigma^{2(m)}_\xi,&(n=1),\\
0,&(n\geq 2).
\end{cases}
\end{align}
On the other hand, from Theorem \ref{thm1}, 
\begin{align*}
\E\left[u_t^{(m)}\right]&=2m(\Omega_0-\Omega_1),\\
  \V\left[u_t^{(m)}\right]&=8\sigma^2(\Omega_0-\Omega_1)+mC_0^{(3)}+2\sum_{k=1}^{q+1}(m-k)C_k^{(3)},\\
 \cov\left[u_t^{(m)}, u_{t\pm n}^{(m)}\right]&=
\begin{cases}
\sum_{k=1}^{q+1}kC_k^{(3)},&(n=1),\\
0,&(n\geq 2)
\end{cases}
\end{align*}
where $C_n^{(3)}$ is in the Theorem \ref{thm1} with $\sigma_\delta=0, f(m)=0$.

Since $\sigma^2$ and $\Omega_i, i=0,\ldots, q$ can be calculated from $r_t^*$, the left side of Equation (\ref{u_rep}) is obtained from $r_t^*$. From the second and third equations of Equation (\ref{u_rep}), letting $\V\left[u_t^{(m)}\right]=:v, \cov\left[u_t^{(m)}, u_{t\pm 1}^{(m)}\right]=:c$,
$
v=(1+\theta^{(m)2}_u)\sigma^{2(m)}_\xi, 
c=\theta^{(m)}_u\sigma^{2(m)}_\xi.
$
Since $\sigma^{2(m)}_\xi>0$ and $v^2-4c^2=8\sigma^2(\Omega_0-\Omega_1)+8m(\Omega_0-\Omega_1)^2\geq 0$,
$
\theta^{(m)}_u=\cfrac{v\pm\sqrt{v^2-4c^2}}{2c}.
$
In order to satisfy the stationary condition of $u^{(m)}_t$:$|\theta^{(m)}_u|<1$, let
$
\theta^{(m)}_u=
\begin{cases}
\cfrac{v-\sqrt{v^2-4c^2}}{2c},&(c>0),\\
\cfrac{v+\sqrt{v^2-4c^2}}{2c}0,&(c<0)
\end{cases}
$ be set. And
$
\sigma^{2(m)}_\xi=\cfrac{c}{\theta^{(m)}_u}.
$
To summarize the above, it is as follows:
\begin{thm}
\label{thm3}
Under the Assumption \ref{assum_true_log_price} and \ref{assum_ma_corp}, when $f(m)=0, \sigma_\delta=0, q\leq m$,
\begin{align*}
u_t^{(m)}&=c_u^{(m)}+\xi_t^{(m)}+\theta^{(m)}_u\xi_{t-1}^{(m)}, \xi_t^{(m)}\sim {\rm WN}(\sigma^{2(m)}_\xi).
\end{align*}
and 
\begin{align*}
c_u^{(m)}&=\E[RV^*_t]-\sigma^2,
\theta^{(m)}_u=\begin{cases}
\cfrac{v-\sqrt{v^2-4c^2}}{2c},&(c>0),\\
\cfrac{v+\sqrt{v^2-4c^2}}{2c}0,&(c<0),
\end{cases}
\sigma^{2(m)}_\xi=\cfrac{c}{\theta^{(m)}_u}.
\end{align*}
where
\begin{align*}
v&=8\sigma^2(\Omega_0-\Omega_1)+mC_0^{(3)}+2\sum_{k=1}^{q+1}(m-k)C_k^{(3)},\\
 c&=\sum_{k=1}^{q+1}kC_k^{(3)}, \sigma^2=m\left(\V\left[r^*_t\right]+2\sum^{q+1}_{i=1}\cov\left[r^*_t, r^*_{t-i/m}\right]\right),\\
C^{(3)}_n&=
\begin{cases}
8(\Omega_0-\Omega_1)^2&(n=0),\\
2(4\Omega_{n}^2+\Omega_{n-1}^2+\Omega_{n+1}^2-4\Omega_{n}\Omega_{n+1}-4\Omega_{n-1}\Omega_{n}+2\Omega_{n-1}\Omega_{n+1})&(1\leq n\leq q+1),
\end{cases}\\
\Omega_i&=
\begin{cases}
-\sum^{q+1-i}_{j=1}j\cov\left[r^*_t, r^*_{t-(i+j)/m}\right]&(i=0,\ldots, q),\\
0&(i\geq q+1).
\end{cases}
\end{align*}
\end{thm}
The parameters of linear state space model (\ref{nw1}), (\ref{nw2}), (\ref{nw3}) and (\ref{nw4}) can be estimated as follows.
\begin{enumerate}
\item Given $ \{\omega^2_i\}_1^p, \{\lambda_i\}_1^p$.
\item Calculate autocovariance of observation return $r_t^*$ and decide $q$ to be $G_n=0, n\geq q+1$.
\item From Equation (\ref{r2sigma}), (\ref{r2Eu}), (\ref{Omega_i}) and (\ref{Omega_0}), letting $f(m)=\sigma_\delta^2=0$ and using observation return $r_t^*$ calculate  $\sigma^2, \E\left[u_t^{(m)}\right](=c_u^{(m)})$ and $\{\Omega_i\}_0^q$.
\item By the method of Meddahi[2003], calculate $c_{IV}, \{\phi_i\}_1^p, \{\theta_i\}_1^p, \sigma^{2(m)}_d$ and $\sigma^2_\eta$ using $\sigma^2, \{\omega^2_i\}_1^p$ and $\{\lambda_i\}_1^p$.
\item From Theorem \ref{thm3}, calculate $\theta_u^{(m)}$ and $\sigma_\xi^{2(m)}$ from $\{\Omega_i\}_0^q$.
\item From NCRV$RV^*_t$ as data and linear state space model (\ref{nw1}), (\ref{nw2}), (\ref{nw3}) and (\ref{nw4}), estimate $\{\omega^2_i\}_1^p, \{\lambda_i\}_1^p$ by QMLE.
\end{enumerate}
\begin{tyu}
Nagakura and Watanabe[2015] estimates the parameters $(\sigma^2, \{\omega^2_i\}_1^p, \{\lambda_i\}_1^p, \sigma^2_\eps, \omega^2_\eps)$ of SR-SARV model and MN from data  $RV_t^*$ and  state space model (\ref{nw1}), (\ref{nw2}), (\ref{nw3}) and (\ref{nw4}) by QMLE. And calculate \\
$(c_{IV}, \{\phi_i\}_1^p, \{\theta_i\}_1^p, c_u^{(m)}, \theta^{(m)}_u, \sigma^{2(m)}_d, \sigma^2_\eta, \sigma^{2(m)}_\xi)$ using estimated parameters of SR-SARV model and MN.

On the other hand, in the above proposal method,  it is possible to calculate $\sigma^2, \Omega_i,\ (i=0,\ldots, q+1)$ and $(c_u^{(m)}, \theta^{(m)}_u, \sigma^{2(m)}_\xi)$  of Equation (\ref{nw2}) from the data $RV_t^*$ and $r_t^*$ by estimating its autocovariance $\cov\left[r^*_t, r^*_{t-(i+j)/m}\right],\ (i=0,\ldots, q)$. This advantage will be described below.

\begin{itemize}
\item Potential theoretical structures:$\sigma^2$(Average of true volatility), $\Omega_i$(Time structure of MN),\\ $(c_u^{(m)}, \theta^{(m)}_u, \sigma^{2(m)}_\xi)$(Structure of bias due to MN) can be calculated from data.
\item The parameters estimated by QMLE from $RV_t^*$ (Only $(\{\omega^2_i\}_1^p, \{\lambda_i\}_1^p)$.) 
\end{itemize}

\end{tyu}
\subsubsection{Model identification and parameter estimation of $u_t^{(m)}$ in case of $f(m)=o\left(m^\frac{1}{2}\right)$}
Here, we assume $f(m)=o\left(m^{\frac{1}{2}}\right)$, $\sigma_\delta=0$ and $q<m$. Furthermore, to make explicit $\displaystyle \cov\left[\int^t_{t-\frac{1}{m}}\sigma(s)ds, \int^{t\pm\frac{n}{m}}_{t\pm\frac{n\pm1}{m}}\sigma(s)ds\right]$ 
 and $\displaystyle \cov\left[\int^{t-1+\frac{i}{m}}_{t-1+\frac{i-1}{m}}\sigma^2(u)du, \int^{t-1\pm n+\frac{j}{m}}_{t-1\pm n+\frac{i\pm1}{m}}\sigma(u)du\right]$ that appear in Theorems \ref{thm1} and \ref{thm2}. These can not be obtained under the assumption \ref{assum_true_log_price}.  Because SR-SARV model defines $\sigma^2(t)$, but it requires $\sigma(t)$ for these calculations. 

So, we use an approximation technique called the delta method. 
Function $g(x,y)$ that depends on two variables, $x,y$, the Taylor series to second order about the point (a, b) is
$$g(x,y)\approx g(a,b)+\left[(x-a)\frac{\partial}{\partial x}+(y-b)\frac{\partial}{\partial y}\right]g(a,b)+\frac{1}{2}\left[(x-a)\frac{\partial}{\partial x}+(y-b)\frac{\partial}{\partial y}\right]^2g(a,b).$$
Let $x, y$ are random variables$X, Y$, $a, b$ are $\E[X], \E[Y]$ and take expected values on both sides,
$$\E[g(X,Y)]\approx g(a, b)+\frac{1}{2}g_{xx}(a,b)\V(X)+\frac{1}{2}g_{yy}(a,b)\V(Y)+g_{xy}(a,b)\cov[X,Y]$$

Let $g(x,y)=\sqrt{xy}$, we have 
\begin{align*}
  \E[\sigma(t+h)\sigma(t)]&\approx \sigma^2+\frac{1}{4\sigma^2}\sum^p_{i=1}\omega_i^2(\exp(-\lambda_ih)-1),\\
  \cov[\sigma(t+h), \sigma(t)]&\approx \frac{1}{4\sigma^2}\sum^p_{i=1}\omega_i^2\exp(-\lambda_ih)-\frac{1}{64\sigma^2}\sum^p_{i=1}\omega_i^2.\\
\text{Similarly,Let $g(x,y)=x\sqrt{y}$, we have}\\
\cov[\sigma^2(t), \sigma(t+h)]&\approx \frac{1}{\sigma}\sum^p_{i=1}\omega^2_i\exp(-\lambda_ih).
\end{align*}
Here, using the approximation $\int^{t\pm \frac{n}{m}}_{t\pm \frac{n\pm 1}{m}} \sigma(s) ds\approx \frac{1}{m}\sigma(t\pm \frac{n}{m})$, we have 
\begin{align*}
  \cov\left[\int^t_{t-\frac{1}{m}}\sigma(s)ds, \int^{t\pm\frac{n}{m}}_{t\pm\frac{n\pm1}{m}}\sigma(s)ds\right]&\approx \frac{1}{m^2}\cov[\sigma(t\pm \tfrac{n}{m}), \sigma(t)],\\
\cov\left[\int^{t-1+\frac{i}{m}}_{t-1+\frac{i-1}{m}}\sigma^2(u)du, \int^{t-1\pm n+\frac{j}{m}}_{t-1\pm n+\frac{i\pm1}{m}}\sigma(u)du\right]&\approx \frac{1}{\sigma m^2}\sum_{k=1}^p\omega^2_k\exp\left\{-\lambda_k(n+\tfrac{j-i}{m})\right\}.
\end{align*}
Then, from Theorem \ref{thm2},
\begin{align*}
  \cov[IV_t, u_{t\pm n}^{(m)}]&=f(m)\sum^m_{i=1}\sum^m_{j=1}\cov\left[\int^{t-1+\frac{i}{m}}_{t-1+\frac{i-1}{m}}\sigma^2(u)du, \int^{t-1\pm n+\frac{j}{m}}_{t-1\pm n+\frac{i\pm1}{m}}\sigma(u)du\right]\\
  &\approx \frac{f(m)}{\sigma m^2}\sum^m_{i=1}\sum^m_{j=1}\sum_{k=1}^p\omega^2_k\exp\left\{-\lambda_k(n+\tfrac{j-i}{m})\right\}\\
  &=\frac{f(m)}{\sigma m^2}\sum_{k=1}^p\omega^2_k\exp(-\lambda_kn)\left[m+\sum^{m-1}_{i=1}i\left\{\exp\left(-\lambda_k\tfrac{m-i}{m}\right)+\exp\left(\lambda_k\tfrac{m-i}{m}\right)\right\}\right]\\
  &=\frac{f(m)}{\sigma m}\sum_{k=1}^p\omega^2_k\left[1+\frac{2}{m}\cosh\left(\lambda_k(1+\tfrac{1}{m})\right)-2\cosh\left(\tfrac{2\lambda_k}{m})\right)+2(1-\tfrac{1}{m})\cosh\left(\tfrac{2\lambda_k}{m})\right)\right]\\
  &=O\bigl(f(m)m^{-1}\bigr).
\end{align*}

Also, the aforementioned approximation 
$$\cov\left[\int^t_{t-\frac{1}{m}}\sigma(s)ds, \int^{t\pm\frac{n}{m}}_{t\pm\frac{n\pm1}{m}}\sigma(s)ds\right]\approx \frac{1}{m^2}\cov[\sigma(t\pm \tfrac{n}{m}), \sigma(t)]=O(\{f(m)\}^2m^{-2}\exp(\lambda_{\rm min} \tfrac{n}{m}))$$
where $\lambda_{\rm min}=\min_{i=1,\ldots, p}\lambda_i$.

Using this, we approximate part $C^{(1)}$ of $\cov\left[u_t^{(m)}, u_{t\pm n}^{(m)}\right]$ in Theorem \ref{thm1}:$4mC_{mn}^{(1)}+4\sum^{m-1}_{k=1}(m-k)(C_{mn+k}^{(1)}+C_{mn-k}^{(1)})$ as follows.
\begin{align*}
  &4mC_{mn}^{(1)}+4\sum^{m-1}_{k=1}(m-k)(C_{mn+k}^{(1)}+C_{mn-k}^{(1)})\\
  &\approx O\bigl(\{f(m)\}^2m^{-2}\bigl\{e^{-\lambda_{\rm min}n}+e^{-\tfrac{\lambda_{\rm min}}{m}}(e^{-\lambda_{\rm min}}-me^{-\tfrac{\lambda_{\rm min}}{m}}+m-1)+e^{\tfrac{\lambda_{\rm min}}{m}}(e^{\lambda_{\rm min}}-me^{\tfrac{\lambda_{\rm min}}{m}}+m-1)\bigr\}\bigr)\\
  &=O\bigl(\{f(m)\}^2m^{-2}\bigr)
\end{align*}

Similarly, $C^{(4)}_n$ and $C^{(5)}_n$ in Theorem \ref{thm2} are as follows:
\begin{align*}
C^{(4)}_n:=\cov\left[r_t^2, e^2_{t+\frac{n}{m}}\right]&=
\begin{cases}
2(f(m))^2\E\left[\left(\int^t_{t-\frac{1}{m}}\sigma(s)ds\right)^2\right]&(n=0, 1)\\
0&(n\geq2)
\end{cases}\nonumber\\
&\approx \begin{cases}
2\left(\frac{\sigma f(m)}{m}\right)^2&(n=0, 1)\\
0&(n\geq2),
\end{cases}\nonumber\\
C^{(5)}_{n}:=\cov\left[r_{t\pm\frac{n}{m}}^2, r_te_t\right]&=
\begin{cases}
f(m)\left(3\E\left[\int^t_{t-\frac{1}{m}}\sigma^2(u)du\int^t_{t-\frac{1}{m}}\sigma(u)du\right]-\cfrac{\sigma^2\E[\sigma(t)]}{m^2}\right)&(n=0)\\
f(m)\cov\left[\int^{t\pm\frac{n}{m}}_{t\pm\frac{n\pm1}{m}}\sigma^2(u)du, \int^t_{t-\frac{1}{m}}\sigma(u)du\right]&(n\geq 1)
\end{cases}\nonumber\\
&\approx O(f(m)m^{-2})
\end{align*}

Finally, applying the above to Theorems \ref{thm1},\ref{thm2}, we have
\begin{align*}
\E \left[u_t^{(m)}\right]&=2m(\Omega_0-\Omega_1)+2(f(m))^2+2f(m)\E\left[\sigma(t)\right],\\
\V \left[u_t^{(m)}\right]&=8(\sigma^2+f(m)\E\left[\sigma(t)\right])(\Omega_0-\Omega_1)-4(3m-1)\cfrac{(f(m))^3}{m^2}\E\left[\sigma(t)\right]\\
&\qquad+mC_0^{(3)}+2\sum^{q+1}_{k=1}(m-k)C_k^{(3)}+o\left(1\right),\\
\cov\left[u_t^{(m)}, u_{t\pm n}^{(m)}\right]&=
\begin{cases}
-4\cfrac{f(m)}{m^2}\E\left[\sigma(t)\right]\left((f(m))^2+m(\Omega_0-\Omega_1)\right)\\
\qquad+\sum^{q+1}_{k=1}kC_k^{(3)}+o\left(1\right),&(n=1),\\
o\left(1\right)&(n\geq 2),
\end{cases}
\end{align*}
where
\begin{align*}
C^{(3)}_n=\begin{cases}
8\left((\Omega_0-\Omega_1)\left(\Omega_0-\Omega_1+\cfrac{(f(m))^2}{m}\right)+\cfrac{(f(m))^4}{m^2}
\right)\\
&(n=0),\\
\gamma_1+\cfrac{4(f(m))^2}{m}&(n=1),\\
\gamma_n&(2\leq n\leq q-1),\\
2(4\Omega_q^2+\Omega_{q-1}^2-4\Omega_{q-1}\Omega_q)&(n=q),\\
2\Omega_q^2&(n=q+1),\\
0&(n\geq q+2),
\end{cases}
\end{align*}
where $\gamma_n=2(4\Omega_{n}^2+\Omega_{n-1}^2+\Omega_{n+1}^2-4\Omega_{n}\Omega_{n+1}-4\Omega_{n-1}\Omega_{n}+2\Omega_{n-1}\Omega_{n+1}),$ and also, 
\begin{align*}
\cov[IV_s, d_t]&=0,\ (\forall s, t), \cov[IV_t, u_{t\pm n}^{(m)}]=o\left(1\right), \cov[d_t, u_{t\pm n}^{(m)}]=o\left(1\right).
\end{align*}
Ignoring $o\left(1\right)$, when $f(m)$ and $\E\left[\sigma(t)\right]$ are given, from previous subsection,   using the data of $r_t^*$, $\Omega_i, i=0,\ldots, q$ can be calculated.  And from the above equation $\V \left[u_t^{(m)}\right], \cov\left[u_t^{(m)}, u_{t\pm 1}^{(m)}\right]$ can be calculated.

Here, $\E[\sigma(t)]$ appears. So, we use next delta method for $\E[\sigma(t)]$.

Random variable $X$, for second order differential function
 $g$, consider the following Taylor expansion around the average $\E[X]$ of $X$:
\begin{align*}
g(X)&\approx g(\E[X])+g'(\E[X])(X-\E[X])+\cfrac{g''(E[X])}{2} (X-\E[X])^2.
\end{align*}
Taking the expected value on both sides, we have
\begin{align*}
\E[g(X)]&\approx g(\E[X])+g'(\E[X])\E[X-\E[X]]+\cfrac{g''(E[X])}{2} \E[(X-\E[X])^2]\\
&=g(\E[X])+\cfrac{g''(E[X])}{2} \V[X].
\end{align*}

We use this to compute $\E[\sigma(t)]$. 
\begin{align*}
\intertext{Let $g(x)=\sqrt{x}$. we have}
\E[\sigma(t)]&=\E\left[g(\sigma^2(t))\right]\nonumber\\
&\approx g\left(\E\left[\sigma^2(t)\right]\right)+\cfrac{g''\left(\E\left[\sigma^2(t)\right]\right)}{2}\V[\sigma^2(t)]\nonumber\\
&=\sqrt{\sigma^2}-\cfrac{1}{2}\left(\cfrac{1}{4\sigma^3}\right)\sum^p_{i=1}\omega_i^2\nonumber\\
&=\sigma-\cfrac{\sum^p_{i=1}\omega_i^2}{8\sigma^3}.
\end{align*}
By this approximation, $\E[\sigma(t)]$ can be represented by $\sigma^2, \{\omega^2_i\}_1^p$. When parameter estimation is performed below, the above expression is used instead of $\E[\sigma(t)]$.

From Equation (\ref{Omega_0}), we have
\begin{align*}
\Omega_0=\cfrac{mG_0-\sigma^2+2m\Omega_1-2f(m)\E[\sigma(t)]-2(f(m))^2}{2m}.
\end{align*}
For $\Omega_0>0$, we restrict the range of $f(m), \E[\sigma(t)]$ as follows:
\begin{align}
\label{res_f}
\cfrac{-\E[\sigma(t)]-\sqrt{\E[\sigma(t)]^2-A}}{2}&\leq f(m)\leq\cfrac{-\E[\sigma(t)]+\sqrt{\E[\sigma(t)]^2-A}}{2},\\
\label{res_es}
\E[\sigma(t)]^2&\geq A,
\end{align}
where $A=-2mG_0-4m\Omega_1+2\sigma^2.$ Then, the parameters of linear state space model (\ref{nw1}), (\ref{nw2}), (\ref{nw3}) and (\ref{nw4}) can be estimated as follows:
\begin{enumerate}
\item Given $\{\omega^2_i\}_1^p, \{\lambda_i\}_1^p$ and $f(m)$ within the range of the Equation (\ref{res_f}) and (\ref{res_es}).
\item Compute the autocovariance of observation return $r_t^*$ and decide $q$ to be $G_n=0, n\geq q+1$.
\item From the observation return $r_t^*$, calculate $\sigma^2, \E\left[u_t^{(m)}\right](=c_u^{(m)})$ and $\{\Omega_i\}_0^q$ by the Equations (\ref{r2sigma}), (\ref{r2Eu}), (\ref{Omega_i}) and (\ref{Omega_0}) 
 with $\sigma_\delta^2=0$.
\item Quote Meddahi[2003], calculate $c_{IV}, \{\phi_i\}_1^p, \{\theta_i\}_1^p, \sigma^{2(m)}_d$ and $\sigma^2_\eta$ using $\sigma^2, \{\omega^2_i\}_1^p$ and $\{\lambda_i\}_1^p$.
\item Similar to Theorem \ref{thm3}, calculate $\theta_u^{(m)}$ and $\sigma_\xi^{2(m)}$ from $\V \left[u_t^{(m)}\right]$ and $\cov\left[u_t^{(m)}, u_{t\pm n}^{(m)}\right]$. Here, these can be represented by $ \{\omega^2_i\}_1^p, \{\lambda_i\}_1^p$ and $f(m)$.
\item From NCRV$RV^*_t$ as data and linear state space model (\ref{nw1}), (\ref{nw2}), (\ref{nw3}) and (\ref{nw4}), estimate $\{\omega^2_i\}_1^p, \{\lambda_i\}_1^p$ and $f(m)$ by QMLE.
\end{enumerate}

\begin{tyu}
Since there are two proposed methods in the above, in order to distinguish, we call the linear state space model (\ref{nw1}), (\ref{nw2}), (\ref{nw3}) and (\ref{nw4}) whose parameters are estimated by the above method  as the weak-f($p$) model and whose parameters estimated by Theorem \ref{thm3} as the zero-f($p$) model. $p$ is the factor number of the SR-SARV model.

Here, we summarize the features of each model.
\begin{description}
\item[NW model:]Assume MN$=$IID noise.\\
Estimation the parameters of $\sigma^2(t)$ and MN by QMLE from data $RV_t^*$. From these, parameters of $IV_t, u^{(m)}_t$ and $d_t^{(m)}$.
\item[zero-f model:]Assume MN$=$normal MA($q$).\\
$\sigma^2, \E[u_t^{(m)}]$ and the parameters of $u^{(m)}_t$ are obtained from the data $r_t^*$. The parameters of $IV_t, d_t^{(m)}$ can be estimated by QMLE in the same way as NW model.
\item[weak-f  model:]Assume MN$=$normal MA($q$)$\ +\ $Correlation with $r_t$ and $m$ is large.\\
From data $r_t^*$ and $RV_t^*$, calculate $\sigma^2$ and $\E[u_t^{(m)}]$. Represent the parameters of  $IV_t, u^{(m)}_t$ and $d_t^{(m)}$ by the parameters of $\sigma^2(t)$ ($\{\omega^2_i\}_1^p, \{\lambda_i\}_1^p$) and $f(m)$. From data $RV_t^*$, estimate $\{\omega^2_i\}_1^p, \{\lambda_i\}_1^p$ and  $f(m)$ by QMLE.
\end{description}
Note that  zero-f and weak-f assume constrains on $f(m)$ and $\sigma_\delta^2$, but no constraints are assumed for estimation of average of $\sigma^2(t)$ and $u_t^{(m)}$.
\end{tyu}

\section{Model comparison}
We compare the IV estimation and 1-ahead prediction accuracy of the NW model and the proposal model, according to Nagakura and Watanabe[2015], by the following method.
\begin{enumerate}
\item IV estimation accuracy (in sample)\\
Let the number of days of usage data be $N$. We estimate the parameters of each model by using all the data of $N$ day, and estimate $\widehat{IV}_{1:N}$ by Kalman Smoother from each estimated model and data $RV_{1:N}^*$.
Evaluation of the estimation accuracy is based on the following MSE, QLIKE (Patton[2015]):
$\displaystyle
{\rm MSE}=\frac{1}{N}\sum^N_{i=1}\left(IV_i-\widehat{IV}_i\right)^2, {\rm QLIKE}=\frac{1}{N}\sum^N_{i=1}\left(\log{\left|\widehat{IV}_i\right|}+\cfrac{IV_i}{\left|\widehat{IV}_i\right|}\right)
$
 and adjusted coefficient of determination $R^2$ of Mincer-Zarnowitz regression:
 $\displaystyle
 IV_t=\hat{a}+\hat{b}\widehat{IV}_t.
 $
MSE is the absolute error loss function, and QLIKE is the relative error loss function. These and Mincer-Zarnowitz regression are proposed as robust evaluation indices in prediction accuracy comparison by Patton[2015].
\item 1-ahead prediction accuracy (out sample) \\
Let $P<N$. Using the data on the $N-P$ day from day 1, estimate the parameters of each model and predict $\widetilde{IV}_{N-P+1}$ by Kalman filter with each estimated model and data $RV^*_{1: N-P}$. Next, shift the data by one day and then estimate the parameter and predict $\widetilde{IV}_{N-P+2}$ from the data $RV^*_{2: N-P+1}.$ Repeat this finally, we predict $\widetilde{IV}_N$ from each estimated model and data $RV^*_{P: N-1}$. Thus, we obtain the 1-ahead predicted sequence $\widetilde{IV}_{N-P+1:N}$.  1-ahead prediction accuracy is also evaluated in the same way as above.
\end{enumerate}
\subsection{Comparison by simulation}
\begin{enumerate}
\item Let $N=$2000, 24 hours a day,  the observation interval is $1$ min $(m=1440)$.
\item Generate $p(t), \sigma^2(t)$ every second with the following Heston model (1-factor).
\begin{align*}
dp(t)&=-\frac{\sigma^2(t)}{2}dt+\sigma(t)dW_1(t),\\
 d\sigma^2(t)&=\kappa(\sigma^2-\sigma^2(t))dt+\gamma \sigma(t)dW_2(t),\\
  dW_1(t)dW_2(t)&=\rho dt.
\end{align*}
The parameters $(\kappa, \sigma^2, \gamma, \rho)$ are set to $(-\log{(0.98)}, 0.5, 0.25, 0)$ by quoting Barndorff-Nielsen and Shephard[2002].
\item  Calculate $IV_t$ from $\sigma^2(t)$.
\item $\eps_t$:From the Assumption \ref{assum_ma_corp}, generate $\eps_t$ as follows:
\begin{align*}
\eps_t&=\eps^{(1)}_t+\eps^{(2)}_t+\delta_t,\\
\eps^{(1)}_t&=fm^{\frac{1-\alpha}{2}}\left[W_1(t)-W_1\left(t-\frac{1}{m}\right)\right],\\
\eps^{(2)}_t&=\zeta_t+\Psi_1\zeta_{t-1}+\Psi_2\zeta_{t-2}, \zeta_t\sim N(0, \sigma_\zeta^2), \\\delta_t&\sim  N(0, \sigma_\delta^2).
\end{align*}
The parameters $(f, \alpha, \Psi_1, \Psi_1, \sigma_\zeta,\sigma_\delta)$ are set to $(0.01, 0.6, 0.3, 0.01017046,10^{-4})$.
\item $p^*_t$:Generate with generated $\eps_t$ and $p(t)$ every 60 seconds:
$p^*_t=p(60t)+\eps_t.$
 \end{enumerate}
We use this $p^*_t$ as data and calculate NCRV from data. The following figure \ref{quasi_ncrv} is a plot of calculated NCRV (red dots) and true IV (gray line).
\begin{figure}[H]
\begin{center}
\includegraphics[width=12cm]{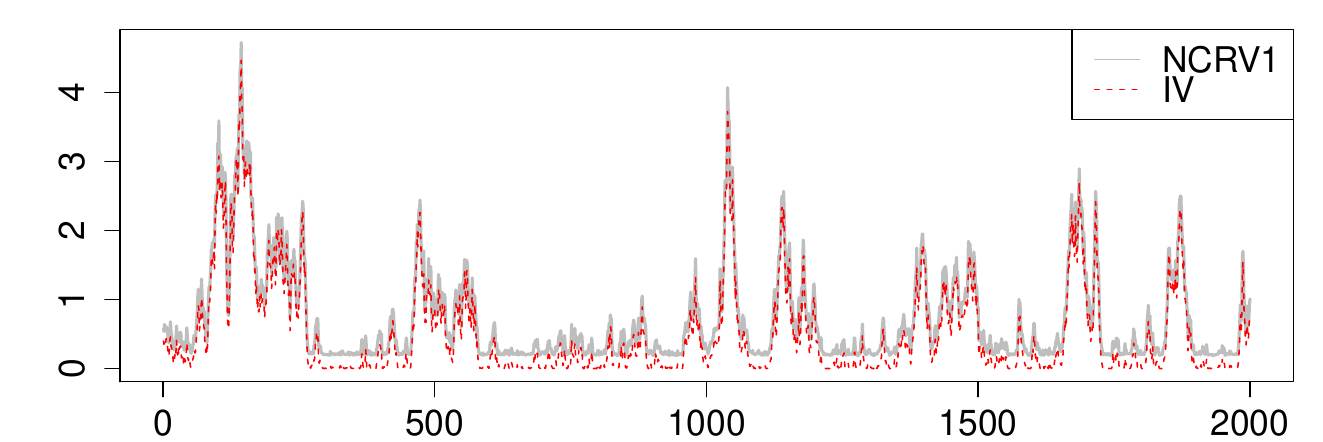}
\caption{NCRV estimated from $r_t^*$ every minute}
\label{quasi_ncrv}
\end{center}
\end{figure}
NCRV is greatly influenced by bias due to MN, and is estimated to be overall larger than IV.
 
\begin{figure}[H]
\begin{center}
\includegraphics[width=12cm]{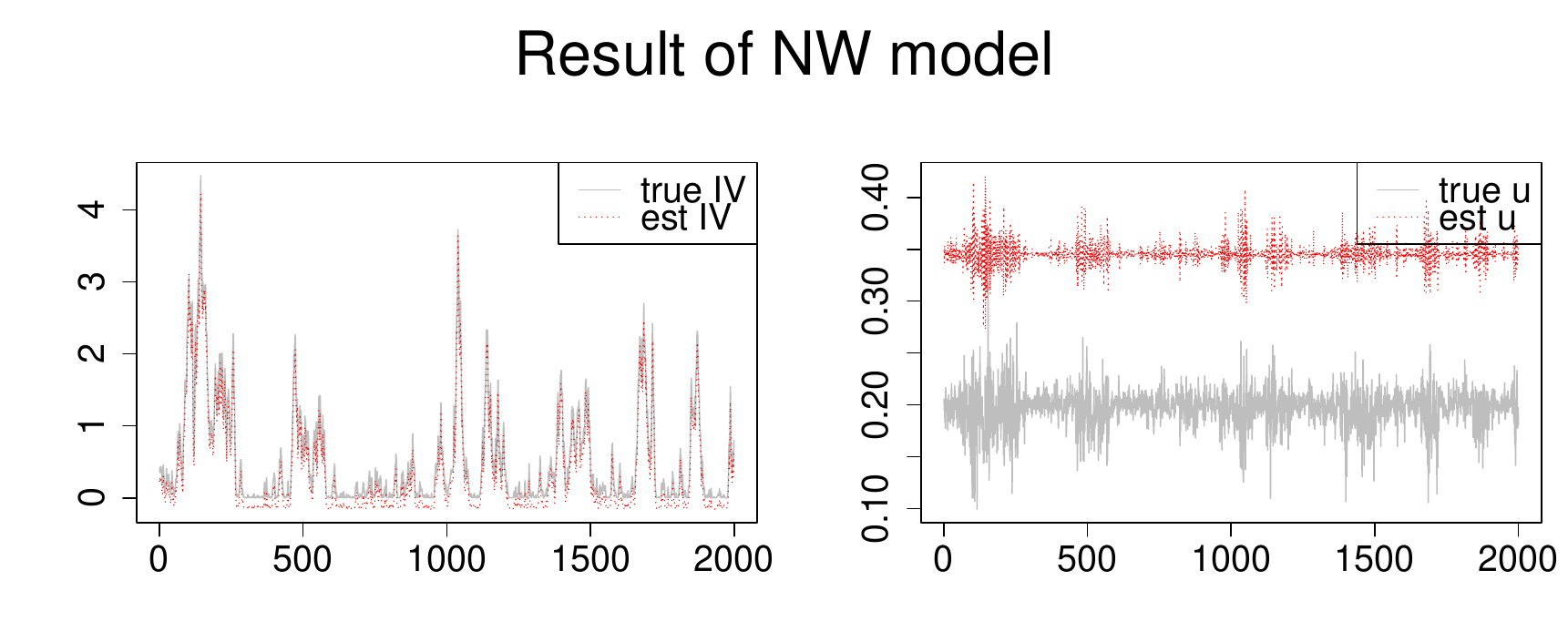}
\caption{Estimation result of NW model ($IV_t$(left), $u_t^{(m)}$(right))}
\label{res_nw}
\end{center}
\end{figure}

\begin{figure}[H]
\begin{center}
\includegraphics[width=12cm]{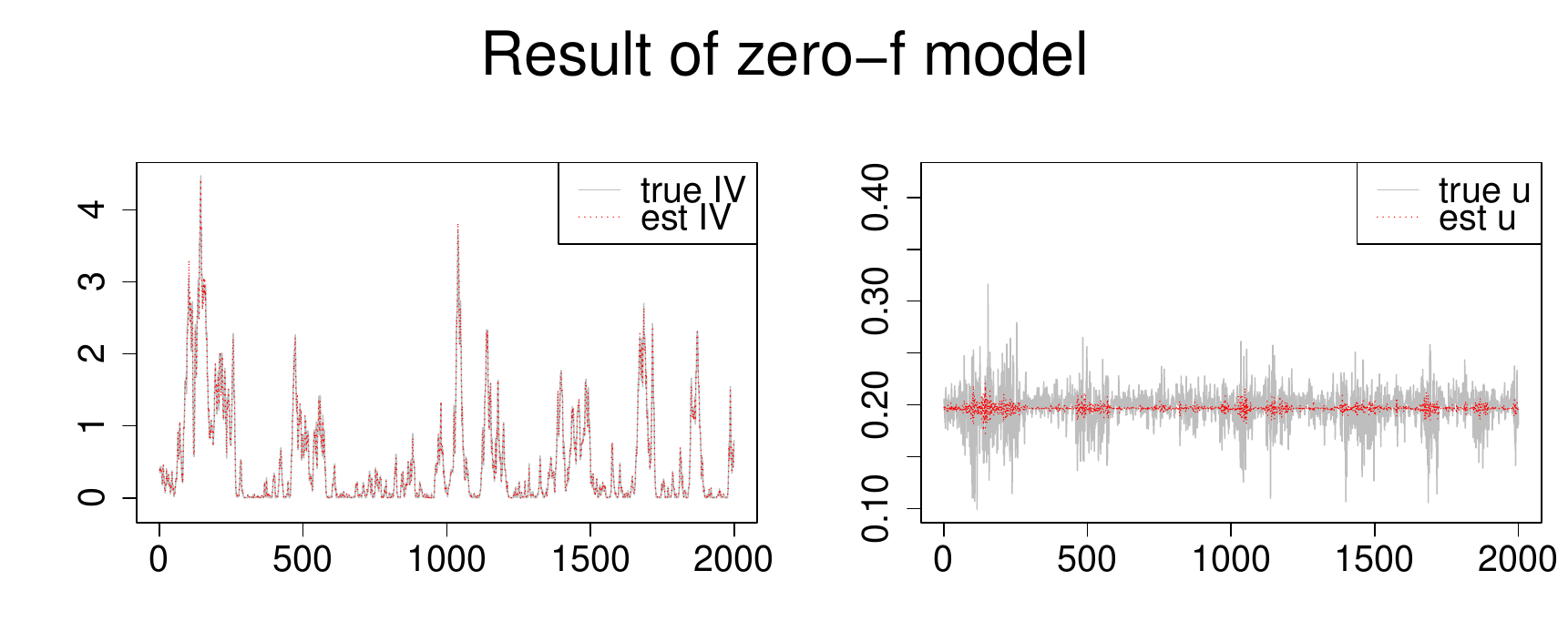}
\caption{Estimation result of zero-f model ($IV_t$(left), $u_t^{(m)}$(right))}
\label{res_zf}
\end{center}
\end{figure}

\begin{figure}[H]
\begin{center}
\includegraphics[width=12cm]{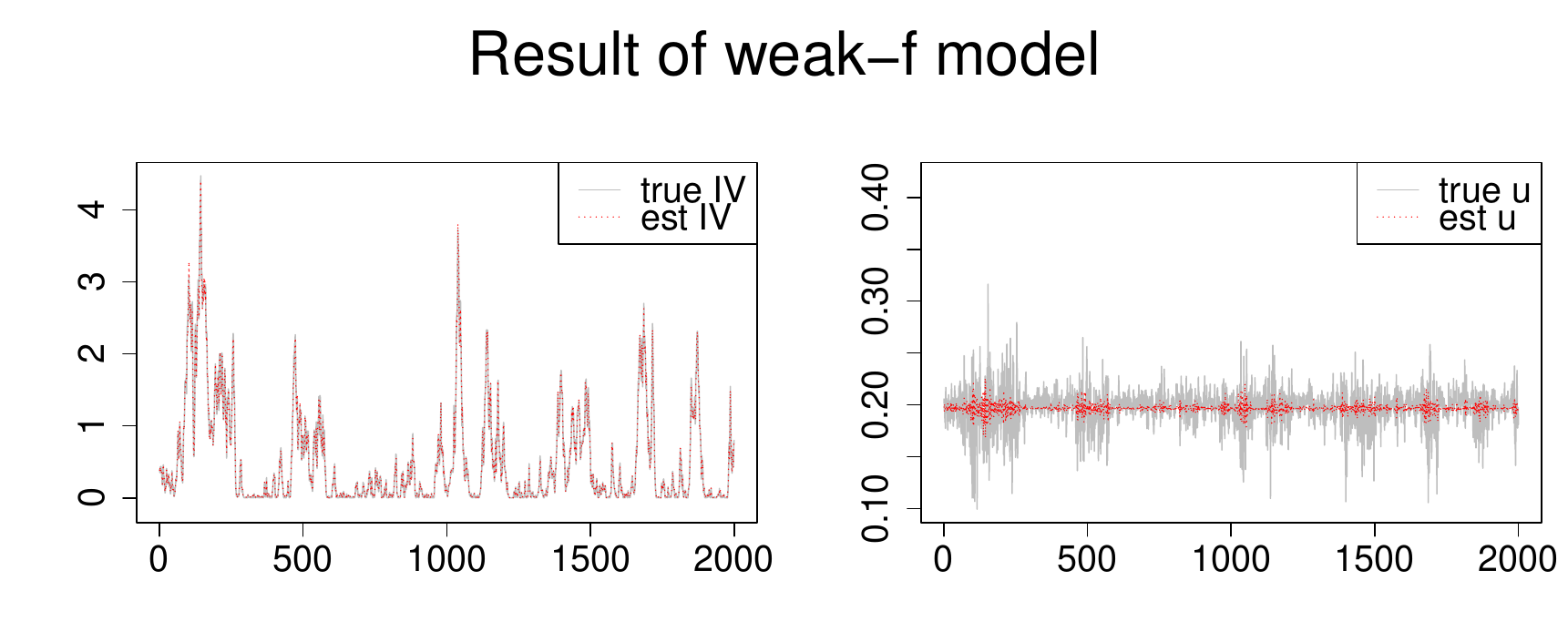}
\caption{Estimation result of weak-f model ($IV_t$(left), $u_t^{(m)}$(right))}
\label{res_wf}
\end{center}
\end{figure}

Estimate the parameters of the NW model and proposal model with the factor number $p=1$, using $p^*_t$ every minute for $N$ days and $RV_t^{*(1440)}$ calculated from data. In the case of NW model, all parameters are estimated  by QMLE according to the method of Nagakura and Watanabe[2015] direct approach (p55) with $RV_t^{*(1440)}$ as data. In the case of proposal model, zero-f and weak-f estimate the parameters  by methods of previous section with $RV_t^{*(1440)}$ and $r_t^*$ as data. The following table \ref{quasi_para} is the estimation result of each model and comparison with a true value.
\begin{table}[H]
\begin{center}
\caption{Estimation result of SR-SARV model and MN}
\begin{tabular}{|l|c|c|c|c|c|c|c|c|}
\hline
&$\sigma^2$&$\omega_1^2$&$\lambda_1$&$\Omega_0\times 10^{4}$&$\Omega_1\times 10^{5}$&$\Omega_2\times 10^{5}$&$f(m)$&$\log{L}$\\ \hline
True&$0.5$&$0.7734$&$0.0202$&$1.510$&$8.093$&$3.122$& -0.0332&-\\ \hline
NW&$ 0.2700$&$0.8220$&$0.0173$&$1.199$&-($0$)&-($0$)&-($0$)&931\\ \hline
zero-f&$0.5132\textcolor{red}{*}$&$0.5899$&$0.0267$&$1.486$&$8.029$\textcolor{red}{*}&$3.072$\textcolor{red}{*}&-($0$)&933\\ \hline
weak-f&$0.5132\textcolor{red}{*}$&$0.7773\textcolor{red}{*}$&$0.0195\textcolor{red}{*}$&$1.491\textcolor{red}{*}$&$8.029$\textcolor{red}{*}&$3.072$\textcolor{red}{*}&-0.0034\textcolor{red}{*}&932\\ \hline
\end{tabular}
\label{quasi_para}
\end{center}
\end{table}Estimate and predict IV according to each estimated model and compare the IV estimation error with the 1-ahead prediction error. Figures \ref{res_nw}, \ref{res_zf} and \ref{res_wf} are plots of the estimation result of $IV_t$ and $u_t^{(m)}$ in the NW, zero-f and weak-f respectively. The gray line is the true value, and the red line is the estimated value in the model.

In NW model, the averages of $IV_t$ and $u_t^{(m)}$ ($\sigma^2$ and $c_u^{(m)}$) are estimated by QMLE. Since this accuracy is bad, $IV_t, u_t^{(m)}$ are shifted from the true value.\footnote{When maximizing the quasi-likelihood function, we tried changing the initial value, but $\sigma^2$ and $c_u^{(m)}$ could not make it converge to  ``good'' value.}
In the proposal model (zero-f model  and weak-f model), these are calculated by Equation (\ref{r2sigma}) and (\ref{r2Eu}) from data $r_t^*$ which is also estimated to be close to the true value. However, the variance of $u_t^{(m)}$ seems to be underestimated, the true variance of  $u_t^{(m)}$ is 3.6641$\times 10^{-4}$, but the value estimated by zero-f model is 1.258$\times 10^{-5}$, weak-f model is 1.819$\times 10^{-5}$. NW model is 1.225$\times 10^{-4}$, which is also underestimated, but it is closest to the true value. This part is a term $4mC_0^{(1)}+8\sum^{m-1}_{k=1}(m-k)C_k^{(1)}+4mC_0^{(2)}+4\sum^{m-1}_{k=1}(m-k)C_k^{(2)}$ in the Theorem \ref{thm1}, where HL.1 contributes to the variance of $u_t^{(m)}$, and zero-f model ignores this term with $f(m)$, so this part is underestimated. In the weak-f model, somehow $f(m)$ is estimated to be smaller than the true value, so it is close to the result of the zero-f model.

The tables \ref{quasi_er} and \ref{quasi_erp} are comparison results of estimation accuracy and prediction accuracy. 
\begin{table}[H]
\begin{center}
\caption{Estimation error}
\begin{tabular}{|c|c|c|c|c|c|}
\hline
&NW($1$)&zero-f($1$)&weak-f($1$)&RK&RK (true $H$)\\ \hline
MSE&0.0232&0.0011\textcolor{red}{*}&0.0011\textcolor{red}{*}&0.0077&0.0057\\ \hline
QLIKE&0.7734&-0.9917&-0.9838&-0.9718&-1.115\textcolor{red}{*}\\ \hhline{|======|}
\multicolumn{6}{|c|}{Mincer-Zarnowitz Regression}\\ \hline
$R^2$&0.9975&0.9976\textcolor{red}{*}&0.9976\textcolor{red}{*}&0.9856&0.9924\\  \hline
$\hat{a}$&0.1457&-0.0031&-0.0034& -0.0089&-0.0056\\
std error&0.0008&0.0009&0.0009&0.0023& 0.0017\\ \hline
$\hat{b}$&1.0067&1.0053&1.0059&0.9796&0.9545\\
std error&0.0011&0.0011&0.0011&0.0026&0.0018\\ \hline
\end{tabular}
\label{quasi_er}
\end{center}
\end{table}
In Table \ref{quasi_er}, RK is added as the IV estimation method. Barndorff-Nielsen et al[2011] constructed an estimator, Realized Kernel $RK_t^{(m)}$ which has asymptotic normality under the existence of MN satisfying HL.1, HL.2 and HL.3:
\begin{align*}
RK_t^{(m)}&=RAC^{(m,0)}+2\sum^{m-1}_{h=1}K\left(\cfrac{h}{H}\right)RAC^{(m,h)},
\intertext{where $RAC^{(m,h)}$ is lag $h$ Realized auto-covariance:}
RAC^{(m,h)}&=\sum^{n-h}_{i=1}r^{(m)}_{t-1+\frac{i}{m}}r^{(m)}_{t-1+\frac{i+h}{m}},
\end{align*}
and Kernel function $K(x) : \mathbb{R}^1\to\mathbb{R}^1$ is $C^2$ and  satisfies $K(0)=1, K'(0)=0$ and appropriate definiteness and integrability. The parameter $H$ is called bandwidth, when $H\propto m^{\frac{3}{5}}$, $RK_t^{(m)}$ has asymptotic normality with convergence rate $m^{\frac{1}{5}}$. Barndorff-Nielsen et al[2011] recommended $K(x)$ be Parzen kernel:
\begin{align*}
K(x)=\begin{cases}
1-6x^2+6x^3&(0\leq x\leq \frac{1}{2}),\\
2(1-x)^3&(\frac{1}{2}x\leq1),\\
0&(x>1),
\end{cases}
\end{align*} 
and they proved bandwidth is optimized as follows: $H=c\xi^{\frac{4}{5}}m^{\frac{3}{5}}, c=3.5134,\xi^2=\cfrac{\V[\eps_t]}{\sqrt{T\int^T_0\sigma^4(t)dt}}$. However, since an unobservable variable is used for $\xi^2$, the optimal value is not obtained. Here, quote Ubukata and Watanabe[2011], estimated every day with $\widehat{\V[\eps_t]}=\cfrac{RV_t^{(m)}}{2m}$ and $\int^T_0\sigma^4(t)dt$ is estimated every day by Realized Quarticity(Barndorff-Nielsen and Shephard[2002]):$\displaystyle RQ_t^{(m)}=\cfrac{m}{3}\sum_{i=1}^mr^{(m)4}_{t-1+\frac{i}{m}}$ with $m=96$. We call $H$ estimated by this method ``rough".  In Table \ref{quasi_er}, ``RK'' using ``rough'' bandwidth $H$ and ``RK (true $H$)'' using the optimal bandwidth $H$ calculated by using true $\sigma^2(t), \eps_t$, which is actually impossible to obtain.

In Table \ref{quasi_er}, for MSE and  $R^2$, zero-f  model and weak-f model are equal and the best. For QLIKE, RK(true $H$) is the minimum and zero-f  model is the second. From these results, it can be seen that the proposal models zero-f model and weak-f model dominates the existing method NW model and RK in the IV estimation accuracy.
\begin{table}[H]
\begin{center}
\caption{1-ahead prediction error}
\begin{tabular}{|c|c|c|c|}
\hline
&NW($1$)&zero-f($1$)&weak-f($1$)\\ \hline
MSE&0.0309&0.0184\textcolor{red}{*}&0.0184\textcolor{red}{*}\\ \hline
QLIKE&0.3432&-0.9829&-1.106\textcolor{red}{*}\\ \hhline{|====|}
\multicolumn{4}{|c|}{Mincer-Zarnowitz Regression}\\ \hline
$R^2$&0.9356&0.9501\textcolor{red}{*}&0.9499\\  \hline
$\hat{a}$& 0.089&0.0046&0.0427\\
std error&0.0078&0.0073&0.0073\\ \hline
$\hat{b}$&0.8973&0.9811&0.9816\\
std error&0.0105&0.0100&0.0101\\ \hline
\end{tabular}
\label{quasi_erp}
\end{center}
\end{table}

In Table \ref{quasi_erp}, for MSE,  zero-f model and weak-f model are equal and minimum. For QLIKE, weak-f model is minimum. For $R^2$, zero-f model is the best. Thus, proposal models zero-f model and weak-f model dominates the  NW model  in the IV 1-ahead prediction accuracy.
\subsection{Model comparision by actual data}
\subsubsection{Data}
Data are taken from the middle price (1749 days) of the dollar-yen rate every 30 seconds (m=2880) from May 1, 2009 to April 29, 2016. \footnote{The data are extracted every 30 seconds from the tick data downloaded from {\url{https://pepperstone.com/en/}}.} The sample size is 4905910. Following Anderson et al[2001], Nagakura, Watanabe[2009, 2015], we remove from the data the trading days that satisfy the following conditions as inactive trading days 
\begin{enumerate}
\item Day with 500 or more missing values.
\item A trading day with more than 1000 zero returns ($r_t^*=0$).
\item A trading day when the rate value does not change for more than 35 minutes.
\end{enumerate} 

The following 184 days were removed from the data. All dates in black are Sundays, and dates in red are other days of the week.\\[0.3cm]
{\scriptsize 2009-05-03, 2009-05-10, 2009-05-17, \textcolor{red}{2009-05-21}, 2009-05-24, 2009-05-31, 2009-06-07, 2009-06-14, 2009-06-21, 2009-06-28, 2009-07-05, 2009-07-12, 2009-07-19, 2009-07-26, 2009-08-02, 2009-08-09, 2009-08-23, 2009-08-30, 2009-09-06, 2009-09-13, 2009-09-20, 2009-09-27, 2009-10-04, 2009-10-11, 2009-10-18, 2009-10-25, 2009-11-01, 2009-11-08, 2009-11-15, 2009-11-22, 2009-12-06, 2009-12-13, 2009-12-20, \textcolor{red}{2009-12-25}, 2009-12-27, 2010-01-03, 2010-01-10, 2010-01-17, 2010-01-24, 2010-01-31, 2010-02-07, 2010-02-21, 2010-02-28, \textcolor{red}{2010-03-03}, 2010-03-07, 2010-03-14, 2010-03-21, 2010-03-28, 2010-04-04, 2010-04-11, 2010-04-18, 2010-04-25, 2010-05-02, 2010-05-09, 2010-05-16, 2010-05-23, 2010-05-30, 2010-06-06, 2010-06-13, 2010-06-20, 2010-06-25, 2010-07-04, 2010-07-11, 2010-07-18, 2010-07-25, 2010-08-01, 2010-08-08, 2010-08-15, 2010-08-22, 2010-08-29, 2010-09-05, 2010-09-12, 2010-09-19, 2010-09-26, 2010-10-03, 2010-10-10, 2010-10-17, 2010-10-24, 2010-10-31, 2010-11-07, 2010-11-14, 2010-11-21, 2010-11-28, 2010-12-05, 2010-12-12, 2010-12-19, 2010-12-26, 2011-01-02, 2011-01-09, 2011-01-16, \textcolor{red}{2011-01-28}, 2011-01-30, 2011-02-06, 2011-02-13, 2011-02-20, 2011-02-27, 2011-03-06, 2011-03-13, 2011-03-20, 2011-03-27, 2011-04-03, 2011-04-10, 2011-04-17, 2011-04-24, 2011-05-01, \textcolor{red}{2011-05-05}, \textcolor{red}{2011-05-06}, 2011-05-08, 2011-05-15, 2011-05-22, 2011-05-29, 2011-06-05, 2011-06-12, 2011-06-19, 2011-06-26, 2011-07-03, 2011-07-10, 2011-07-17, 2011-07-24, 2011-08-07, 2011-08-14, 2011-08-21, 2011-08-28, 2011-09-04, 2011-09-11, 2011-09-18, 2011-09-25, 2011-10-02, 2011-10-09, 2011-10-16, 2011-10-23, 2011-10-30, 2011-11-06, 2011-11-13, 2011-11-20, 2011-11-27, 2011-12-04, 2011-12-11, 2011-12-18, 2011-12-25, 2012-01-08, 2012-01-15, 2012-01-17, 2012-01-22, 2012-01-29, 2012-01-31, 2012-02-05, 2012-02-12, 2012-02-19, 2012-02-26, \textcolor{red}{2012-02-29}, 2012-03-04, 2012-03-11, 2012-03-18, 2012-03-25, 2012-04-01, 2012-04-08, 2012-04-15, 2012-04-22, 2012-04-29, 2012-05-06, \textcolor{red}{2012-05-11}, 2012-05-13, 2012-05-20, 2012-05-27, 2012-07-01, 2012-07-08, 2012-07-15, 2012-07-22, 2012-07-29, 2012-11-18, 2012-12-25, \textcolor{red}{2013-01-01}, \textcolor{red}{2013-12-25}, \textcolor{red}{2014-01-01}, \textcolor{red}{2014-09-30}, 2014-11-09, \textcolor{red}{2014-11-26}, \textcolor{red}{2014-12-02}, \textcolor{red}{2014-12-03}, \textcolor{red}{2014-12-25}, \textcolor{red}{2015-01-01}, 2015-05-31, \textcolor{red}{2015-12-25}}\\[0.3cm]

\begin{figure}[H]
\centering
\includegraphics[width=10cm]{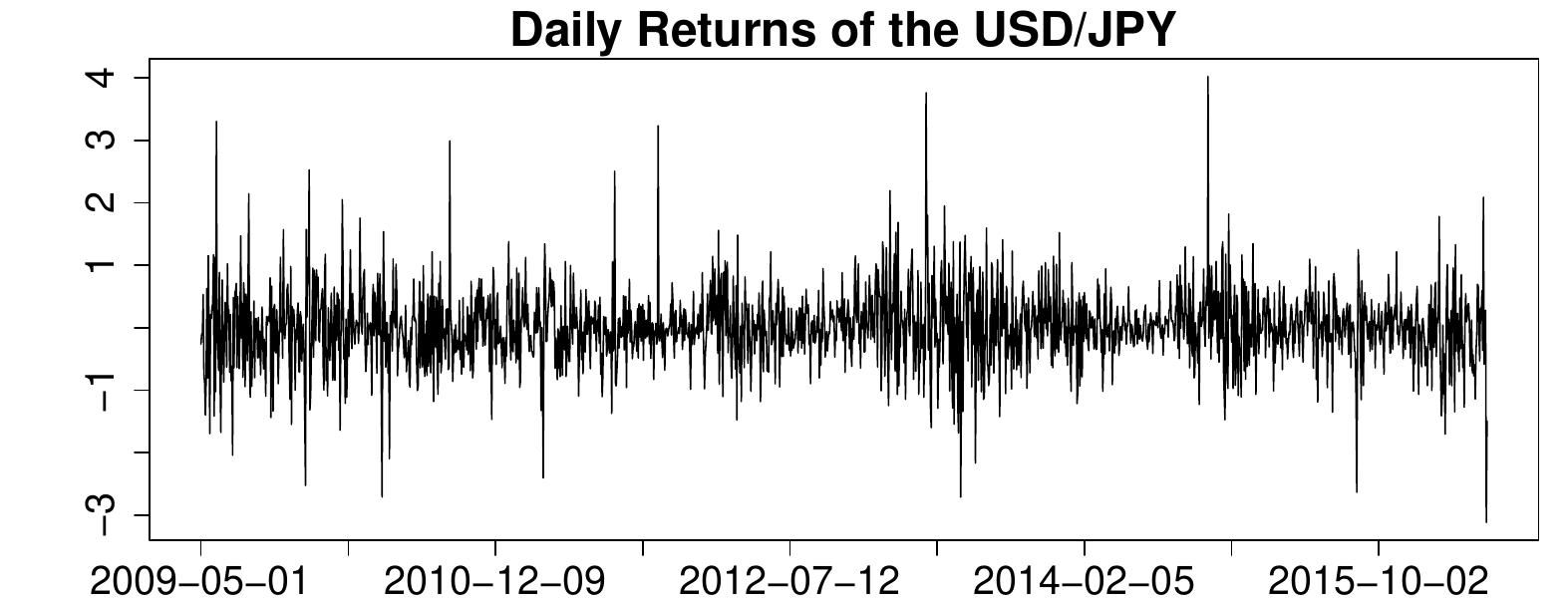}
\caption{Plot of the daily returns}
\label{daily_return}
\end{figure}

The next Flg. \ref{daily_return} is a plot of the daily returns for the data period used.
From the data, we computed the observed returns of equation (\ref{obsret}) every 30 seconds ($m=2880$), 1 minute ($m=1440$), 5 minutes ($m=288$), 10 minutes ($m=144$) and 30 minutes ($m=48$) multiplied by 100. Table \ref{summary_r} is the basic statistics of the computed observation returns.
\begin{center}
\begin{threeparttable}[H]
\caption{Basic statistics of the computed observation returns}
\begin{tabular}{|l|c|c|c|c|c|}
\hline
&0.5min&1min&5min&10min&30min\\ \hline
$m$&2880&1440&288&144&48\\ \hline
Mean$\times 10^4$& 0.0151&  0.0302& 0.1509& 0.2865&0.8681 \\ \hline
Variance$\times 10^4$& 2.345& 3.913&15.94&29.96 &86.96 \\ \hline
SAC(1)&-0.1609\textcolor{blue}{*}& -0.1075\textcolor{blue}{*}& -0.0455\textcolor{blue}{*}&-0.0196\textcolor{blue}{*} & -0.0229\textcolor{blue}{*}\\ \hline
SAC(2)&-0.0065\textcolor{blue}{*} &-0.0137 \textcolor{blue}{*}&-0.0050\textcolor{blue}{*} & -0.0120\textcolor{blue}{*}&\ 0.0015 \\ \hline
SAC(3)& -0.0144\textcolor{blue}{*}&0.0021\textcolor{blue}{*} & -0.0063\textcolor{blue}{*}&-0.0044 &\ 0.0011 \\ \hline
SAC(4)&\ 0.0043\textcolor{blue}{*} & -0.0003& \ 0.0028&\ 0.0045 &\ 0.0008 \\ \hline
SAC(5)& -0.0086\textcolor{blue}{*}& -0.0029\textcolor{blue}{*}& -0.0021\textcolor{blue}{*}&-0.0068\textcolor{blue}{*} &-0.0050 \\ \hline
SAC(6)&-0.0017\textcolor{blue}{*} & \ 0.0013& -0.0115\textcolor{blue}{*}&\ 0.0052 \textcolor{blue}{*}&-0.0023 \\ \hline
SAC(7)&0.0062\textcolor{blue}{*}& -0.0036\textcolor{blue}{*}&\ 0.0023 &-0.0014 &\ 0.0005 \\ \hline
SAC(8)& -0.0118\textcolor{blue}{*}&-0.0039 \textcolor{blue}{*}&\ 0.0086\textcolor{blue}{*} &-0.0041 &-0.0032 \\ \hline
SAC(9)&\ 0.0113 \textcolor{blue}{*}&-0.0010 &\ 0.0003&-0.0014 &-0.0025 \\ \hline
SAC(10)&-0.0059 \textcolor{blue}{*}&-0.0058 \textcolor{blue}{*}& -0.0065\textcolor{blue}{*}&\ 0.0003 &\ 0.0056 \\ \hline
\end{tabular}
\begin{tablenotes}
\item Mean$\times 10^4$ is $10^4$ times the sample mean, Variance$\times 10^4$ is $10^4$ times the sample unbiased variance, and SAC($n$) is the sample autocorrelation coefficient at lag $n$.
\end{tablenotes}
\label{summary_r}
\end{threeparttable}
\end{center}
In all cases, the mean is almost equal to zero. The larger $m$ is, the smaller the mean and variance are. As for the autocorrelation, SAC(1) is relatively large, while those after SAC(2) are quite small. In addition, the larger $m$ is, the larger the first-order autocorrelation is. Therefore, for each autocorrelation at $m$, we performed the $k$-order Ljung-Box test with $k=1,2,\ldots,1000$, and all the autocorrelations are significant even at the significance level of 1$\%$. In addition, $k$-order autocorrelations were tested for each $k$. The SAC \textcolor{blue}{*} are the values that are significant at the 1$\%$ level of significance in this test. The next table\ref{acf_mn} is the autocorrelation $\Omega_i/\Omega_0$ calculated from the autocovariance $\Omega_i,\ i=0, 1, \ldots, q$ of MN calculated from the Theorem\ref{thm3}.
\begin{table}[H]
\begin{center}
\caption{Estimation of MN autocorrelation}
\label{acf_mn}
\begin{tabular}{|c|c|c|c|c|c|}
\hline
&0.5min&1min&5min&10min\\ \hline
$m$&2880&1440&288&144\\ \hline
$\hat{q}+1$&22&16&13&6\\ \hline
$\Omega_0\times 10^{-5}$&6.4925&12.666&22.171&22.816\\ \hline
$\Omega_1/\Omega_0$&0.3102&0.5634&0.5313&0.5271\\ \hline
$\Omega_2/\Omega_0$&0.2019&0.4589&0.3966&0.3719\\ \hline
$\Omega_3/\Omega_0$&0.1172&0.3969&0.2988&0.2520\\ \hline
$\Omega_4/\Omega_0$&0.0845&0.3413&0.2473&0.1761\\ \hline
$\Omega_5/\Omega_0$&0.0362&0.2868&0.1751&0.0806\\ \hline
$\Omega_6/\Omega_0$&0.0190&0.2413&0.1186&None(0)\\ \hline
$\Omega_7/\Omega_0$&0.0083&0.1918&0.1468&None(0)\\ \hline
$\Omega_8/\Omega_0$&-0.0248&0.1535&0.1579&None(0)\\ \hline
$\Omega_9/\Omega_0$&-0.0152&0.1274&0.1061&None(0)\\ \hline
$\Omega_{10}/\Omega_0$&-0.0468& 0.1043&0.0518&None(0)\\ \hline
$\Omega_{11}/\Omega_0$&-0.0569&0.0992&0.0456&None(0)\\ \hline
$\Omega_{12}/\Omega_0$&-0.0786&0.0620&0.0521&None(0)\\ \hline
$\Omega_{13}/\Omega_0$&-0.1113&0.0406&None(0)&None(0)\\ \hline
$\Omega_{14}/\Omega_0$&-0.1087&0.0359&None(0)&None(0)\\ \hline
$\Omega_{15}/\Omega_0$&-0.1130&0.0279&None(0)&None(0)\\ \hline
$\Omega_{16}/\Omega_0$&-0.1277&None(0)&None(0)&None(0)\\ \hline
$\Omega_{17}/\Omega_0$&-0.1346&None(0)&None(0)&None(0)\\ \hline
$\Omega_{18}/\Omega_0$&-0.0850&None(0)&None(0)&None(0)\\ \hline
$\Omega_{19}/\Omega_0$&-0.1006&None(0)&None(0)&None(0)\\ \hline
$\Omega_{20}/\Omega_0$&-0.0640&None(0)&None(0)&None(0)\\ \hline
$\Omega_{21}/\Omega_0$&-0.0561&None(0)&None(0)&None(0)\\ \hline
\end{tabular}
\end{center}
\end{table}

For $\hat{q}+1$, the starting point of $k$ where SAC($k$) took three consecutive non-significant values in this test is considered as the truncation of the autocorrelation, and this $k-1$ is set as $\hat{q}+1$. With this $\hat{q}$, we determined the MA order $q$ of MN.  The larger $m$ is, the larger $q$ is. The variance appears to decrease with $m$. With the exception of $m=2880$, the autocorrelation for each order is larger for larger $m$. 

From these observed returns, we calculated each RV using the Formula (\ref{rv}). The following figure shows a plot of the calculated RVs.
\begin{figure}[H]
\begin{center}
\includegraphics[width=15cm]{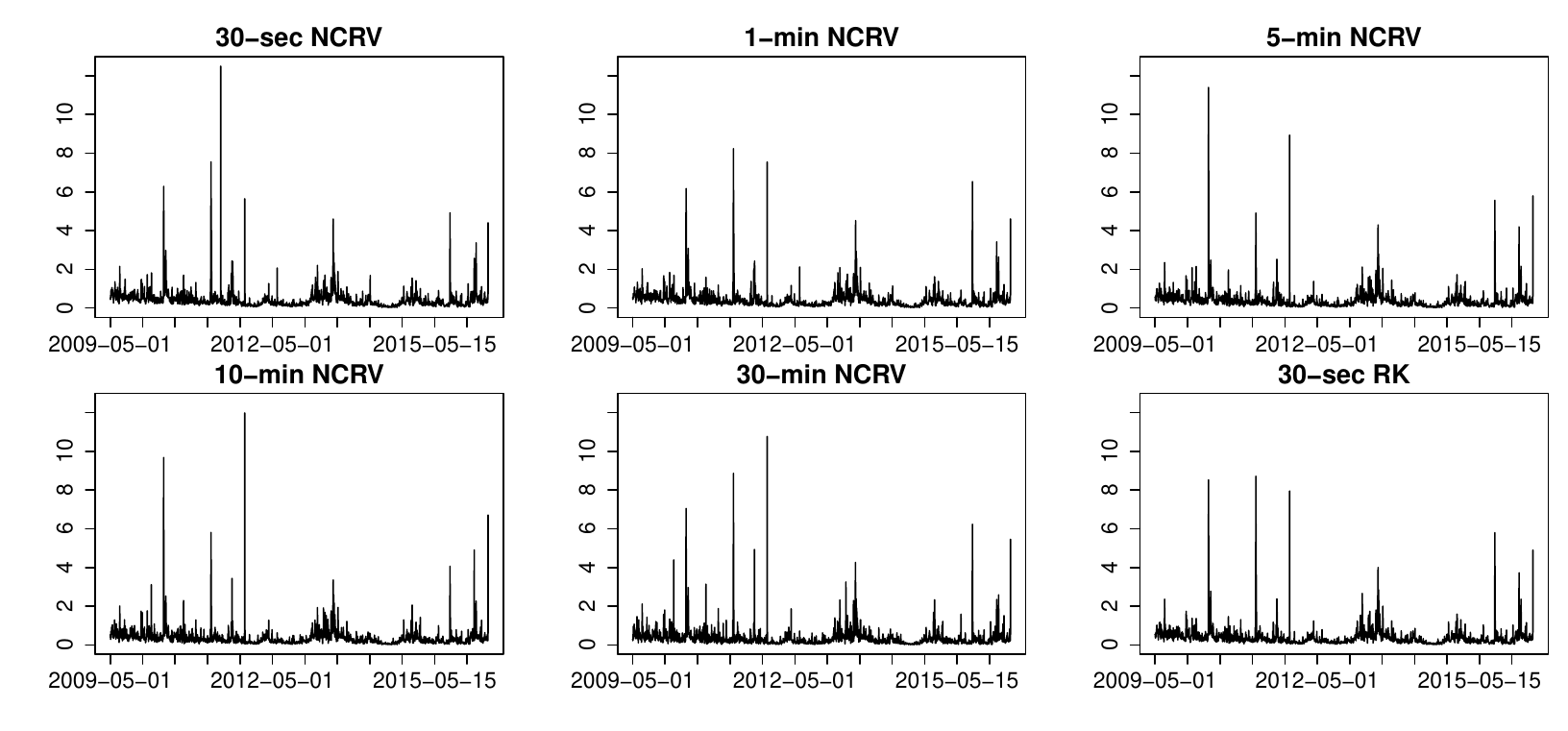}
\caption{Top row from left, $RV_t^{*(2880)}, RV_t^{*(1440)}, RV_t^{*(288)}$. Bottom row, from left $RV_t^{*(144)}, RV_t^{*(48)}, RK_t^{(2880)}$.}
\label{rvs}
\end{center}
\end{figure}
Table \ref{summary_rv} is the basic statistics for each calculated RV and RK.
\begin{center}
\begin{threeparttable}[H]
\caption{RV Basic Statistics}
\begin{tabular}{|l|c|c|c|c|c||c|}
\hline
&0.5min&1min&5min&10min&30min&RK\\ \hline
$m$&2880&1440&288&144&48&2880\\ \hline
Mean&0.4263& 0.4146& 0.3896& 0.3765&0.3602&0.3875 \\ \hline
Variance&0.3132&0.2596&0.2849&0.3023&0.3106&0.2670 \\ \hline
SAC(1)&0.350&0.431&0.344&0.270&0.257&0.3774\\ \hline
SAC(2)&0.262&0.314&0.200&0.167&0.129&0.2629 \\ \hline
SAC(3)&0.191&0.228&0.168&0.147&0.140&0.1874 \\ \hline
SAC(4)&0.161&0.191&0.137&0.113&0.096&0.1483 \\ \hline
SAC(5)&0.137&0.159&0.119&0.099&0.081&0.1259  \\ \hline
SAC(6)&0.128&0.146&0.117&0.104&0.083&0.1232  \\ \hline
SAC(7)&0.116&0.130&0.113&0.101&0.079&0.1103 \\ \hline
SAC(8)&0.119&0.138&0.111&0.090&0.063&0.1111 \\ \hline
SAC(9)&0.151&0.173&0.153&0.125&0.108&0.1562 \\ \hline
SAC(10)&0.148&0.178&0.168&0.141&0.118&0.1647 \\ \hline
\end{tabular}
\label{summary_rv}
\end{threeparttable}
\end{center}
As seen in Fig.\ref{mnacf}, we can confirm that the larger $m$ is, the larger the average of RV is.

\subsubsection{Comparison method}
Estimation errors are obtained by estimating the parameters of the model and IV using all the data. The 1-ahead forecasting error uses 1500 days of data, estimates the parameters of the model, and forecasts IV 1-ahead. This is done 500 times, shifting the data interval by one day. 
Since IV is unobservable, we use the Realized Kernel of Barndorff-Nielsen et al [2011] as a proxy variable:

Barndorff-Nielsen et al[2011] constructed an estimator Realized Kernel $RK_t^{(m)}$ with asymptotic normality in the presence of MNs satisfying HL.1, HL.2 and HL.3:
\begin{align*}
RK_t^{(m)}&=RAC^{(m,0)}+2\sum^{m-1}_{h=1}K\left(\cfrac{h}{H}\right)RAC^{(m,h)}.
\intertext{where $RAC^{(m,h)}$ is the $h$-order Realized Auto-covariance,}
RAC^{(m,h)}&=\sum^{n-h}_{i=1}r^{(m)}_{t-1+\frac{i}{m}}r^{(m)}_{t-1+\frac{i+h}{m}},
\end{align*}
$K(x)$:$\mathbb{R}^1\to\mathbb{R}^1$ is a Kernel function, which satisfies appropriate non-negative definiteness and integrability in the $C^2$ class, $K(0)=1, K'(0)=0$. The parameter $H$ is the bandwidth, and when $H\propto m^{\frac{3}{5}}$, $RK_t^{(m)}$ is asymptotically normal with the convergence rate $m^{\frac{1}{5}}$. Barndorff-Nielsen et al [2011], $K(x)$ is a Parzen kernel:
\begin{align*}
K(x)=\begin{cases}
1-6x^2+6x^3&(0\leq x\leq \frac{1}{2}),\\
2(1-x)^3&(\frac{1}{2}<x\leq1),\\
0&(x>1)
\end{cases}
\end{align*}
The optimal parameters recommended are $H=c\xi^{\frac{4}{5}}m^{\frac{3}{5}}, c=3.5134,\xi^2=\cfrac{\V[\eps_t]}{\sqrt{T\int^T_0\sigma^4(t)dt}}$. Here, following Ubukata, Watanabe [2011], $\widehat{\V[\eps_t]}=\cfrac{RV_t^{(m)}}{2m}$ is estimated daily from $\frac{RV_t^{(m)}}{2m}$, $\int^T_0\sigma^4(t)dt$ is estimated daily from the Reaized Quarticity(Barndorff-Nielsen, Shephard [2002]): $\displaystyle RQ_t^{(m)}=\cfrac{m}{3}\sum_{i=1}^mr^{(m)4}_{t-1+\frac{i}{m}}$, estimated daily with $m=96$ and We use $RK_t^{(2880)}$, calculated from the observation returns every 30 seconds, as a proxy variable for $IV_t$.

The following tables\ref{1f_par} and \ref{2f_par} are the estimated SR-SARV model and MN parameters for each model for $p=1$ and $2$, respectively.
\begin{table}[H]
\begin{center}
\caption{1-factor model:Parameter estimation results for SR-SARV-model and MN
}
\footnotesize
\begin{tabular}{|c|c|c|c|c|c|c|c|c|c|c|c|c|}
\hline
$m$					&\multicolumn{3}{|c|}{$m=2880$}&\multicolumn{3}{|c|}{$m=1440$}&\multicolumn{3}{|c|}{$m=288$}&\multicolumn{3}{|c|}{$m=144$}\\ \hline
model                		& NW&zero-f&weak-f&NW&zero-f&weak-f& NW&zero-f&weak-f&NW&zero-f&weak-f\\ \hline
$\sigma^2$       		& 0.3099&0.4176&0.4176&0.3755&0.4043&0.4043 & 0.3351&0.4003&0.4003&  0.3589& 0.4027&0.4027 \\ \hline
$\omega_1^2$			&0.1306& 0.5764& 0.1350&0.1671&0.4206&0.1434 & 0.1167&0.5022&  0.1337&0.0689&0.5965& 0.0863 \\ \hline
$\lambda_1$			&0.1698& 2.237&0.1690 & 0.2015&1.722&0.2538  &0.0546&2.105&0.3230&0.0681&2.688&0.1860 \\ \hline
$\Omega_0\times10^{4}$ &0.1530&0.6492&0.8753&0.1585&1.267&1.653 &0.0903&0.2171&4.274&0.0722&1.262&6.1468\\ \hline
$f(m)$     				&0&0&-0.1502&0&0& -0.1268 &0&0&-0.1428&0&0& -0.1642 \\ \hline
$\log{L}$         		        &-1322&-1374&-1324&-1096&-1165&-1095&-1268&-1296&-1259&-1349&-1378&-1348 \\ \hline
\end{tabular}
\label{1f_par}
  \end{center}
\end{table}
%1つにもどす
\begin{table}[H]
\begin{center} 
\caption{2-factor model:Parameter estimation results for SR-SARV-model and MN}
\footnotesize
\begin{tabular}{|c|c|c|c|c|c|c|c|c|c|c|c|c|}
\hline
$m$					&\multicolumn{3}{|c|}{$m=2880$}&\multicolumn{3}{|c|}{$m=1440$}&\multicolumn{3}{|c|}{$m=288$}&\multicolumn{3}{|c|}{$m=144$}\\ \hline
model           	        		& NW&zero-f&weak-f&NW&zero-f&weak-f& NW&zero-f&weak-f&NW&zero-f&weak-f\\ \hline
$\sigma^2$			&0.3067&0.4176&0.4176&0.3552 & 0.4043&0.4043&0.3764 &0.4003&0.4003& 0.3713&0.4027&0.4027\\ \hline
$\omega_1^2$ 			& 0.0378& 0.0851&0.0379 &0.0396&0.0983&0.0397 & 0.0411&0.0503&0.0411&0.0356&0.0455&0.0356\\ \hline
$\lambda_1$      		& 0.0168&0.0886&0.0169  &0.0207&0.1291&0.0208 & 0.0236&0.0357&0.0236&0.0207&0.0340&0.0207\\ \hline
$\omega_2^2$   		&0.1194&0.7017&0.1193  &0.1233&0.4607&0.1233  & 0.1624&0.5225&0.1618&0.1093&0.6810&0.2319\\ \hline
$\lambda_2$      		&0.5451&4.863&  0.5459 &0.5643&4.321&0.5641  &   1.156&3.182&1.154&0.9419&4.278&1.081\\ \hline
$\Omega_0\times10^{4}$ &0.1934&0.6492&0.9361 & 0.1872&1.614&1.774  & 0.1865&2.171&4.779&0.1053&1.262&6.750\\ \hline
$f(m)$				&0&0&-0.1768&0&0&-0.1515&0&0&-0.1599&0&0&-0.1712\\ \hline
$\log{L}$            		 &-1311&-1318&-1311 & -1084&-1093&-1084 & -1242&-1244&-1242& -1337&-1339&-1337\\ \hline
\end{tabular}
\label{2f_par}
\end{center}
\end{table}
For each model, the parameters $\sigma^2, \omega_1^2, \lambda_1$ of the SR-SARV model have similar values regardless of $m$, which is consistent with the theory that the parameters of the SR-SARV model are independent of $m$. 

Although there seems to be no regularity in the zero-f model with respect to $\Omega_0$, the NW model takes similar values regardless of $p, m$. This is consistent with the results of Nagakura and Watanabe [2009, 2015]. On the other hand, the weak-f model increases as $m$ decreases, and is estimated to be larger than the NW model. Furthermore, $\Omega_0$ is always estimated larger for the weak-f model than for the zero-f model. This suggests that the method of theorem ref{thm3} underestimates $\Omega_0$. Also, $f(m)$ are all negative, this implies a negative correlation between returns $r_t$ and MN $\eps_t$.

The following Tables \ref{1f_ss_par} and \ref{2f_ss_par} are the estimation results of the State Space Form for each model with $p=1, 2$, respectively.
\begin{table}[H]
\begin{center}  
\caption{Parameter estimation results for 1-factor model State Space Form}
\footnotesize
\begin{tabular}{|c|c|c|c|c|c|c|c|c|c|c|c|c|}
\hline
$m$&\multicolumn{3}{|c|}{$m=2880$}&\multicolumn{3}{|c|}{$m=1440$}&\multicolumn{3}{|c|}{$m=288$}&\multicolumn{3}{|c|}{$m=144$}\\ \hline
model                   & NW&zero-f&weak-f&NW&zero-f&weak-f& NW&zero-f&weak-f&NW&zero-f&weak-f\\ \hline
$c_{IV}$       &0.0482&0.3732&0.0650& 0.0685&0.3350&0.0906 & 0.0178& 0.3516& 0.1105&0.0236&0.3753&0.0683 \\ \hline
$\phi_1$  	    &0.8437&0.1067&0.8444&0.8174&0.1713& 0.7758&0.9468&0.1217&0.7239&0.9341&0.0680&0.8301\\ \hline
$\theta_1$     &0.2675&0.2092&0.2675&0.2673&0.2278&0.2669 &0.2679&0.2145&0.2663& 0.2678&0.1914&.2674 \\ \hline
$\sigma^2_\eta$  &0.0233&0.2811&0.0207&0.0278&0.2179&0.0353 &0.0075&0.2494&0.03934&0.0054&0.2716&0.0166 \\ \hline
$c_u$         &0.0881&0.0087&0.0087&0.0456& 0.0103&0.0103     &0.0520&-0.0106&-0.0107& 0.0208&-0.0261&-0.0261 \\ \hline
$\theta_u\times10^{4}$&0.8680&0.2887&1.734&1.736&0.2230&3.466&8.685&0.7968&17.32&17.40&0.6486&34.69 \\ \hline
$\sigma_\xi^2$          &0.2019&0.0002&0.1809&0.1344&0.0002& 0.1291 &0.2068&0.0003&0.1639&0.2375&0.0003&0.2169 \\ \hline
$\sigma_d^2\times10^{4}$&1.574&5.213&2.150&3.834&8.128&4.262 &15.90&45.92&20.41&27.49&104.8&34.51 \\ \hline
\end{tabular}
\label{1f_ss_par}
\end{center}
\end{table}
\begin{table}[H]
\begin{center}
\caption{Parameter estimation results for 2-factor model State Space Form}
\footnotesize
\begin{tabular}{|c|c|c|c|c|c|c|c|c|c|c|c|c|}
\hline
$m$&\multicolumn{3}{|c|}{$m=2880$}&\multicolumn{3}{|c|}{$m=1440$}&\multicolumn{3}{|c|}{$m=288$}&\multicolumn{3}{|c|}{$m=144$}\\ \hline
model                   & NW&zero-f&weak-f&NW&zero-f&weak-f  & NW&zero-f&weak-f&NW&zero-f&weak-f\\ \hline
$c_{IV}$       &0.0021& 0.0351&0.0029&0.0031&0.0483&0.0036&0.0060&0.0134&0.0064&0.0046&0.01328&0.0050\\ \hline
$\phi_1$   &1.563&0.9229&1.562& 1.548&0.8921&1.547&1.291&1.006&1.291&1.369&0.9804&1.371\\ \hline
$\phi_2$   &-0.5700&-0.0070& -0.5695&-0.5570&-0.0116&-0.5567&-0.3071& -0.0400&-0.3078&-0.3819&-0.0134&-0.3841\\ \hline
$\theta_1$      &-0.6817& -0.6605&-0.6815&-0.6744&-0.5551&-0.6742 &-0.6683&-0.7225&-0.6681&-0.6661&-0.7639&-0.6658\\ \hline
$\theta_2$      &-0.2490&-0.0955&-0.2489&-0.2467&-0.0949&-0.2467 &-0.2284&-0.1547&-0.2284&-0.2349&-0.1234&-0.2350\\ \hline
$\sigma^2_\eta$   &0.0509&0.2630&0.0509&0.0538&0.2027&0.0538   &0.0905&0.2364&0.0901& 0.0587&0.2567&0.0579\\ \hline
$c_u$     &0.1114& 0.0087&0.0087&0.0539&0.0103&0.0103 &0.0107&-0.0107&-0.0107&0.0030&-0.0261&-0.0261\\ \hline
$\theta_u\times10^{4}$ &0.8679&0.2887&1.736&1.736&0.2229&3.475&8.691&0.7968&17.30&17.41&0.6486&34.66\\ \hline
$\sigma_\xi^2$             &0.1730&0.0002&0.1729&0.1155&0.0002&0.1155    &0.1267&0.0003&0.1269&0.1809&0.0003&0.1813\\ \hline
$\sigma_d^2\times10^{4}$ &1.745&1.527&1.527&4.171&1.527&1.527&23.96&1.527&1.527&39.26&1.527&1.527\\ \hline
\end{tabular}
\label{2f_ss_par}
\end{center}
\end{table}
As in the SR-SARV model, the parameters $c_{IV}, \phi_1, \phi_2, \theta_1, \theta_2, \sigma_\eta^2$ of IV have similar values independent of $m$. 

The bias and estimation error parameters $c_u, \theta_u, \sigma_\xi^2, \sigma_d^2$ seem to depend on $m$, but they take similar values for $p$. The $c_u$ of the NW model increases as $m$ increases. This is consistent with the results of the volatility signature plot (Fig \ref{mnacf}). The $c_u$ of the zero-f and weak-f models also increases with the value of $m$ except for $m=2880$, while $\sigma_\xi^2$ conversely decreases. The $\theta_u$ of the NW and weak-f models are larger for larger $m$. 

The $\sigma_d^2$ of the NW model, zero-f and weak-f models for $p=1$ increases as $m$ decreases. This is consistent with the asymptotics of RV. On the other hand, the same value is obtained for all $m$ for  zero-f and weak-f models $p=2$. For the NW model and the zero-f and weak-f models for $p=1$, it is a natural interpretation that the variance of estimation error $\sigma_d^2$ increases as the sample size $m$ gets smaller. Meanwhile, it is interesting to note that the $sigma_d^2$ of the zero-f and weak-f models for $p=2$ take the same value of 1.527. This value is smaller than in any other case.

The following tables \ref{summary_1fiv1},  \ref{summary_2fiv1} are the basic statistics of the estimated IV for each model for $p=1, 2$.
\begin{center}
\begin{threeparttable}[H]
\caption{IV basic statistics of 1-factor model}
\begin{tabular}{|c|c|c|c|c|c|c|}
\hline
&\multicolumn{3}{|c|}{$m=2880$}&\multicolumn{3}{|c|}{$m=1440$}\\ \hline
model                   & NW&zero-f&weak-f&NW&zero-f&weak-f\\ \hline
Mean    &0.3347&0.4176&0.4175&0.3694&0.4043&0.4041 \\ \hline
Variance&0.0790&0.3117&0.0840& 0.0892&0.2576&0.0940 \\ \hline
SAC(1)&0.9650&0.3513&0.9615&0.9493&0.4343&0.9352 \\ \hline
SAC(2)&0.8838& 0.2621&0.8731&0.8365&0.3153&0.7977 \\ \hline
SAC(3)&0.7866&0.1910&0.7695&0.7116&0.2288&0.6558 \\ \hline
SAC(4)& 0.6927&0.1609&0.6715&0.6013&0.1918&0.5391 \\ \hline
SAC(5)&0.6112&0.1375&0.5883&0.5144&0.1599& 0.4535 \\ \hline
$\V\left[u_t\right]$&0.1593&1.830$\times 10^{-7}$&0.1540&0.1015&7.406$\times 10^{-7}$&0.0922 \\ \hline
$\V[IV_t]/\V[RV_t^*]$&0.2525&0.9953&0.2681& 0.3435&0.9920&0.3620\\ \hline
$\V[u_t]/\V[RV_t^*]$&0.5087&5.856$\times 10^{-7}$&0.4918&0.3909&2.852$\times 10^{-6}$&0.3553\\ \hline
$\corr[r_t, \eps_t]$&0            &0        &-0.2588&0&0& -0.2240\\ \hline
\end{tabular}
\label{summary_1fiv1}
\end{threeparttable}
\begin{threeparttable}[H]
\begin{tabular}{|c|c|c|c|c|c|c|}
\hline
&\multicolumn{3}{|c|}{$m=288$}&\multicolumn{3}{|c|}{$m=144$}\\ \hline
model                   & NW&zero-f&weak-f&NW&zero-f&weak-f\\ \hline
Mean& 0.3372&0.4003&0.4002&0.3605&0.4026&0.4023\\ \hline
Variance&0.0564&0.2751& 0.0756&0.1031&0.2811&0.0488 \\ \hline
SAC(1)&0.9873&0.3562&0.9209&0.8980&0.2896&0.9673\\ \hline
SAC(2)&0.9584&0.2023&0.7606& 0.7190&0.1795& 0.8942\\ \hline
SAC(3)&0.9202&0.1701&  0.6087& 0.5722&0.1590&0.8085 \\ \hline
SAC(4)&0.8774&0.1393&0.4942&0.4745& 0.1246& 0.7254 \\ \hline
SAC(5)&0.8333&0.1207&0.4175&0.4128&0.1095&0.6522 \\ \hline
$\V\left[u_t\right]$& 0.1818& 6.491$\times 10^{-7}$& 0.1205&0.0746&3.340$\times 10^{-6}$&0.1810 \\ \hline
$\V[IV_t]/\V[RV_t^*]$&0.1982&0.9656&0.2653&0.1395&0.9298&0.1614\\ \hline
$\V[u_t]/\V[RV_t^*]$&0.6383&2.270$\times 10^{-6}$& 0.4231&0.7042&1.417$\times 10^{-6}$&0.5987\\ \hline
$\corr[r_t, \eps_t]$&0&0&-0.3377&0&0& -0.4822\\ \hline
\end{tabular}
\label{summary_1fiv2}
\end{threeparttable}
\end{center}
\begin{center}
\begin{threeparttable}[H]
\caption{IV basic statistics of 2-factor model(1)}
\begin{tabular}{|c|c|c|c|c|c|c|}
\hline
$m$&\multicolumn{3}{|c|}{$m=2880$}&\multicolumn{3}{|c|}{$m=1440$}\\ \hline
model                   & NW&zero-f&weak-f&NW&zero-f&weak-f\\ \hline
Mean & 0.3693&0.4175&0.4177&0.3758&0.4042&0.4043\\ \hline
Variance &0.0917&0.3114&0.0917&0.1031& 0.2572&0.1031\\ \hline
SAC(1) &0.9164&0.3516&0.9164&0.8980&0.4347&0.8980\\ \hline
SAC(2) &0.7621&0.2630&0.7621&0.7190&0.3170&0.7191\\ \hline
SAC(3) &0.6273&0.1917&0.6274&0.5722&0.2301&0.5723\\ \hline
SAC(4) &0.5325&0.1616&0.5326&0.4744& 0.1932&0.4746\\ \hline
SAC(5) &0.4713&0.1382&0.4715&0.4128&0.1611&0.4130\\ \hline
$\V\left[u_t\right]$ &0.1243&1.654$\times 10^{-6}$&0.1243&0.0747&5.649$\times 10^{-6}$&0.0746\\ \hline
$\V[IV_t]/\V[RV_t^*]$& 0.2927&0.9943&0.2927&0.3971&0.9905& 0.3971\\ \hline
$\V[u_t]/\V[RV_t^*]$&0.3971   &5.282$\times 10^{-6}$&0.3968&0.2875&2.175$\times 10^{-5}$&0.2873\\ \hline
$\corr[r_t, \eps_t]$&0               &0          &-0.3213&0&0&-0.2862\\ \hline
\end{tabular}
\label{summary_2fiv1}
\end{threeparttable}
\begin{threeparttable}[H]
\begin{tabular}{|c|c|c|c|c|c|c|}
\hline
$m$&\multicolumn{3}{|c|}{$m=288$}&\multicolumn{3}{|c|}{$m=144$}\\ \hline
model                   & NW&zero-f&weak-f&NW&zero-f&weak-f\\ \hline
Mean &0.3788&0.4003&0.4003&0.3733& 0.4026&0.4026\\ \hline
Variance &0.1016&0.2741&0.1012&0.0665&0.2778&0.0659\\ \hline
SAC(1) &0.7897&0.3571&0.7905&0.8720&0.2929&0.8739\\ \hline
SAC(2) &0.5431&0.2064&0.5445&0.6843&0.1796&0.6883\\ \hline
SAC(3) &0.4287&0.1737&0.4301& 0.5672&0.1590& 0.5716\\ \hline
SAC(4) &0.3771&0.1431&0.3784&0.5041&0.1246&0.5083\\ \hline
SAC(5) &0.3488&0.1244&0.3500&0.4702&0.1095&0.4743\\ \hline
$\V\left[u_t\right]$ &0.0721&7.127$\times 10^{-6}$&0.0722&0.1279&3.340$\times 10^{-6}$&0.1284\\ \hline
$\V[IV_t]/\V[RV_t^*]$&0.3566&0.9622&0.3554&0.2199&0.9188&0.2179\\ \hline
$\V[u_t]/\V[RV_t^*]$&0.2530&2.502$\times 10^{-5}$&0.2537&0.4229&1.105$\times 10^{-5}$&0.4247\\ \hline
$\corr[r_t, \eps_t]$&0&0&-0.3938&0&0&-0.5153\\ \hline
\end{tabular}
\label{summary_2fiv2}
\end{threeparttable}
\end{center}

In the zero-f model, $\V\left[u_t\right]$ is very small, and the result is like a parallel shift of NCRV by $c_u$, so that the auto-covariance is almost the same as that of the original NCRV. On the other hand, the values of the NW model and the weak-f model are similar, and their autocovariances after the first order are larger than those of the original NCRV. This is because the variance of NCRV is larger than that of IV due to the variation of $u_t$. $\V[IV_t]/\V[RV_t^*], \V[u_t]/\V[RV_t^*]$ is the ratio of biases due to IV and MN in NCRV. For $p=1$, the ratio of IV between the NW model and the weak-f model is less than half, and decreases with $m$ except for $m=1440$. For $p=2$, the ratio of IVs of the NW model and the weak-f model behaves similarly to that for $p=1$, but is slightly larger. For the zero-f model, the ratio of IV is almost unity. In other words, the variation of bias due to MN autocorrelation is quite small.  The $\corr[r_t, \eps_t]$ is obtained from the parameter estimates of the weak-f model by the Equation (\ref{covreps}) $\corr[r_t, \eps_t]=\cfrac{f(m)\E[\sigma(t)]}{\sqrt{m\ sigma^2\left(\cfrac{(f(m))^2}{m}+\Omega_0\right)}}$ is the correlation between returns and MN.  The absolute value of $p=2$ is larger than $p=1$. Furthermore, the absolute value decreases as $m$ increases except for $m=2880$.
  
The following figure\ref{result_of_model} shows the $\widehat{IV}_t$ estimation results for each model at each $m$ (top, solid line: $RV_t^{*(m)}$, dashed line: $\widehat{IV}_t$), $\hat{u}_t^{(m)}$ (middle), $\widetilde{IV}_t$ one period ahead (bottom).

\begin{figure}[H]
\begin{center}
\includegraphics[width=12cm,height=12.5cm]{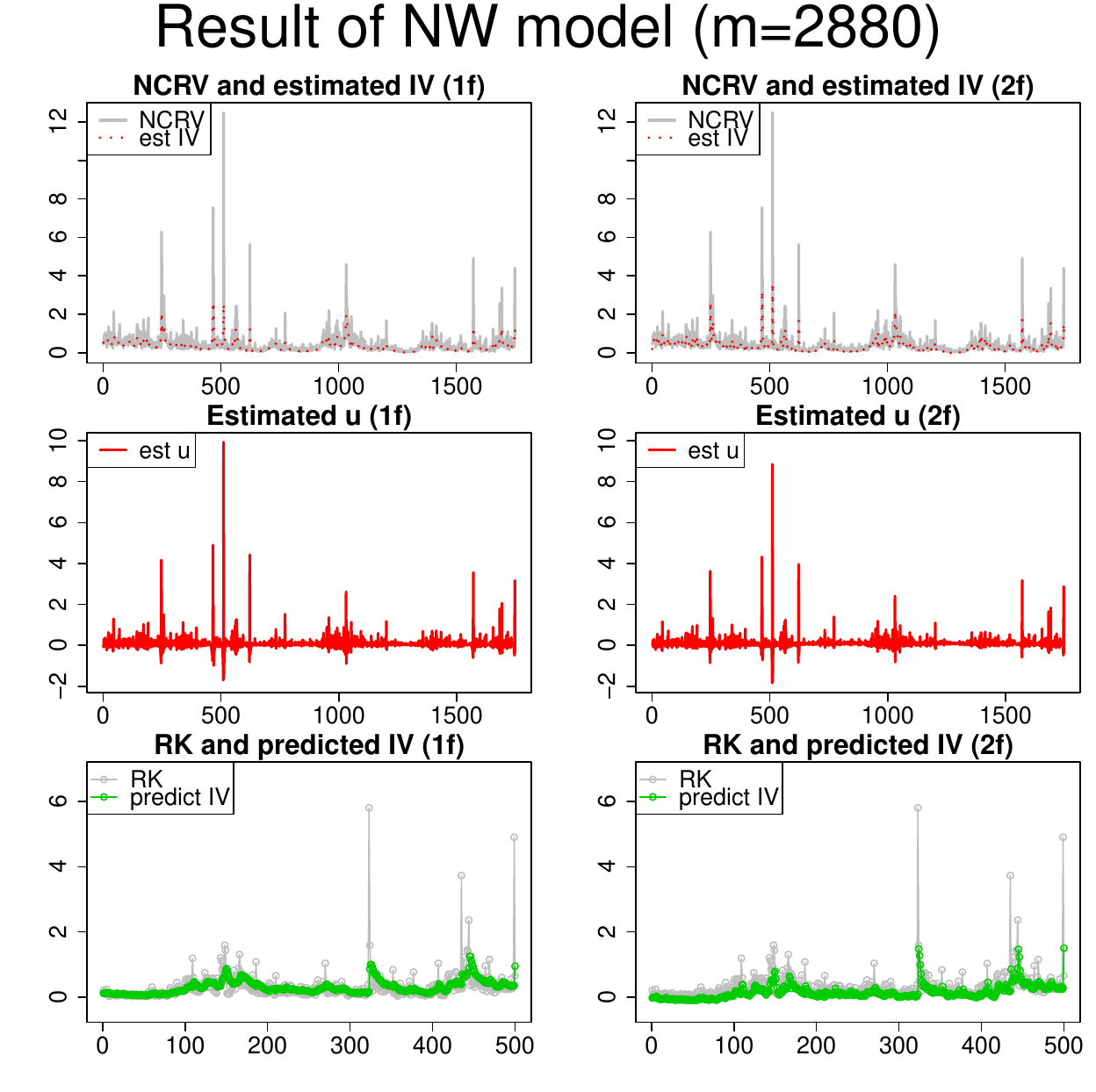}
%\subcaption{NW model ($m=2880$)}
\includegraphics[width=12cm,height=12.5cm]{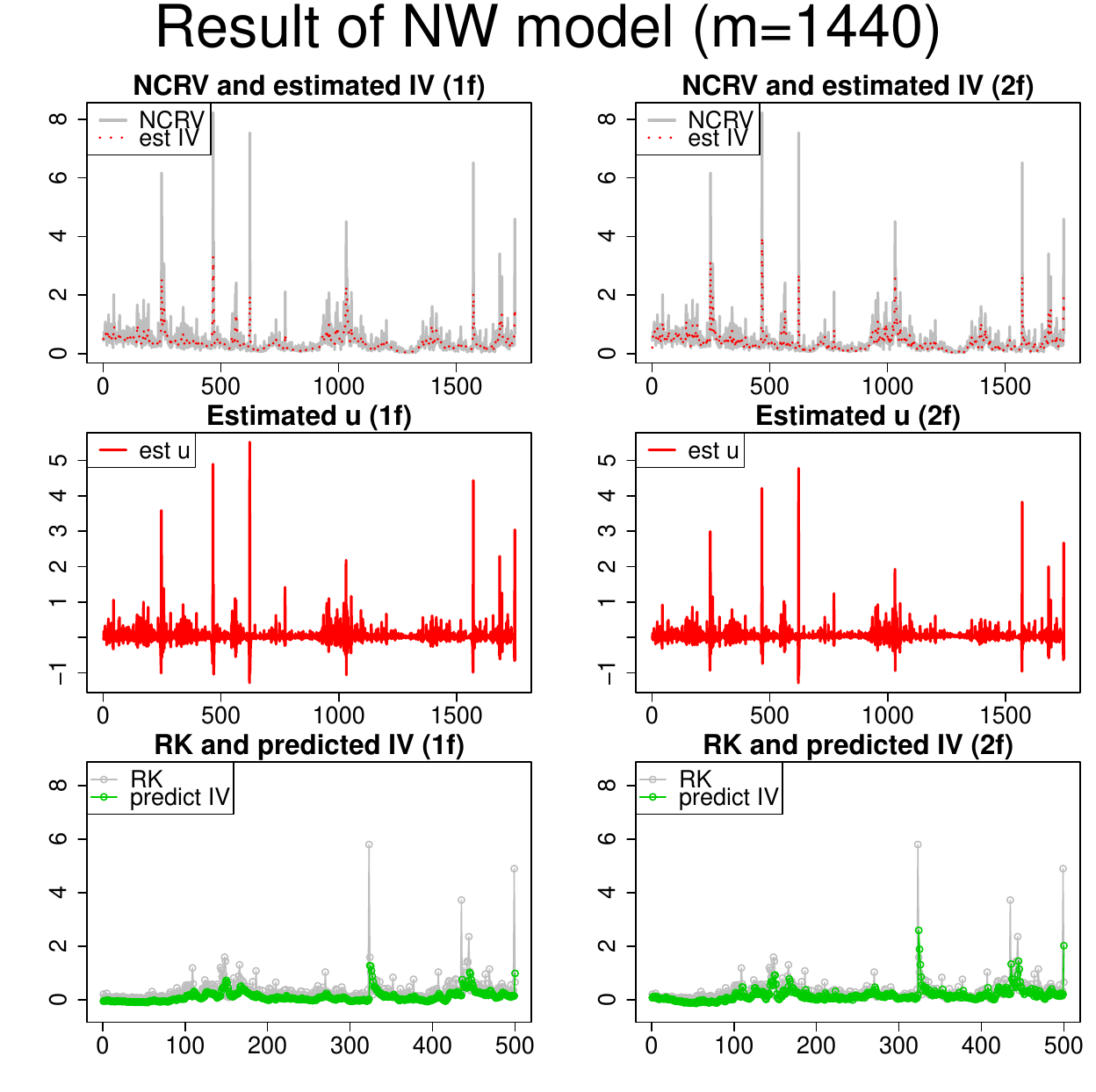}
%\subcaption{NW model ($m=1440$)}
\end{center}
\end{figure}
\newpage
\begin{figure}[H]
\begin{center}
\includegraphics[width=12cm,height=12.5cm]{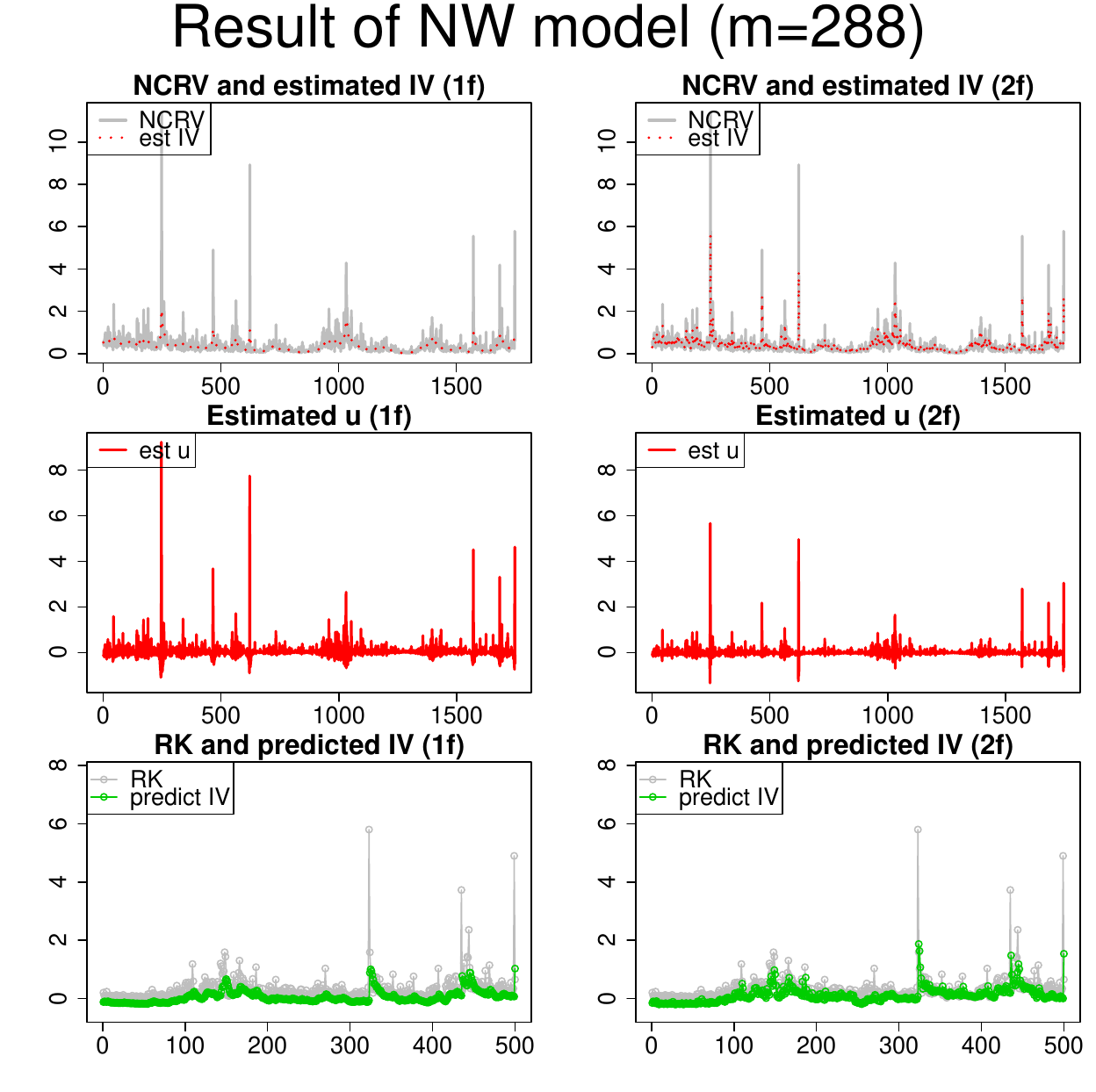}
%\subcaption{NW model ($m=288$)}
\includegraphics[width=12cm,height=12.5cm]{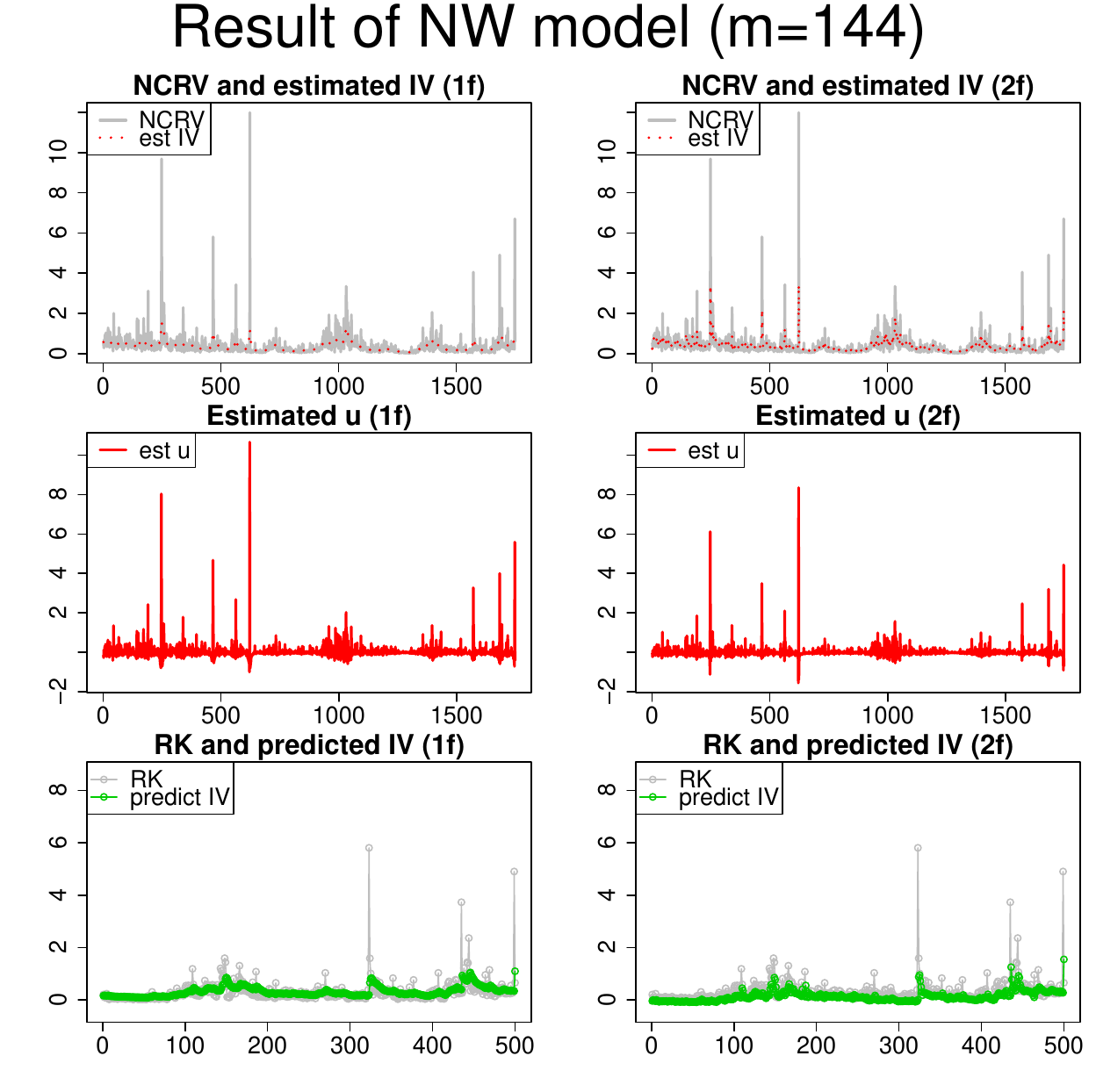}
%\subcaption{NW model ($m=144$)}
\end{center}
\end{figure}
\newpage%
\begin{figure}[H]
\begin{center}
\includegraphics[width=12cm,height=12.5cm]{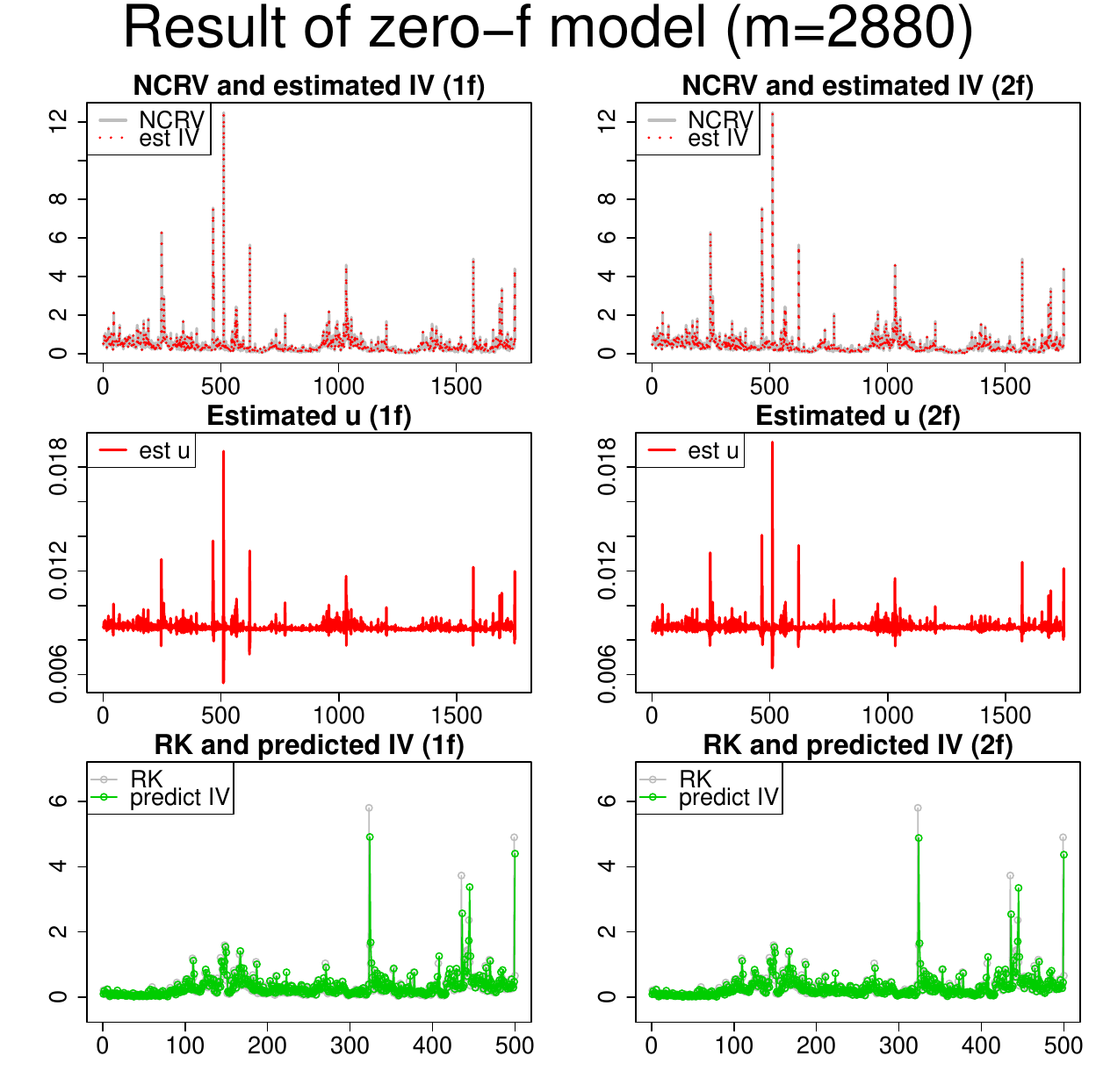}
%\subcaption{zero-f model ($m=2880$)}
\includegraphics[width=12cm,height=12.5cm]{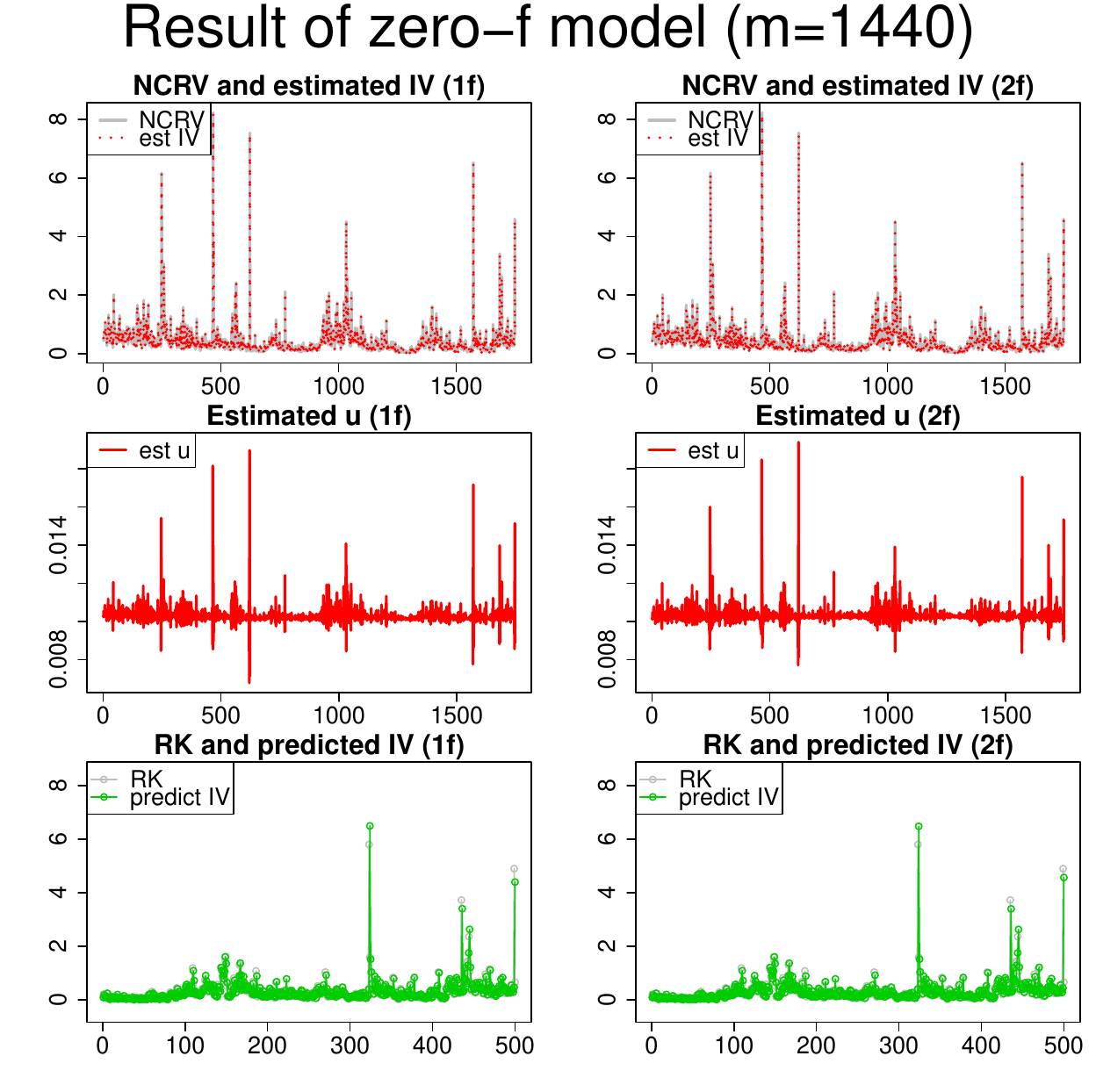}
%\subcaption{zero-f model ($m=1440$)}
\end{center}
\end{figure}
\newpage
\begin{figure}[H]
\begin{center}
\includegraphics[width=12cm,height=12.5cm]{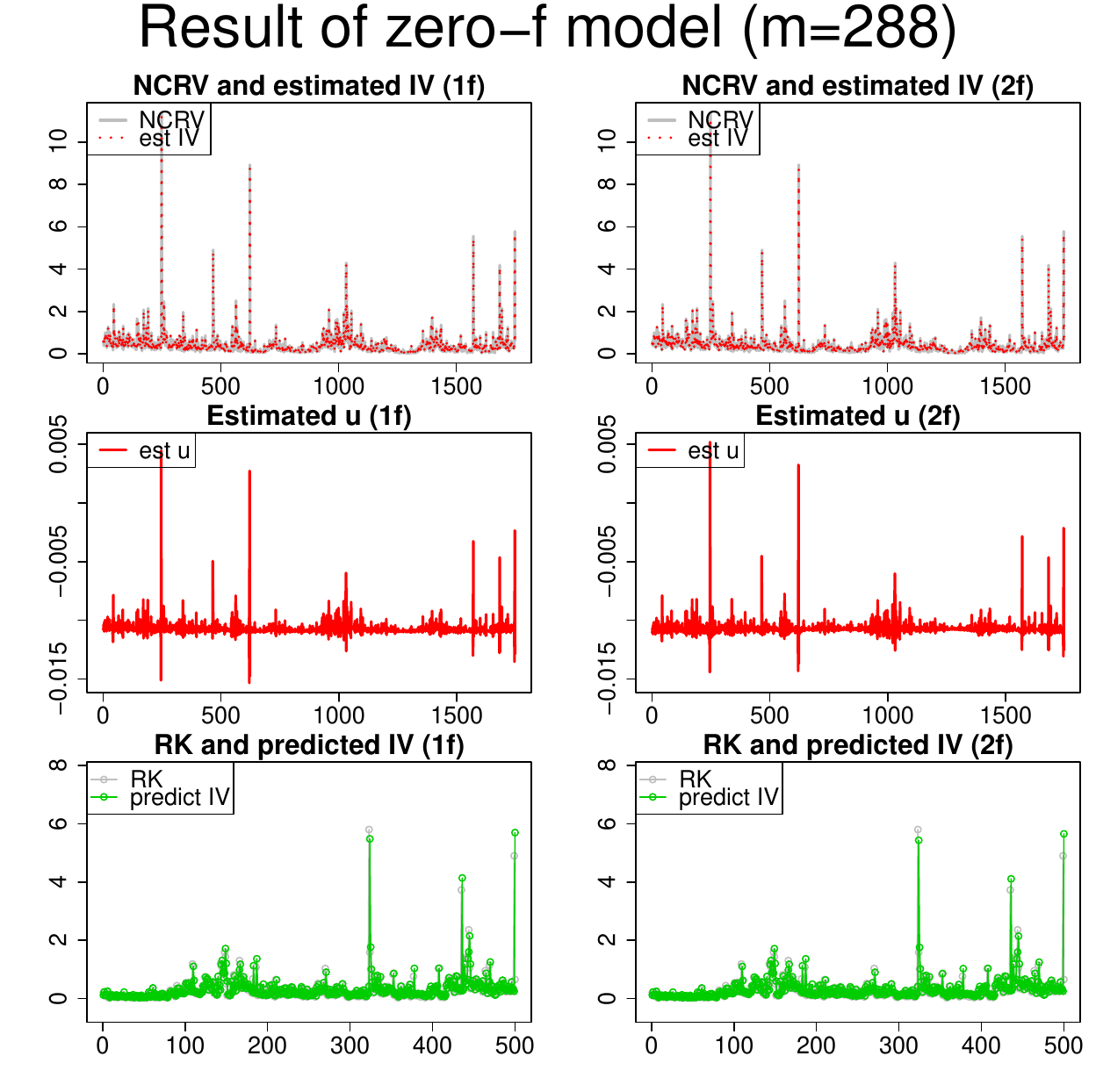}
%\subcaption{zero-f model ($m=288$)}
\includegraphics[width=12cm,height=12.5cm]{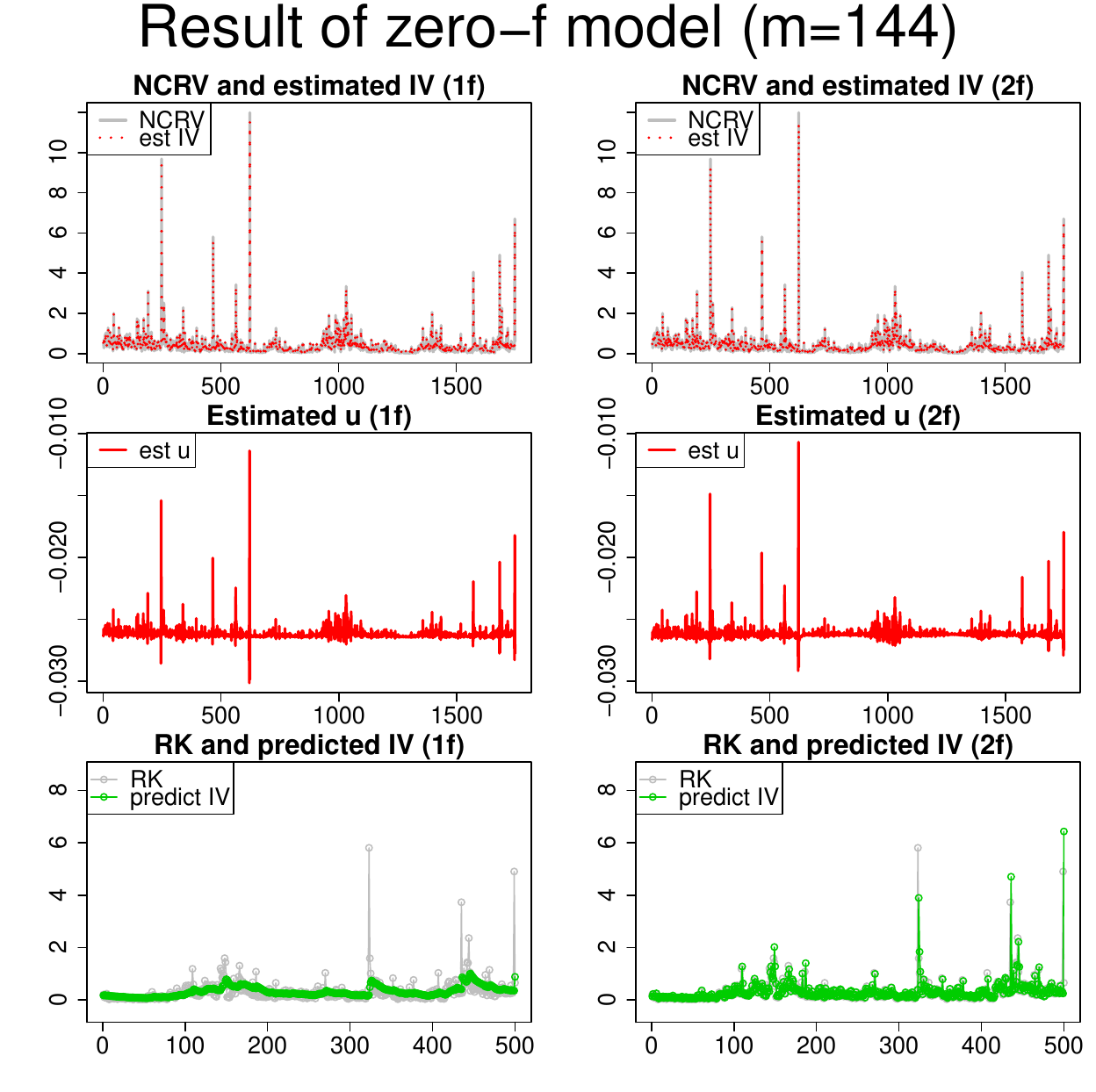}
%\subcaption{zero-f model ($m=144$)}
\end{center}
\end{figure}
\newpage%
\begin{figure}[H]
\begin{center}
\includegraphics[width=12cm,height=12.5cm]{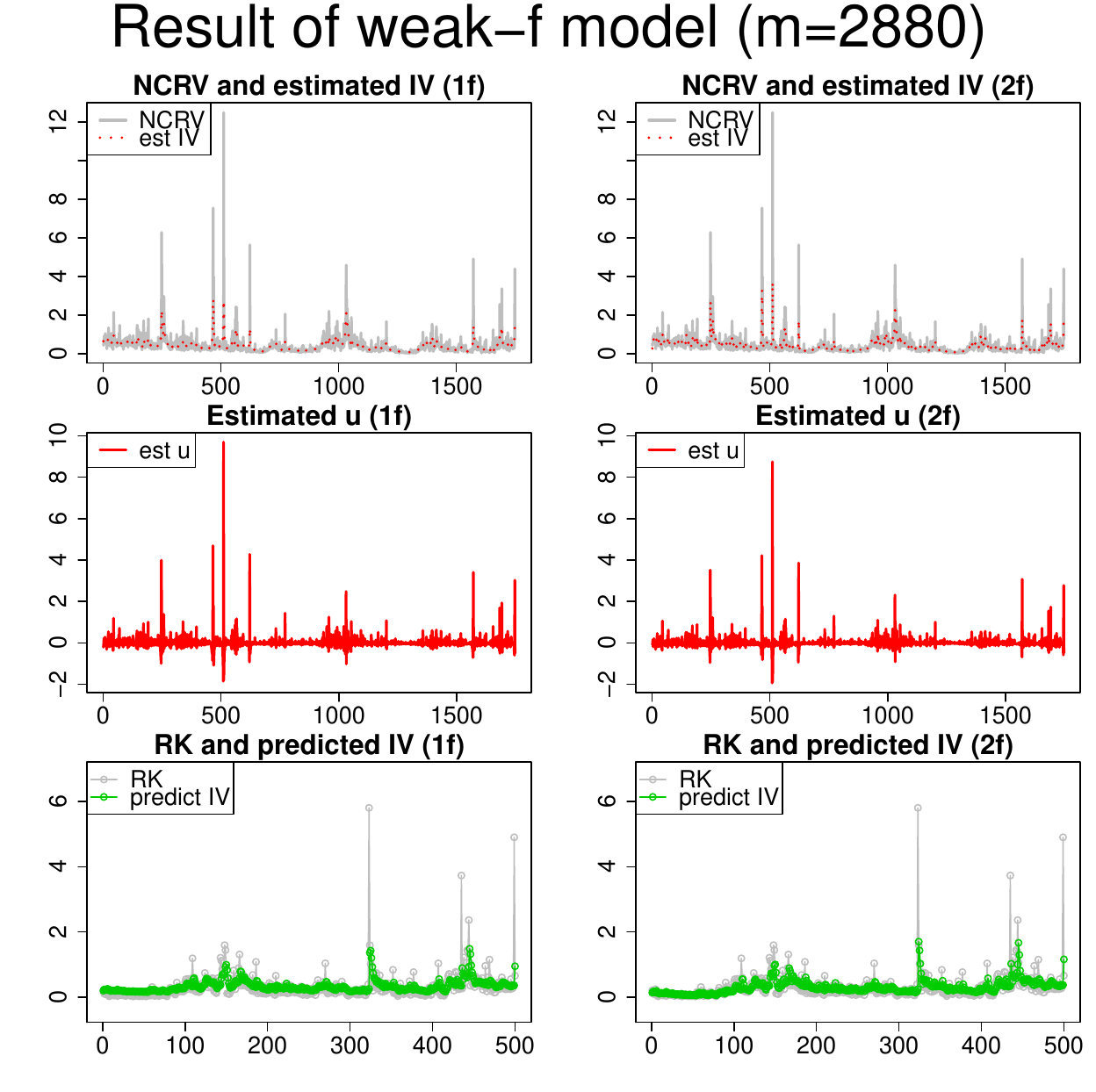}
%\subcaption{weak-f model ($m=2880$)}
\includegraphics[width=12cm,height=12.5cm]{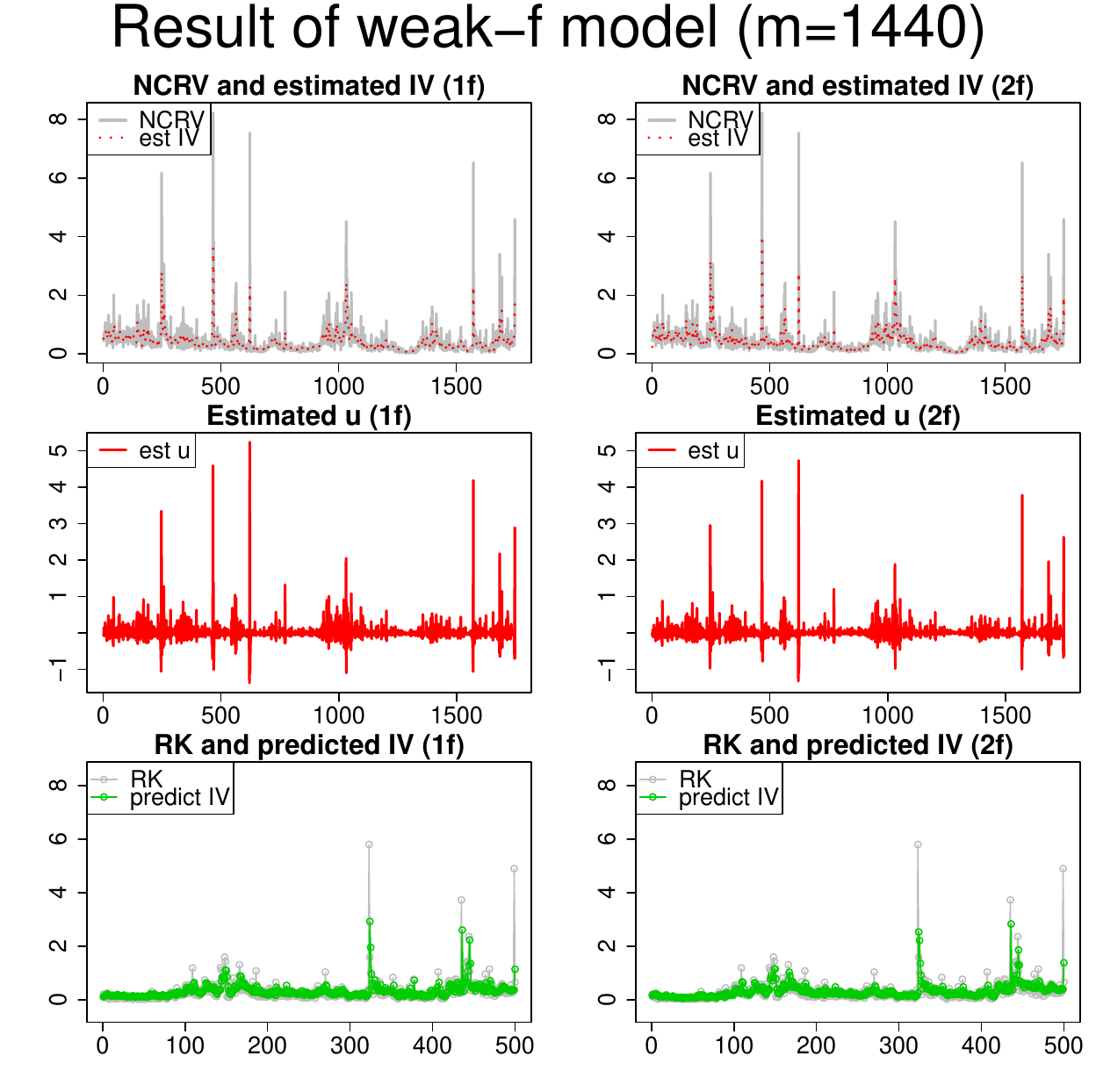}
%\subcaption{weak-f model ($m=1440$)}
\end{center}
\end{figure}
\newpage
\begin{figure}[H]
\begin{center}
\includegraphics[width=12cm,height=12.5cm]{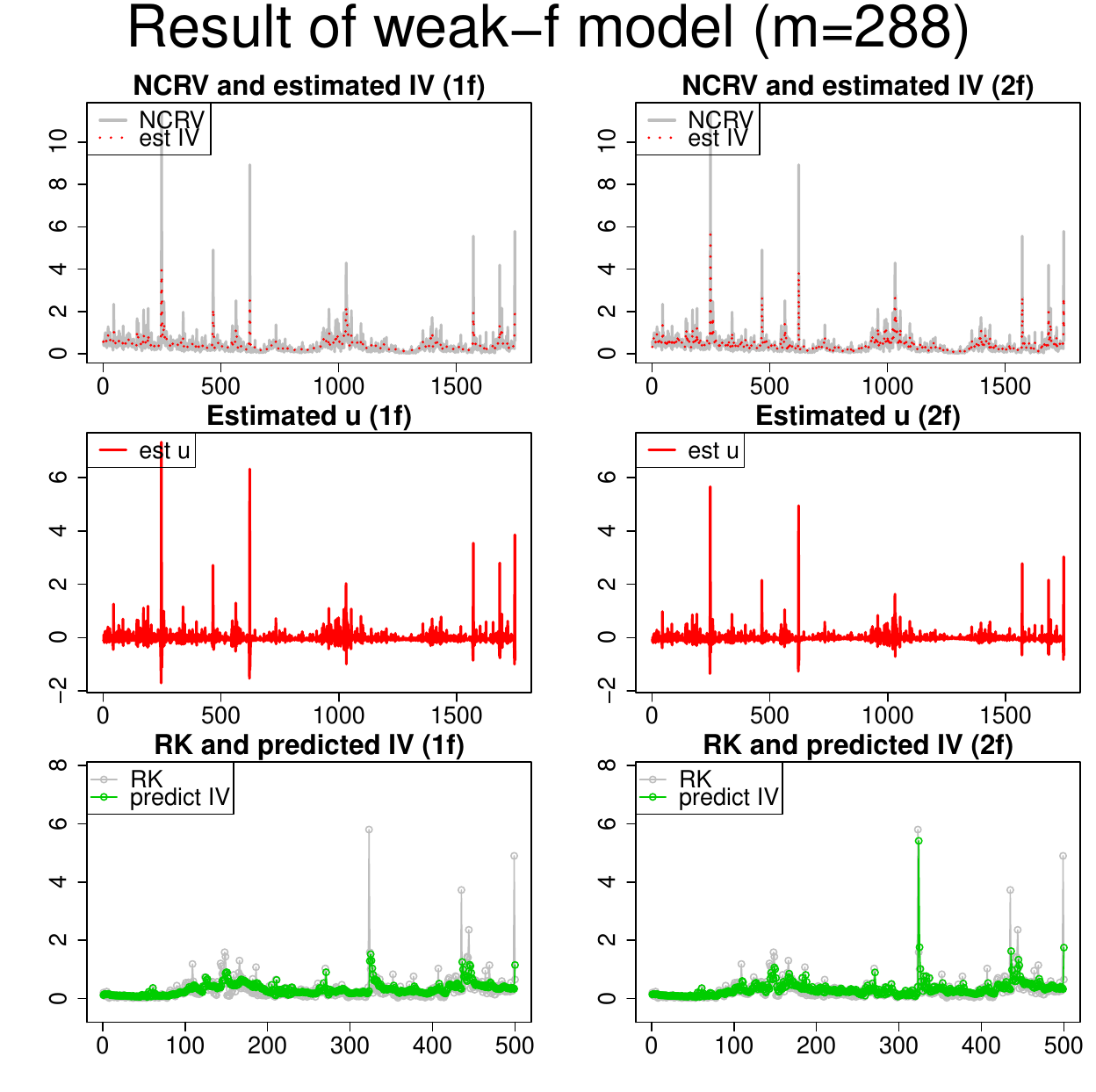}
%\subcaption{weak-f model ($m=288$)}
\includegraphics[width=12cm,height=12.5cm]{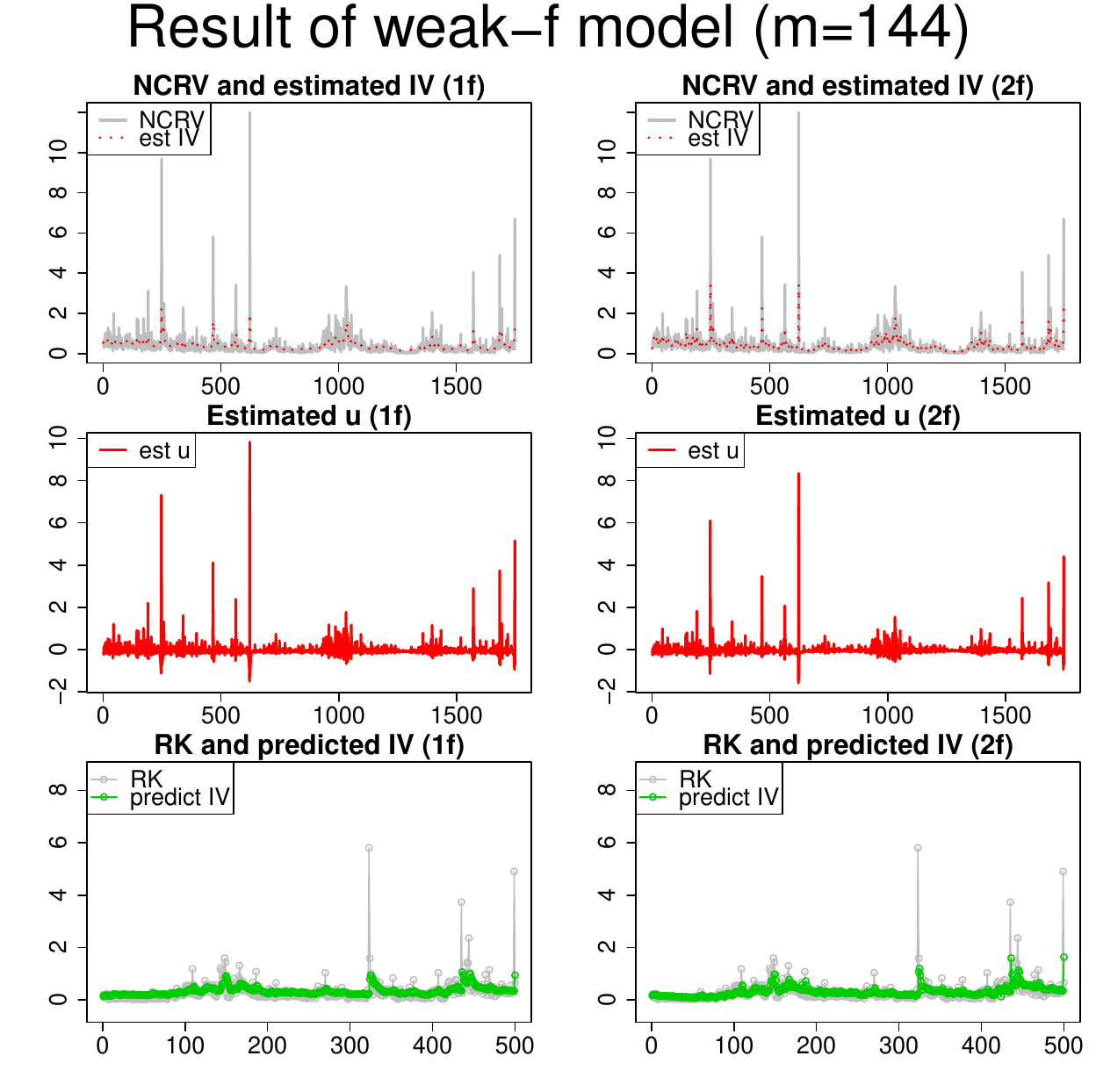}
%\subcaption{weak-f model ($m=144$)}
\end{center}
\caption{Estimation results for each model, one period ahead}
\label{result_of_model}
\end{figure}

The following table shows the IV estimation and one period ahead forecasting errors for each model with $IV_t=RK_t ^{(2880)}$ for each model. The best value is marked with \textcolor{red}{*}, the second and third best values are marked with \textcolor{red}{**} and \textcolor{red}{***}, respectively.

\begin{table}[H]
\begin{center}
\caption{$m=2880$:IV estimation error($IV_t=RK_t^{(2880)}$)}
\begin{tabular}{|c|c|c|c|c|c|c|}
\hline
&NW(1)&zero-f(1)&weak-f(1)&NW(2)&zero-f(2)&weak-f(2)\\ \hline
MSE &0.1513&0.1029\textcolor{red}{**}&0.1462&0.1311 &0.1028\textcolor{red}{*}& 0.1269\textcolor{red}{***}\\ \hline
QLIKE&-0.1475&-0.2893\textcolor{red}{**}&-0.1866&0.2473&-0.2982\textcolor{red}{*}&-0.2053 \textcolor{red}{***}\\ \hhline{|=======|}
\multicolumn{7}{|c|}{Mincer-Zarnowiz Regression}\\ \hline
$R^2$    &0.4614&0.6822\textcolor{red}{**}&0.4710&0.5529&0.6823\textcolor{red}{*}&0.5530 \textcolor{red}{***}\\  \hline
$\hat{a}$&-0.0304&0.0682&-0.1236&-0.0121&0.0680&-0.1430\\
std error& 0.0141&0.0087&0.0157&0.0119&0.0087&0.0140\\ \hline
$\hat{b}$&1.248&0.7645&1.224&1.269&0.7650&1.269\\
std error&0.0322&0.0124&0.0310&0.0279&0.0125&0.0273\\ \hline
\end{tabular}
\label{2880_er}
\end{center}
\begin{center}
\caption{$m=1440$:IV estimation error($IV_t=RK_t^{(2880)}$)}
\begin{tabular}{|c|c|c|c|c|c|c|}
\hline
&NW(1)&zero-f(1)&weak-f(1)&NW(2)&zero-f(2)&weak-f(2)\\ \hline
MSE &0.1174&0.0107\textcolor{red}{*}&0.1079&0.0900 \textcolor{red}{***}&0.0107\textcolor{red}{*}&0.0896\textcolor{red}{**}\\ \hline
QLIKE&-0.2217&-0.2959\textcolor{red}{*}&-0.2143&-0.2259&-0.2958\textcolor{red}{**}&-0.2309\textcolor{red}{***} \\ \hhline{|=======|}
\multicolumn{7}{|c|}{Mincer-Zarnowiz Regression}\\ \hline
$R^2$    & 0.5995&0.9609\textcolor{red}{*}&0.6390& 0.7156 \textcolor{red}{***}&0.9608\textcolor{red}{**}&0.7155\\  \hline
$\hat{a}$&-0.1074&-0.0160&-0.1571&-0.1034& -0.0162&-0.1630\\
std error&0.0124&0.0031& 0.0122&0.0099&0.0031& 0.0106\\ \hline
$\hat{b}$& 1.340&0.9981&1.347&1.361&0.9988&1.361\\
std error&0.0261& 0.0048&0.0242&0.0205& 0.0048&0.0205\\ \hline
\end{tabular}
\label{1440_er}
\end{center}
\begin{center}
\caption{$m=288$:IV estimation error($IV_t=RK_t^{(2880)}$)}
\begin{tabular}{|c|c|c|c|c|c|c|}
\hline
&NW(1)&zero-f(1)&weak-f(1)&NW(2)&zero-f(2)&weak-f(2)\\ \hline
MSE &0.1436&0.0198 \textcolor{red}{**}&0.1139&0.0400 \textcolor{red}{***}&0.0197 \textcolor{red}{*}&0.0728\\ \hline
QLIKE&-0.1905& -0.2965 \textcolor{red}{**}&-0.1999&-0.2193&-0.2970 \textcolor{red}{*}&-0.2360 \textcolor{red}{***}\\ \hhline{|=======|}
\multicolumn{7}{|c|}{Mincer-Zarnowiz Regression}\\ \hline
$R^2$    &0.5101&0.9290 \textcolor{red}{**}&0.6483&0.8708 \textcolor{red}{***}&0.9291 \textcolor{red}{*}&0.8073\\  \hline
$\hat{a}$&-0.1797&0.0072&-0.2182&-0.0212&0.0066&-0.1966\\
std error&0.0158&0.0041& 0.0129& 0.0058&0.0041&0.0087\\ \hline
$\hat{b}$&1.445&0.9496&1.513&1.158&0.9514&1.459\\
std error&0.0338& 0.0062& 0.0266&0.0106&0.0063&0.0170\\ \hline
\end{tabular}
\label{288_er}
\end{center}
\begin{center}
\caption{$m=144$:IV estimation error($IV_t=RK_t^{(2880)}$)}
\begin{tabular}{|c|c|c|c|c|c|c|}
\hline
&NW(1)&zero-f(1)&weak-f(1)&NW(2)&zero-f(2)&weak-f(2)\\ \hline
MSE &0.1830&0.0284\textcolor{red}{**}&0.1539&0.1106\textcolor{red}{***}&0.0279\textcolor{red}{*}&0.1109\\ \hline
QLIKE&-0.1670&-0.2852\textcolor{red}{**}&-0.1639&-0.2205\textcolor{red}{***}&-0.2867\textcolor{red}{*}&-0.2006\\ \hhline{|=======|}
\multicolumn{7}{|c|}{Mincer-Zarnowiz Regression}\\ \hline
$R^2$    &0.3580&0.9002\textcolor{red}{**}&0.5034&0.7039\textcolor{red}{***}& 0.9009\textcolor{red}{*}&0.7011\\  \hline
$\hat{a}$&-0.1481& 0.0151&-0.2804&-0.2403&0.0128&-0.2913\\
std error& 0.0198&0.0049&0.0180&0.0118&0.0048&0.0125\\ \hline
$\hat{b}$&1.506&0.9247&1.660&1.681&0.9306&1.686\\
std error&0.0482&0.0073&0.0394&0.0260&0.0073&0.0263\\ \hline
\end{tabular}
\label{144_er}
\end{center}
\end{table}
From the table \ref{2880_er}, \ref{1440_er}, \ref{288_er}, \ref{144_er}, we know the following.
\begin{itemize}
\item IV estimation error, zero-f(2) and zero-f(1) are in first and second place.
\item NW(2) and weak-f(2) are often in the third place, but the scores of NW(2) and weak-f(2) are close. For $m=2880, 1440$, weak-f(2) tends to be dominant, and for $m=288, 144$, NW tends to be dominant.
\item $p=1$, weak-f(1) is more dominant than NW(1) in many cases.
\item zero-f(1) and zero-f(2) have almost the same value.
\end{itemize}
For IV estimation, zero-f(2) seems to be the most dominant. 
\begin{table}[H]
\begin{center}
\caption{$m=2880$:IV 1-ahead prediction error($IV_t=RK_t^{(2880)}$)}
\begin{tabular}{|c|c|c|c|c|c|c|}
\hline
&NW(1)&zero-f(1)&weak-f(1)&NW(2)&zero-f(2)&weak-f(2)\\ \hline
MSE &0.1778 \textcolor{red}{***}&0.2458& 0.1746\textcolor{red}{**}&0.1992&0.2449& 0.1720\textcolor{red}{*}\\ \hline
QLIKE&-0.3113\textcolor{red}{**}&-0.3064 \textcolor{red}{***}&-0.2931&25.14&-0.2342&-0.3331\textcolor{red}{*}\\ \hhline{|=======|}
\multicolumn{7}{|c|}{Mincer-Zarnowiz Regression}\\ \hline
$R^2$    & 0.1198&0.1159& 0.1283 \textcolor{red}{***}&0.1414\textcolor{red}{**}&0.1155&0.1435\textcolor{red}{*}\\  \hline
$\hat{a}$&0.0794& 0.1893&0.0126&0.1933&0.1957& 0.05719\\
std error&0.0337&0.0241&0.0394&0.02270& 0.0236& 0.0333\\ \hline
$\hat{b}$&0.7722&0.3687&0.8989&0.7846&0.3692&0.8330\\
std error&0.0929&0.0452&0.1041&0.0860&0.0453& 0.0905\\ \hline
\end{tabular}
\label{2880_erp}
\end{center}
\begin{center}
\caption{$m=1440$:IV 1-ahead prediction error($IV_t=RK_t^{(2880)}$)}
\begin{tabular}{|c|c|c|c|c|c|c|}
\hline
&NW(1)&zero-f(1)&weak-f(1)&NW(2)&zero-f(2)&weak-f(2)\\ \hline
MSE &0.2150&0.2730&0.1874\textcolor{red}{*}&0.2039 \textcolor{red}{***}&0.2760& 0.1881\textcolor{red}{**}\\ \hline
QLIKE&10.56&-0.1957 \textcolor{red}{***}&-0.3248\textcolor{red}{**}&1.739&-0.1547&-0.3365\textcolor{red}{*}\\ \hhline{|=======|}
\multicolumn{7}{|c|}{Mincer-Zarnowiz Regression}\\ \hline
$R^2$    &0.1307\textcolor{red}{*}&0.10851&0.1210&0.1242 \textcolor{red}{***}&0.1063&0.1243\textcolor{red}{**}\\  \hline
$\hat{a}$&0.2193&0.2067&0.1184&0.2059&0.2085&0.1256\\
std error&0.0215&0.0232&0.0298&0.0226&0.0232&0.0289\\ \hline
$\hat{b}$&0.8403&0.3251&0.5902&0.6211&0.3200&0.5786\\
std error&0.0963&0.0414&0.0706&0.0733&0.0411&0.0682\\ \hline
\end{tabular}
\label{1440_erp}
\end{center}
\begin{center}
\caption{$m=288$:IV 1-ahead prediction error($IV_t=RK_t^{(2880)}$)}
\begin{tabular}{|c|c|c|c|c|c|c|}
\hline
&NW(1)&zero-f(1)&weak-f(1)&NW(2)&zero-f(2)&weak-f(2)\\ \hline
MSE &0.2429&0.2922& 0.1785\textcolor{red}{*}&0.2230 \textcolor{red}{***}&0.2898&0.2046\textcolor{red}{**}\\ \hline
QLIKE&4.489&-0.3067&-0.3176\textcolor{red}{**}&5.121&-0.3074 \textcolor{red}{***}&-0.3424\textcolor{red}{*}\\ \hhline{|=======|}
\multicolumn{7}{|c|}{Mincer-Zarnowiz Regression}\\ \hline
$R^2$    &0.1252\textcolor{red}{*}&0.0851&0.1185 \textcolor{red}{***}&0.1208\textcolor{red}{**}&0.0860&0.1104\\  \hline
$\hat{a}$&0.2697&0.2172&0.0686&0.2403&0.2163&0.1564\\
std error&0.0193&0.0236&0.0350&0.0206&0.02367&0.0272\\ \hline
$\hat{b}$& 0.7836&0.2852&0.7527&0.6125&0.2883&0.4770\\
std error&0.0920& 0.0414&0.0912&0.0734&0.0416&0.0601\\ \hline
\end{tabular}
\label{288_erp}
\end{center}
\begin{center}
\caption{$m=144$:IV 1-ahead prediction error($IV_t=RK_t^{(2880)}$)}
\begin{tabular}{|c|c|c|c|c|c|c|}
\hline
&NW(1)&zero-f(1)&weak-f(1)&NW(2)&zero-f(2)&weak-f(2)\\ \hline
MSE &0.1763\textcolor{red}{*}&0.1785&0.1768\textcolor{red}{**}&0.2074&0.2970&0.1772\textcolor{red}{***}\\ \hline
QLIKE&-0.3113\textcolor{red}{***}& -0.3044&-0.2749&7.029&-0.3157\textcolor{red}{*}&-0.3120\textcolor{red}{**}\\ \hhline{|=======|}
\multicolumn{7}{|c|}{Mincer-Zarnowiz Regression}\\ \hline
$R^2$    &0.1199\textcolor{red}{**}&0.1111&0.1177&0.1213\textcolor{red}{*}&0.0743&0.1185\textcolor{red}{***}\\  \hline
$\hat{a}$&0.0402&0.0518&-0.0340&0.2011&0.2224&0.0517\\
std error&0.0377& 0.0378&0.0461&0.0230& 0.0238&0.0367\\ \hline
$\hat{b}$&0.8547&0.8137&1.017&0.8135&0.2695&0.8098\\
std error&0.1028&0.1022&0.1237&0.0973&0.0420&0.0981\\ \hline
\end{tabular}
\label{144_erp}
\end{center}
\end{table}
From the tables \ref{2880_erp}, \ref{1440_erp}, \ref{288_erp}, and \ref{144_erp}, we know the following.
\begin{itemize}
\item weak-f(1) and weak-f(2) are often ranked 1st and 2nd.
\item NW(1) and NW(2) have the next best scores in terms of the number of times they are ranked first and second, and NW(2) in terms of the number of times they are ranked third, but their QLIKE is often quite large.
\item The scores of zero-f(1) and zero-f(2) are not good.
\end{itemize}
For IV one period ahead forecasts, weak-f(2) seems to be superior.

The zero-f model has by far the best score in IV estimation, but the one-period ahead forecasting is quite poor. The weak-f model is superior in one period ahead prediction, and the score of the weak-f model is not bad in IV estimation. The NW model has a score similar to that of the weak-f model in IV estimation, but is clearly inferior in one period ahead prediction.

\section{Conclusion}
 In this paper, we relaxed the assumptions of Nagakura and Watanabe[2009, 2015], and extended their model assuming Hansen and Lunde[2006] type dependent MN that auto-correlate and correlate with return. Under these assumptions, we derived  the properties of the bias due to MN $u_t^{(m)}$. 
Based on these, we proposed the parameters of $u_t^{m}$ and dependent MN estimation method from NCRV and observation return data. 

In model comparison by simulation, proposal model was superior to the existing method in the accuracy of parameter estimation, IV estimation and 1-ahead prediction.

In  model comparison by actual data on the dollar-yen rate, we newly found the following
\begin{enumerate}
\item The variance of the MN is smaller for larger $m$.
\item The persistence of the autocorrelation of the MN is longer for larger $m$.
\item For data with observation intervals longer than a minute, the autocorrelation of each order is larger for larger $m$. 
\item The correlation between MN and returns is negative, and the absolute value of the correlation is smaller for larger $m$ for data with observation intervals longer than minutes. 
\item The mean of the bias due to MN is larger for larger $m$, but its variance is not.
\item The variation of the bias due to the autocorrelation of the MNs is quite small.
\end{enumerate}

\end{document}